\documentclass[12pt, a4paper]{article}
\usepackage{authblk}
\usepackage{float}
\usepackage{tabls}
\usepackage{graphicx}
\usepackage{parskip}
\usepackage{amsmath,amsfonts}
\usepackage{setspace} %for setting line spacing

\usepackage{lscape} %for using landscape 
\usepackage[comma]{natbib} %for more flexible referencing

\usepackage{url} %for website addresses
\usepackage{dcolumn} %for decimal point alignment
\usepackage{booktabs} %for professionaly looking tables
\usepackage{tabularx} % for creating tables
\usepackage{multirow} % columns spanning multiple rows
\usepackage[in]{fullpage} % for smaller margins
\usepackage[multiple, stable, bottom]{footmisc}% for multiple footnotes at the same place
\usepackage{appendix} % to create separate appendices for each section
\usepackage{titlesec}
\usepackage{changepage}
\usepackage{xcolor}
\usepackage{longtable,pdflscape,booktabs} % to extend tables over multiple pages
\usepackage{enumerate}

\usepackage[textsize=tiny]{todonotes} % todos and comments

\newcolumntype{d}[1]{D{.}{\cdot}{#1}}
\newcolumntype{.}{D{.}{.}{-1}}
\newcolumntype{,}{D{,}{,}{2}}
{\def\sym#1{\ifmmode^{#1}\else\(^{#1}\)\fi}
\raggedbottom % To prevent white spaces between paragraphs

\widowpenalty=10000
\clubpenalty=10000
\setcounter{page}{1}

\title{THE ECONOMIC EFFECT OF GAINING A NEW QUALIFICATION IN LATER LIFE\thanks{\scriptsize{Corresponding author: Anna Zhu, RMIT University. Email: anna.zhu@rmit.edu.au. \newline \newline We thank Tim Robinson, Hayley Fisher, Bruce Bradbury and numerous seminar and conference participants for helpful comments. The authors would like to thank Tessa Loriggio, Prabath Abeysekara, Michael Duffield, and Yin-King Fok for their excellent research assistance. \newline \newline Zhu acknowledges the support of the Australian Research Council Linkage Project (LP170100472). This paper uses unit record data from the Household, Income and Labour Dynamics in Australia (HILDA) Survey. The HILDA Project was initiated and is funded by the Australian Government Department of Social Services (DSS), and is managed by the Melbourne Institute of Applied Economic and Social Research (Melbourne Institute). The findings and views reported in this paper, however, are those of the authors and should not be attributed to either DSS or the Melbourne Institute.}}} \vspace{0.2cm} \author{Finn Lattimore$^1$\thanks{\scriptsize{This work was performed while the author was working at the Gradient Institute. Views expressed in this paper are those of the author and do not reflect those of the Reserve Bank of Australia.}}\vspace{0.2cm} ,  Daniel Steinberg$^2$ and Anna Zhu$^3$ \\\small {$^1$Reserve Bank of Australia} \vspace{0.2cm} \\ {$^2$Gradient Institute} \vspace{0.2cm} \\ {$^3$RMIT University, IZA}}

\begin{document}

\defcitealias{department2021}{DESE, 2021}
\defcitealias{department2019a}{DESE, 2019a}
\defcitealias{department2019b}{DESE, 2019b}
\defcitealias{department2022a}{DESE, 2022a}
\defcitealias{department2022b}{DESE, 2022b}

\maketitle

\begin{abstract}
  
Pursuing educational qualifications later in life is an increasingly common
phenomenon within OECD countries since technological change and automation
continues to drive the evolution of skills needed in many professions. We focus
on the causal impacts to economic returns of degrees completed later in life,
where motivations and capabilities to acquire additional education may be
distinct from education in early years. We find that completing an additional
degree leads to more than \$3000 (AUD, 2019) extra income per year compared to
those who do not complete additional study. For outcomes, treatment and
controls we use the extremely rich and nationally representative longitudinal
data from the Household Income and Labour Dynamics Australia survey (HILDA). To
take full advantage of the complexity and richness of this data we use a
Machine Learning (ML) based methodology for causal effect estimation. We are
also able to use ML to discover sources of heterogeneity in the effects of
gaining additional qualifications. For example, those younger than 45 years of
age when obtaining additional qualifications tend to reap more benefits (as
much as \$50 per week more) than others.

\end{abstract}



\clearpage
\onehalfspacing

\section{Introduction}

Pursuing educational qualifications later in life is an increasingly common
phenomenon within OECD countries \citep{oecd2016}. Technological change and
automation continues to drive the evolution of skills needed in many
professions, or to oust the human workforce in others. This is particularly
true for middle-income workers performing routine tasks
\citep{autor2008,acemoglu2011}. Also at the lower end of the
income-distribution, such as among welfare recipients, governments are
increasingly trying to promote the idea of life-long learning.

This paper contributes to understanding one efficacy dimension of these policy
and individual choices by estimating the causal effects on earnings and by
focusing on mature-age students. We add to previous work on the returns to
education for `younger students'. Previous research points to positive and
significant wage premiums for younger cohorts with more education, ranging
between 5 and 13\% \citep{angrist1991,harmon2003,machin2006} or even higher
than 15\% as in the case of \cite{harmon1995}. The wage returns to education
may be more uncertain for older students as they face higher opportunity costs
to study and need to navigate a more fragmented system in the postsecondary
education setting.

We also add to the literature that investigates the economic returns for
mature-age learners at community or training colleges \citep{jacobson2005, chesters2015, 
zeiden2015, polidano2016, xu2016, belfield2017, dynarski2016, dynarski2018,
mountjoy2022}. The evidence on the labour market returns to vocational and
community college education is strong and positive, particularly for female
students \citep{belfield2017, zeiden2015, perales2017}. The results are even
stronger once authors account for the different earnings-growth profiles of
students and non-students before undertaking the degree \citep{dynarski2016,
dynarski2018}.

By focusing on the one institutional setting -- the community or training
college -- the results of such studies may not be generalisable to the entire
mature-age education market, such as to students who seek different degree
types or who study at different institutions \citep{belfield2017model,
mountjoy2022}. We add to this literature by estimating the returns across all
formal degree-types (post-graduate degrees, training certificates, diplomas
etc), and spanning all subjects and institutions at which the study took place.
This means we analyse the effects for a group of students with a larger span of
demographic and socio-economic background characteristics. The broad remit of
students that we analyse also allows our study to compliment studies that
evaluate government-run training programs, which tend to enrol low-productivity
workers \citep{ashenfelter1978, ashenfelter1985, bloom1990, leigh1990,
raaum2002, jacobson2005, card2018, knaus2022}. 

We contribute the first evidence in systematically identifying which groups of
mature-age students tend to benefit more from further education. We also
compliment previous studies that already find significant heterogeneity by
degree-type, institutional setting, and by the background characteristics of
the student \citep{blanden2012, zeiden2015, polidano2016, dorsett2016, xu2016,
belfield2017,perales2017, bockerman2019}. A benefit of a systematic,
data-driven approach to heterogeneity analysis is that it can reduce the risk
of overlooking important sub-populations compared to less data-driven
approaches \citep{athey2017, knaus2021}. 

A key challenge in estimating the causal returns to later-life education is
that factors that enable mature-age learners to pursue and complete a
qualification may also be precursors to later-life success. Moreover, the
drivers of degree completion may be numerous and related to other variables in
complex, unknown ways. We use a machine learning (ML) based methodology in this
work since it allows us to intensively control for many confounding factors, as
well as discover sources of treatment heterogeneity. ML algorithms also
automatically discover nonlinear relationships that may be unknown to the
researcher. For high-dimensional and complex datasets such as we use in this
research, these methodological abilities are crucial in
reducing bias from model mis-specification and confounding (e.g.~selection into
treatment), and reducing variance from correlation/collinearity.  

We adapt ML tools for causal inference purposes. We recognise that, as with all statistical models, we make assumptions when we use ML techniques for causal inference, and these need to be tested. One key assumption is that the controls included in the ML models sufficiently account for selection into treatment. We propose to undertake a replication exercise where we compare the results of the ML model with that of baseline models, using Ordinary Least Squares (OLS) and Fixed Effects. We also contrast the selected control variables in the ML model with those that were manually selected in \cite{chesters2015}, and comment on the potential biases from manual variable selection. We have chosen this published work because it uses the same data (HILDA) and examines the same topic.

The results show that an additional degree in later-life increases total future
earnings by more than an average of \$3,000 per year compared to those who do
not complete any further study. We consistently estimate this causal effect
using a selection-on-observables strategy based on T-learner, Doubly Robust and
Bayesian models. The estimate is based on 19 years of detailed nationally
representative Australian data from the Household Income and Labour Dynamics
Australia (HILDA) survey. Two dimensions of these data are important. The first
is that they contain a wealth of information about each respondent. For
example, we begin with more than 3,400 variables per observation, including
information about the respondents' demographic and socio-economic background,
and on their attitudes and preferences. Access to this broad range of
information means that by controlling for them, we can potentially proxy for
unobservable differences between those who do and do not obtain a new
qualification. Secondly, this dataset contains many variables that are highly
correlated, so we require a systematic approach to reduce such information
redundancy -- something that ML models are adept at.

Our ML approach also identifies new sub-populations for which the treatment
effects are different. We document that the starting homeloan amount and
employment aspirations are significant factors related to the extent of gain
from further study. We also find that the starting levels of and pre-study
trends in personal and household income are hugely important. Age and mental
health variables also account for variation in estimated effects. All of these
variables are consistently selected as being significant for prediction out of
the 3,400 features within the HILDA data. This selection is consistent across
different ML models (which includes linear and non-linear model classes) and
across numerous boostrap draws of the original sample.

Previous studies have found that individuals who seek a futher degree tend to
have slower-growing earnings in the period before their study starts compared
to similar individuals who do not seek further study \citep{jacobson2005,
dynarski2016, dynarski2018}. By accounting for dynamic selection into obtaining
a further degree, we can be confident that we compare the earnings paths of
mature-age students to the paths of similar non-students who displayed the same
earnings (and other) paths before study began. In this paper, we explicitly
control for the trajectories of socio-economic and demographic circumstances
before study starts. Standard fixed effects estimation would miss these dynamic
confounders. We find that our ML estimates are significantly smaller than the
size of the standard fixed effects results. We also estimate lower returns
compared to Ordinary Least Squares (OLS) models. We document the additional
confounder variables that we include in our models but are usually omitted from
standard OLS specifications. These variables suggest there is significant
selection into mature-age students who undertake a further degree.

We adapt ML models for the purpose of estimating causal effects. Standard
off-the-shelf ML models are better suited to predictive purposes. When
obtaining a prediction, off-the-shelf ML models can find generalisable patterns
and minimise overfitting issues, though the use of cross-validation, because
the true outcomes are observed. This means that we can optimize a
goodness-of-fit criterion. Causal parameters, however, are not observed in the
data, which means we cannot directly train and evaluate our models. 

In this paper, we take the difference between the two optimal outcome models,
which can achieve the optimum bias-variance trade-off point for the conditional
average treatment effect. Specifically, we model the response surfaces for two
conditional mean equations -- one using the treatment observations and another
using the control observations. We estimate these equations with ML methods
such as the T-learner and Doubly Robust. Here, we employ both linear (LASSO and
Ridge) and non-linear (Gradient Boosted Regression) model classes. We compare
and evaluate their comparative performance using nested cross-validation. We
then test the statistical significance of our causal parameters by examining
the distribution of the estimates through bootstrapping. Last, we use a variety
of Bayesian ML models following the formulation presented in \citet{hahn2020}
that reduce effect estimation bias within the Bayesian paradigm. These models
have several properties that may be desirable, such as the ability directly
parameterise heterogeneous prognostic and treatment models.

\section{Context: Higher education and Vocational study in Australia}

Mature-age education in Australia is among the highest in the world. In 2014, Australia’s participation in vocational education by those aged 25-64 was the highest among OECD countries. The tertiary education rate for those aged 30-64 was the second highest \citep{perales2017}. Mature-age Australians are increasingly enrolling in university or college to change employers, change careers, gain extra skills, improve their promotion prospects and earning capability or search for better work/life balance. Redundancy and unemployment have also been driving forces for individuals to return to education later in life \citep{coelli2012}.

The increase in mature-age learners accessing higher education has in part been driven by government policy. In 2009, the Australian government adopted a national target of at least 40\% of 25-34-year-olds having attained a qualification at bachelor level or above by 2025 \citep{oshea2015}. This was part of a policy that transitioned Australia to a demand-driven system \citep{ua2020}. The policy had a large effect on access to higher education, as it removed the cap on the number of university student places. By 2017, 39\% of 25-34-year-olds had a bachelor’s degree or higher \citep{caruso2018}.

While the initial uptake of university places in the demand-driven system was strong, especially among mature-age students\footnote{Between 2010 and 2012, growth in mature-age enrolments in undergraduate courses doubled for the 30-39 age group and tripled for the 40+ age group.} \citep{ua2019}, growth in undergraduate enrolments slowed since 2012. In 2018, mature-age enrolments even dropped below the previous year. The 40+ age group showed the worst growth, receding by 10\%, while the 25-29’s and 30-39’s showed growth of around -4\% \citep{ua2020}. The decline of enrolments coincided with the freezing of the Commonwealth Grant Scheme (CGS) which capped funding at 2017 levels, effectively ending the demand-driven system \citep{ua2020}. 

Access to Commonwealth Supported Places (CSPs) have since been limited to 2017 levels, with cap raises from 2020 subject to performance measures \citep{uasystem}. As a proportion of the working age population, mature-age students also participated less in vocational education and training (VET) over the same period. It appears the introduction of the demand-driven system also increased VET participation between 2010 and 2012, before continuing its decline \citep{atkinson2016}. Total VET enrolments since 2018 stabilised, with 2019 and 2020 enrolments slightly above 2018 levels\footnote{Total VET enrolments 2016-2020.} \citep{ncver2021}. The impact of COVID-19 on 2021 enrolments is yet to be fully determined. So far, VET enrolments for the first half of 2021 are well above the previous 4 years across all age groups, with $\sim$1 million enrolments in 2021 compared to $\sim$870 thousand enrolments in 2017\footnote{Government funded program enrolments Jan-June 2017-2021.} \citep{ncver2021}.

The cost of a bachelor’s degree for domestic students in Australia is the sixth highest among OECD countries \citep{ua2020}. In 2018, the average annual cost of a bachelor’s degree was around \$5,000 in Australia, about half of the top 2 most expensive countries where it costs around \$9,000 in the US and \$12,000 in the UK\footnote{Values are in US dollars.}. VET and TAFE courses in Australia cost a minimum of \$4,000 per year on average while post-graduate courses cost a minimum of \$20,000 per year on average\footnote{Values are in Australian dollars.} \citep{studies2018}.

Mature-age students can cover the cost of further study themselves or they can receive support from the government. Students at university or approved higher education providers can access financial support from the Higher Education Loan Program (HELP) scheme, which provides income-contingent loans. This allows students to defer their tuition fees until their earnings reach the compulsory repayment threshold, upon which repayments are deducted from their pay throughout the year at a set rate. Postgraduate students can access the Commonwealth Supported Place (CSP) scheme, which subsidises tuition fees for those studying at public universities and some private higher education providers. However, most CSPs are for undergraduate study. 

FEE-HELP is the HELP scheme available to full-fee paying students who don’t qualify for a CSP i.e., post-graduate students. VET Students Loans (formerly VET FEE-HELP) are also part of the HELP scheme and are available to students undertaking vocational education and training (VET) courses outside of higher education \citep{ua2020}. CSPs and HELP loans are withdrawn from students who fail half of their subjects, assessed on a yearly or half-yearly basis depending on the level of study.\footnote{Yearly at bachelor level and per trimester for courses lower than bachelor level.}

\section{Data}

We use data from the Household Income and Labour Dynamics Australia (HILDA)
survey. These data are rich, and we exploit the full set of background
information on individuals (beginning with more than 3,400 variables per
observation).

HILDA covers a long time span of 19 years, starting in 2001. We use the 2019 release. This means we observe respondents annually from 2001 to 2019.

\subsection{Sample exclusions}

Our main analysis sample contains respondents who were 25 years or above in 2001. This allows us to focus on individuals who obtain a further education – beyond that acquired in their previous degree. 

Our main analysis focuses on measuring the impact of further education using
wave 19 outcomes. Here, the feature inputs to the models are taken from the
individuals in 2001. We delete any individuals who were ‘currently studying’ in 2001. This also ensures that our features, which are defined in 2001 are not
contaminated by the impacts of studying but clearly precede the study spell of
interest. These sample exclusions result in 7,359 respondents being dropped
because they are below the age of 25 in 2001 and a further 1,387 respondents
being dropped because they were studying in 2001.

We then restrict the sample to those who are present in both 2001 and 2019. This ensures that we observe base characteristics and outcomes for every person in our analysis sample. This results in a further 5,727 respondents being dropped from the sample. Our analysis sample has 5,441 observations. More details of our main analysis sample and data can be found in the Online Appendix Document 1.\footnote{For sensitivity analysis, a second sample of respondents are examined. They are slightly younger when they began study, their feature values are taken in the two years before study began and their outcomes are measured four years after their study began. In this second sample, there are 1,814 individuals who started and completed a further educational degree, and 60,945 person-wave control observations who never completed a further degree. We detail our second approach in the Online Appendix Document 2.}

\subsection{Outcomes}

We measure outcomes in 2019 across the groups of individuals who did and did not get re-educated. We use annual earnings to measure the economic returns to education. We also analyse outcomes related to the labour market such as employment, changes in earnings, changes in occupation, industry, and jobs.\footnote{A second approach is to use outcomes measured four years after the start of a study spell. For sensitivity analysis, we repeat our main estimations using this second approach. Here, as many individuals in our dataset never started a further degree i.e. they are in our control group, we assign a time stamp to them for every year the control person theoretically could have started to study. We do this for every year from 2003 to 2019. This implies that control group individuals can be duplicated multiple times in the dataset. We then measure the control individuals' outcomes 4 years after their theoretical time stamp.}

\subsection{Treatment}

We define further education as an individual who obtains a further degree in a formal, structured educational program. These programs must be delivered by a certified training, teaching or research institution. Thus, we do not analyse informal on-line degrees (such as Coursera degrees). We also do not consider on-the-job training as obtaining further education.

Our treatment variable is a binary variable that takes the value of 1 if an individual has obtained an additional degree anytime between wave 2 (2002) and wave 17 (2017). As we analyse outcomes in 2019, this means we calculate the average returns between 2 years and up to 17 after course completion. We delete any respondent who obtained a qualification after wave 17. This allows us to analyse outcomes at least two years after course completion.

HILDA documents formal degree attainment in two ways. The first is to ask respondents, in every, wave what is their highest level of education. The second way is to ask respondents, in every wave, if they have acquired an additional educational degree since the last time they were interviewed.

We utilise both these questions to construct our measure of further education. Using the first question, we compare if the highest level of education in 2019 differs from that in 2001. If there has been an upgrade in educational qualification between these two years, we set the treatment indicator to be one and zero otherwise. This question, however, only captures upgrades in education; it fails to capture additional qualifications that are at the same level or below as the degree acquired previously by the respondent. We rely on the second survey question to fill this gap.

These two survey questions thus capture any additional qualification obtained from 2002 to 2017, inclusive. Additional qualifications refer to the following types of degrees: Trade certificates or apprenticeships; Teaching or nursing qualifications, Certificate I to IV, Associate degrees, Diplomas (2-year and 3-year fulltime), Graduate Certificates, Bachelor, Honours, Masters and Doctorate degrees.

\subsection{Covariates/features}

We define our covariates, or features as they are known in machine learning
parlance, using 2001 as the base year. Since we delete any respondents who were
currently studying in 2001, we ensure that all features were defined before a
respondent begins further study.\footnote{We also test the sensitivity of our results to using feature inputs that are taken from the individuals closer to the timing of their study, namely two years before study began. Here, we use both the year and the two years preceding the start of a study spell to define our features. This allows us to capture both level and growth values in the features.}

A unique approach to our feature selection strategy is that we use all the
information available to us from the HILDA survey in 2001. This means that we
have more than 3,400 raw variables per observation. Before using the features
in a ML model, we delete any features that are identifiers or otherwise deemed
irrelevant for explaining the outcome.

In order to reduce redundancy in this vast amount of information, we next apply a supervised Machine learning model to predict outcomes 5 years ahead of 2001 i.e., in 2006. We then select the top 100 variables that are most predictive of the outcome in 2006.\footnote{Confounders are features that both have an impact on the outcome and on the treatment. \cite{cherno2018} suggest including the union of features kept in the two structural equations (outcome on features and treatment on features). Here, we only include the features that predict the outcome equation because including features that are only predictive of the treatment can erroneously pick up instrumental variables (see \cite{pearl2012class} for a discussion of this issue). } These variables are listed in Table \ref{tab:sumstats}. 

\subsection{Missing variables from the baseline model}

As part of a replication exercise, we constrast the results from the ML model with published work using Ordinary Least Squares (OLS) and Fixed Effects models. We also contrast the features selected in the ML model with an approach that manually selects the variables as in the case of \cite{chesters2015}. We call this the `baseline' model.

As a descriptive exercise, Table \ref{tab:mlvars} presents the features that were `missed' by the baseline model. In the baseline model, we included features such as age, gender, state of residence, household weekly earnings, highest level of education attained, and current work schedule. This collection of variables have been informed by theory or previous empirical results.  

The data-driven model identifies more salient variables compared to the baseline model. Additional variables include employment conditions such as work schedule, casual employment, firm size, tenure or years unemployed; financial measures such as weekly wage, investment income and mortgage debt; health measures such as limited vigorous activity and tobacco expenses; and work-life preferences related to working hours and child care. 

We identify variables as missing from the baseline model if those variables explain the residual variation in the outcome. Specifically, we regress the residuals from the baseline models (without the treatment included) on the features included in the data-driven model and train a LASSO model to highlight the salient variables that were missed. The variables that are chosen are listed in Table \ref{tab:mlvars}. We also document how these variables are correlated to the outcome and to the treatment in order to give us a sense of the direction of the bias their omission may induce. 

Most of the omitted variables bias the OLS estimates is upwards.\footnote{Exceptions include casual employment status, the presence of a past doctorate qualification, years unemployed, parental child care and dividend and business income.} The upward bias is consistent with the ML-models estimating an economic returns on obtaining a new qualification that is significantly smaller than the returns from an OLS model or a Difference-in-Difference - Fixed Effects (DD-FE)  model. In the DD-FE model, we use the same 5,441 individuals as the other methods but they are followed over two waves: 2001 and 2019 (i.e. there are 10,882 person-wave observations). We control for individual and wave fixed-effects. 

Figure \ref{fig:method} displays the estimated returns from six different
models. The first three bars show significantly higher returns based on the OLS
(no controls), OLS (with controls) and the DD-FE models compared to the last
three bars, which are based on the ML models -- Gradient Boosted Regression,
Doubly Robust and Bayesian Causal Forest. We discuss these methods in more
detail below. 

It is important to highlight that our approach to identifying missing variables
from the baseline model is a descriptive one. As previously mentioned,
the ML algorithm randomly selects variables that are highly correlated thus we
may have missed out on reporting the label of important variables omitted from
the baseline model.

\section{Descriptive Figures and Tables}

We calculate the average returns to degree completion for mature-age students who completed degrees between 2002 and 2017. The window in which study and degree-completion took place is noticeably large. However, sample size limitations with our survey data mean that it is not feasible to run an ML analysis, disaggregated by the timing-of-completion.

In order to obtain some insights into the potential heterogeneity over time, we present a series of descriptive graphs in this section. Here, our aim is not to present any causal analysis but to describe which groups studied earlier in the time period (and thus had more time to accumulate returns). These graphs can also point to the potential different factors driving study across the time period, and different effects on earnings depending on how much time has elapsed since completion. 

Figure \ref{fig:yearcomp} presents the distribution of degree completion over time. There is a steep decline in degree-completion proportions over time. This is likely to reflect the aging profile of HILDA survey respondents and that further study is disproportionately higher among the younger cohorts (25-44 year olds) (See Figure \ref{fig:degbyage}).

Over time, Figure \ref{fig:yearcompdeg} shows that the composition of degrees completed has shifted. Among those who completed a degree in later years, compared to those who completed a degree in the earlier period, a higher percentage completed a Certificate III or IV, Diploma or Advanced Diploma as opposed to a lower-level degree (Certificate I or II or below). In all years, the most frequently completed degrees are Cert 3 or 4, Associate degrees, Diplomas and Advanced Diplomas. 

%removed graph below, samples are too small.
%Figure \ref{fig:yearcompgen} shows that the gender composition for degree completions between 2002 and 2017, inclusive. Females tend to be more likely to complete degrees than males and this gender disparity has slightly increased over time. However, there is a stark difference in the types of degrees completed across gender. 

%Figure \ref{fig:yearcompdeggen} shows the distribution across degree type over time, and further broken down by gender. Males are more likely to complete a Certificate III or IV, Diploma or Advanced Diploma, than females. By contrast, females are more likely to complete a Certificate I or II or a degree below this level. Females, however, as also more likely to complete a degree at the level of Bachelor or higher, than males. 

%This trend that has increased across time with a sharp discontinuity occurring in 2012. 
The predominance of Cert 3 or 4 degrees is common across gender. Although, Figure \ref{fig:degbysex} shows the distribution of degrees is more heavily skewed towards these degrees for men then they are for women. 

Figure \ref{fig:yearearnempl} shows an increase in both average earnings and employment overtime between 2002 and 2017. Despite the upward trajectory, these outcomes show more volatility following 2008. This is likely to reflect the smaller samples in the later years of the survey. In our main analysis we average the returns over time as the samples within each year are inadequate to draw inference about heterogeneity across time.

%Figure \ref{fig:sexearnempl} breaks these outcomes down by gender. Among females, the proportion employed increased up to 2012 to rise above the proportion of males employed. Following 2012, female employment trended down while male employment trended up. While female average earnings were below male average earnings through most of the period, the catch up to male earnings subsided in 2012 also. 

\section{Method}

We aim to estimate the causal impact of obtaining a new qualification. Our empirical challenge is a missing data one in the sense that we do not observe the counterfactual outcome for each person -- what would have their income been if they had/had not obtained a new qualification?

We use capitalisation to denote random variables, where $Y \in \mathbb{R}^+$ is
the outcome variable, $T \in \{0, 1\}$ is the binary treatment indicator, and
$X \in \mathcal{X}$ are the conditioning variables (which can be a mix of
continuous or categorical in type). Small case is used to denote realisations
of these random variables, e.g. $y$, $t$ and $x$, and we may use a subscript
for an individual realisation, e.g. $y_i$ for individual $i$ from a sample of
size $n$.

Under the potential outcomes framework of \cite{imbens2015}, $Y(0)$ and $Y(1)$ denote the outcomes we would have observed if treatment were set to zero ($T=0$) or one ($T=1$), respectively. In reality, we only observe the potential outcome that corresponds to the realised treatment,
\begin{align}
Y = T \cdot Y(1) + (1-T) \cdot Y(0).
\end{align}

The missing data problem (or the lack of counterfactuals) is especially problematic when the treated group is different from the control group in ways that also affect outcomes. Such selection issues mean that we cannot simply take the difference in the average of the non-missing values of $Y(0)$ and $Y(1)$.

To address the missing data problem, we turn to a range of ML-based techniques.
Standard ML tools are purposed to predict, but our aim is to estimate the
causal parameter. These are different aims, and so we have to adapt the ML
tools. We may potentially bias our causal parameter of interest if we were to
use the off-the-shelf tools. For example, if we were to select the important
confounders using an ML model to predict the outcome $Y$, then we may
undervalue the importance of variables that are highly correlated to the
treatment $T$ but only weakly predictive of $Y$ \citep{cherno2018}.

We approach filling the missing data indirectly with three types of ML models
that have been specially adapted to causal inference. They are: the T-Learner,
Doubly Robust and Bayesian models. For all our models, we require the following identification assumptions.

\subsubsection*{Identification assumptions}

To interpret the estimated parameter as a causal relationship, the following assumptions are needed:

\begin{enumerate}
  \item \textbf{Conditional independence} (or conditional ignorability/exogeneity or conditional unconfoundedness) Rubin (1980): $Y(0)$ and $Y(1)$ are independent of $T$ conditional on $X$; i.e. $\left\{Y(0), Y(1)\right\} \perp T~|~X$.  
\end{enumerate}

This assumption requires that the treatment assignment is independent of the two potential outcomes. Practically, this amounts to assuming that components of the observable characteristics available in our data, or flexible combinations of them, can proxy for unobservable characteristics. Otherwise, unobservable confounding bias remains.

A benefit of using all the features the HILDA dataset has to offer is that we may minimise unobserved confounding effects. Specifically, we rely on the 3,400 features and complex interactions between them as well as flexible functional forms to proxy for components of this unobserved heterogeneity. For example, while we do not observe ability or aptitude directly, we may capture components of it with other measures that are observed in HILDA such as past educational attainment or the long list of income and other sources of income variables (see Table 1 for a list of the features). 

The reader is likely to conceptualise other dimensions of unobserved heterogeneity that may not be captured in Table 1. There are two likely scenarios in this case. First, HILDA may not be exhaustive enough, even with its existing richness, to capture all dimensions of unobserved heterogeneity. As a result, our estimates may be biased. 

Another potential scenario is that the source of unobserved heterogeneity in question (or some components of it) is still captured but modelled under the guise of another variable label. Variables that are highly correlated with each other are unlikely to be simultaneously included in the model. This is because the ML algorithm, in attempting to reduce the amount of information redundancy, may have randomly dropped one or more of those correlated variables. 

\begin{enumerate}
  \setcounter{enumi}{1}

  \item \textbf{Stable Unit Treatment Value Assumption} (SUTVA) or
    counterfactual consistency: $Y = Y(0) + T \cdot (Y(1) - Y(0))$. 

\end{enumerate}  

Assumption 2 ensures that there is no interference, no spill-over effects, and no hidden variation between treated and non-treated observations. SUTVA may be violated if individuals who complete further education influence the labour market outcomes of those who do not complete further education. For example, if the former group absorb resources that would otherwise be channelled to the latter group. Alternatively, the former group may be more competitive in the labour market and reduce the probability of promotions or job-finding for the latter group. As those who complete further education are a relatively small group, it is unlikely that these general equilibrium effects would occur. 

\begin{enumerate}
  \setcounter{enumi}{2}
  \item \textbf{Overlap Assumption} or common support or positivity -- no
    subpopulation defined by $X = x$ is entirely located in the treatment or
    control group, hence the treatment probability needs to be bounded away
    from zero and one.
\end{enumerate}

The overlap is an important assumption because counterfactual extrapolation
using the predictive models, 
\begin{align}
  \mathbb{E}[Y|X{=}x, T{=}1] &\approx \mu_1(x) \quad \textrm{and} \label{eq:T1} \\
  \mathbb{E}[Y|X{=}x, T{=}0] &\approx \mu_0(x) \label{eq:T0}
\end{align}
is likely to perform best for treatment and control subpopulations that have a
large degree of overlap in $\mathcal{X}$. If the treatment and control groups
had no common support in $\mathcal{X}$, we would be pushing our counterfactual
estimators to predict into regions with no support in the training data, and
therefore we would have no means by which to evaluate their performance.

This means the optimum bias-variance trade-off point for the conditional
average treatment effect may not align with the optimum bias-variance trade-off
point for the separate $\mu_1(x)$ and $\mu_0(x)$ models. Since, ultimately we
are interested in the CATEs (as opposed to the predictive accuracy of the
individual conditional mean functions), this can mean that we have biased
CATEs.

\begin{enumerate}
  \setcounter{enumi}{3}
  \item \textbf{Exogeneity of covariates (features)} -- the features included in the conditioning set are not affected by the treatment. 
\end{enumerate}

To ensure this, we define all of our features at a time point before any
individual started studying. Specifically, we use the first wave of HILDA (in
2001) to define our features. We only look at those individuals who completed
further education in 2002 onwards. Furthermore, we delete any individuals who
were currently studying in 2001 to ensure the features cannot reflect
downstream effects of current study. 

With the strong ignorability and overlap assumptions in place, treatment effect estimation reduces to estimating two response surfaces – one for treatment and one for control.

\subsection{T-Learner model}

The first adaptation of ML models for causal estimation is the T-learner approach. We aim to measure the amount by which the
response $Y$ would differ between hypothetical worlds in which the treatment
was set to $T=1$ versus $T=0$, and to estimate this across subpopulations
defined by attributes $X$. 

The T-learner is a two-step approach where the conditional mean functions defined in Equations (\ref{eq:T1}) and (\ref{eq:T0}) are estimated separately with any generic machine learning algorithm.

Machine learning methods are well suited to find generalizable predictive patterns, and we employ a range of model classes including linear (LASSO and Ridge) and non-linear (Gradient Boosted Regression). Once we obtain the two conditional mean functions, for each observation, we can predict the outcome under treatment and control by plugging each observation into both functions. Taking the difference between the two outcomes results in the Conditional Average Treatment Effect (CATE).

To show this, we define our parameter of interest, the CATE, which is formally defined as:
\begin{align}
\tau (x) = \mathbb{E}[Y(1) - Y(0) | X{=}x],
\end{align}
which, with the assumptions outlined previously, is equivalent to taking the
difference between two conditional mean functions $\mu _1(x) - \mu _0(x)$:
\begin{align}
\tau (x) & = \mu _1(x) - \mu _0(x) \nonumber \\
  & \approx \mathbb{E}[Y | T {=} 1, X {=} x] - \mathbb{E}[Y | T {=} 0, X {=} x] \nonumber \\
& = \mathbb{E}[Y(1) - Y(0) | X {=} x]. \label{eq:indTE}
\end{align}
In this estimation, we are not interested in the coefficients from regressing
$Y$ on $X$. What we require is a good approximation of the function $\tau(x)$,
and hence good estimates from $\mu_1(x)$ and $\mu_0(x)$, which is within the
perview of machine learning methods.

A benefit of our set-up is that when we take the difference between the two conditional mean functions, we coincidently find the optimum bias-variance trade-off point for the conditional average treatment effect. This means that we have an indirect way to obtain the best prediction of the CATE through two predictive equations, where we observe the true outcomes (and thus are able to regularise).

In practice, however, this indirect way of minimising the mean squared error
for each separate function to proxy for the minimum mean squared error of the
treatment effect can be problematic. See, for example,
\cite{kunzel2019,kennedy2020} for settings when the T-learner is not the
optimal choice. One potential estimation problem arises when there are fewer
treated individuals than control individuals and the individual regression
functions are non-smooth. In this instance the response surfaces can be
difficult to estimate them in isolation, and the T-learner does not exploit the
shared information between treatment and control observations. For example, if
$X$ relates to $Y$ in the same fashion for treated and control observations the
T-learner cannot utilise this information. As a result, the estimate $\mu_1$
tends to over smooth the function; in contrast, the estimate $\mu_0$
regularises to a lesser degree because there are more control observations.
This means a na\"ive plug-in estimator of the CATE that simply takes the
difference between $\mu_1 - \mu_0$ will be a poor and overly complex estimator
of the true difference. It will tend to overstate the presence of heterogeneous
treatment effects. We turn to other ML models to address this potential
problem.

\subsection{Doubly Robust model}

The second approach is the Doubly Robust learner (DR-learner). It is similar to
the T-learner in that it separately models the treatment and control surfaces,
but it uses additional information from a propensity score model. In this case
the propensity score model is a machine learning classifier that attempts
to estimate the treatment assignment process,
\begin{align}
  \mathbb{E}[T{=}1|X{=}x] &= \mathbb{P}(T{=}1|X{=}x) \approx \rho(x),
\end{align}
where $\rho(x)$ as a probabilistic machine learning classifier. This allows
information about the students' background, and the nature and complexity of
their situation that may have led them to pursue further education to be
incorporated into the model. Thus, the doubly robust approach can improve upon
the T-learner approach because it can reduce misspecification error either
through a correctly specified propensity score model or through correctly
specified outcome equations. Another feature of the Doubly Robust approach is
that it places a higher weight on observations in the area where the relative
count of treatment and control observations is more balanced (i.e. the area of
overlap). This may allow better extrapolations of the predicted outcomes within
the region of overlap. The ATE is estimated from three separate estimators, 
\begin{equation}
  \hat{ATE} = \frac{1}{n} \sum_{i=1}^{n}
    \left[\frac{t_i(y_i - \mu_1(x_i))} {\rho(x_i)} + \mu_1(x_i) \right]
    - \frac{1}{n} \sum_{i=1}^{n} 
    \left[ \frac{(1-t_i)(y_i - \mu_0(x_i))}{1-\rho(x_i)} + \mu_0(x_i) \right] \label{eq:DR}
\end{equation}
% where:
% \begin{itemize}
%   \item ${\rho}(x_i)$ is an estimation of $\mathbb{E}[T{=}1|X{=}x_i]$ -- the propensity score, using logistic regression,
%   \item $\mu_1(x_i)$ is an estimation of $\mathbb{E}[Y|X{=}x_i, T{=}1]$ using any regression model,
%   \item $\mu_0(x_i)$ is an estimation of $\mathbb{E}[Y|X{=}x_i,T{=}0]$ using any regression model.
% \end{itemize}
Previously, with the T-learner, we were just estimating $\mu_0(x)$ and
$\mu_1(x)$. With the DR-learner, we augment $\mu_0(x)$ and $\mu_1(x)$. For
example, for the treated observations, we augment $\mu_1(x)$ by multiplying the
prediction error by the inverse propensity scores. This up-weights those who
get treated but who are statistically similar to the control observations. We
then apply this same augmentation to the $\mu_0(x)$ for the control
observations.

\subsection{Bayesian Models}

The third approach is to use Bayesian models. We follow the general formulation
presented by \cite{hahn2020} that suggests a predictive model of the following
form,
\begin{align}
  \mathbb{E}[Y|X{=}x_i, T{=}t_i] \approx \mu_0(x_i, \rho(x_i)) 
    + \tau(x_i)\cdot t_i, \label{eq:bayesmods}
\end{align}
where $\mathbb{E}[T=1|X{=}x_i] \approx \rho(x_i)$ is the propensity score of
individual $i$ for the treatment. The component $\mu_0(x_i, \rho(x_i))$ is
known as the `prognostic' effect, and is the impact of the control variates,
$X$, on the outcome without the treatment. Then we are left with $\tau(x_i)$,
which is the individual treatment effect,
\begin{align*}
  \mathbb{E}[Y|X{=}x_i, T{=}1] - \mathbb{E}[Y|X{=}x_i, T{=}0] &\approx 
    \left[\mu_0(x_i, \rho(x_i)) + \tau(x_i)\right] - \mu_0(x_i, \rho(x_i)), \\
    &=\tau(x_i).
\end{align*}
Average treatment effect is then just simply estimated as,
\begin{align*}
  \hat{ATE} = \frac{1}{n}\sum^n_{i=1} \tau(x_i).
\end{align*}
The advantage of this approach are manifold. From a Bayesian perspective, it
allows us to place explicit and separate priors on the prognostic and treatment
components of the models. For example, it may be sensible to expect the
prognostic component to be flexible and strongly predictive of the outcome,
while me may expect that the treatment component is relatively simple and small
in magnitude \citep{hahn2020}. Furthermore, this separation of model components
and inclusion of the propensity score minimises bias in the form of
regularisation induced confounding (RIC) which is discussed in more detain in
\citep{hahn2018, hahn2020}. And finally, it is a very natural way to estimate
heterogeneous treatment effects, since we can parameterise $\tau(x_i)$
directly as an additive effect on $\mu_0$, rather than having to separately 
parameterise control and treatment surfaces.

We explore three different model classes for $\mu_0$ and $\tau$, the first is a
linear model for both prognostic and treatment models, the next uses a Gaussian
process (GP), and lastly we use Bayesian additive regression trees (BART). We 
detail these models in the following sections.

\subsubsection*{Hierarchical Linear Model}

The first Bayesian model uses linear prognostic and treatment components 
from Equation~(\ref{eq:bayesmods}),
\begin{align*}
y_i &\sim \mathcal{N}\!\left(
  \mu_0(x_i, \rho(x_i)) + \tau(x_i)\cdot t_i, 
  \sigma^2\right) 
  \quad\textrm{where}, \\
\mu_0(x_i, \rho(x_i)) &= w_0 + w_x^\top x_i + w_\rho \rho(x_i), \\
\tau(x_i) &= w_t + w_{tx}^\top x_i.
\end{align*}
We have used the following hierarchical priors,
\begin{align*}
  \{\lambda_0, \lambda_x, \lambda_\rho\} &\sim \textrm{Uniform}(0, 100) \\
  \{\lambda_t, \lambda_{tx}\} &\sim \textrm{Uniform}(0, 1000) \\
  \sigma &\sim \textrm{HalfCauchy}(25) \\
  w_0 &\sim \mathcal{N}(0,\lambda_0^2) \\
  w_x &\sim \mathcal{N}(0,\lambda_x^2 \textrm{I}_d) \\
  w_\rho &\sim \mathcal{N}(0,\lambda_\rho^2) \\
  w_t &\sim \mathcal{N}(0,\lambda_t^2) \\
  w_{tx} &\sim \mathcal{N}(0,\lambda^2_{tx} \textrm{I}_d),
\end{align*}
where $I_d$ is the identity matrix of dimension $d$, which is the number of
control factors. The propensity score, $\rho(x_i)$, is obtained from a logistic
regression model. We also tested a gradient boosted classifier
\citep{friedman2001} for this using five-fold nested cross validation.  It did
not seem to be more performant than the logistic model on held-out log-loss
score.

For model inference, we use the no U-turn MCMC sampler \citep{hoffman2014} in
the numpyro software package \citep{bingham2019, phan2019}. The choice of an
uniform improper and non-informative prior over the regression weight scales,
$\lambda_*$, is motivated by the advice in \citet{gelman2006} where we desire a
non-informative prior that admits large values. We choose a broader 
prior for the treatment component of the model to minimise bias as suggested by
\cite{hahn2020}. We first burn in the Markov chain for 30,000 samples, then draw
1000 samples from the posterior parameters to approximate the ATE,
\begin{align}
  \hat{ATE} = \frac{1}{Sn} \sum^S_{s=1} \sum^n_{i=1} \tau^{(s)}(x_i),
  \label{eq:bayesATE}
\end{align}
where $(s)$ denotes a sample from the posterior parameters has been used to 
construct a random realisation of the treatment model component, and $S = 1000$.

\subsubsection*{Gaussian Process Regression}

Gaussian process (GP) regression can be viewed as a non-linear generalisation
of Bayesian linear regression that makes use of the kernel trick
\citep{williams2006, bishop2006}. Another way of understanding a GP is that is
parameterises a distribution over functions (response surfaces) directly,
rather than model weights as is the case with Bayesian linear regression.

Say we have the regression function, $\mathbb{E}[Y|X{=}x_i] = f(x_i)$, a
Gaussian process models the covariance of $f(x)$ directly using a kernel 
function,
\begin{align*}
  \mathbb{E}[f(x_i) \cdot f(x_j)] &= k(x_i, x_j) \qquad \textrm{or}, \\
  \mathbb{E}[Y_i \cdot Y_j] &= k(x_i, x_j) + \sigma^2 \delta_{ij},
\end{align*}
where $\delta_{ij}$ is a Kroneker delta, and is one iff $i=j$, otherwise zero.
This formulation also assumes $\mathbb{E}[Y] = \mathbb{E}[f(x)] = 0$ for 
simplicity -- and can be used directly if the outcomes are transformed to be
zero mean, or we can model an additional mean function (see \citet{williams2006}
for details). The Gaussian process can be written as,
\begin{align*}
  \mathbf{y} \sim \mathcal{N}(\mathbf{0}, \mathbf{K} + \sigma \mathbf{I}_n),
\end{align*}
where $\mathbf{y} = [y_1, \ldots, y_i, \ldots, y_n]^\top$ is the vector of all
outcome samples, $\mathbf{K}$ is the covariance matrix with elements
$\mathbf{K}_{ij} = k(x_i, x_j)$, and $\mathbf{I}_n$ the $n$-dimensional
identity matrix. 

To implement the functional relationship in Equation (\ref{eq:bayesmods}) in a
Gaussian process, we create the kernel function over $\langle x, t \rangle$
pairs,
\begin{align*}
  k(\langle x_i, t_i \rangle, \langle x_j, t_j \rangle) 
    = \sigma^2_{\mu_0} 
      k_{\mu_0}(\langle x_i, \rho(x_i) \rangle, \langle x_j, \rho(x_j) \rangle) 
    + t_i t_j \cdot [\sigma^2_\tau k_\tau(x_i, x_j) + \tau_0].
\end{align*}
Here $k_{\mu_0}$ and $k_\tau$ are the prognostic and treatment kernels
respectively, $\sigma_{\mu_0}$ and $\sigma_\tau$ allow us to scale the
contribution of these kernels to the functional relationships learned, and
$\tau_0$ permits a constant treatment effect. This induces the functional
relationship we want; $f(x_i, t_i) = \mu_0(x_i, \rho(x_i)) + \tau(x_i) \cdot
t_i$. We use the same propensity model for $\rho(x_i)$ as the linear model 
previously.

We have chosen isotropic Mat\'{e}rn $\frac{3}{2}$ kernel functions for
$k_{\mu_0}$ and $k_\tau$, 
\begin{align*}
  k_{\nu=3/2}(x_i, x_j) = 
    \left(1 + \frac{\sqrt{3}|x_i - x_j|}{l} \right)
    \exp\left(\frac{-\sqrt{3}|x_i - x_j|}{l}\right),
\end{align*}
where $l$ is the length scale parameter, and controls the width of the kernel
function. Smaller length scales allow for more high-frequency variation in the
resulting function $f(x_i)$. The Mat\'{e}rn kernel is a stationary and
isotropic kernel, but does not have excessive smoothness assumptions on the
functional forms it can learn -- this kernel leads to the response surface
being at least once differentiable \citep{williams2006}. A Gaussian process
with this kernel can learn non-linear and interaction-style relationships
between input features and the outcome. Our composite kernel is not
necessarily stationary however, as we have included a non-stationary term, $t_i
t_j$. 

A-priori, we expect reasonably smooth variation $\mathbb{E}[y_i \cdot y_j]$ so
we choose a long length-scale for the prognostic kernel function, $l_{\mu_0}$ =
10, and an amplitude, $\sigma^2_{\mu_0} = 1$. We expect an even smoother
relationship with less contribution for the treatment, and set the
corresponding kernel parameters as; $l_\tau = 50$, $\sigma^2_\tau = 0.1$ and
$\tau_0 = .001$. These parameters are then optimised using the maximum
likelihood type-II procedure outlined in Section 5.4.1 of \citet{williams2006}.

The ATE is then approximated as,
\begin{align*}
  \hat{ATE} = \frac{1}{Sn} \sum^S_{s=1}\sum^n_{i=1}
    f^{(s)}_*(x_i, 1) - f^{(s)}_*(x_i, 0),
\end{align*}
where $f^{(s)}_*(x_i, t)$ are samples from the Gaussian process posterior
predictive distribution\footnote{See Equations (2.22)-(2.24)
of \citet{williams2006}.} with kernel inputs $k_*(\langle x_i, t \rangle, \langle x_i, t
\rangle)$, which is equivalent to sampling from the distribution over
$\tau(\cdot)$. We use $S=100$ samples.

\subsubsection*{Bayesian Causal Forests}
The last Bayesian model we use is the Bayesian causal forest introduced In
\citet{hahn2020}. Broadly it models the prognostic and treatment components As
Bayesian additive regression trees (BART),
\begin{align*}
y_i &\sim \mathcal{N}\!\left(
  \mu_0(x_i, \rho(x_i)) + \tau(x_i)\cdot t_i, 
  \sigma^2\right) 
  \quad\textrm{where}, \\
\mu_0(x_i, \rho(x_i)) &= \textrm{BART}(x_i, \rho(x_i)), \\
\tau(x_i) &= \textrm{BART}(x_i).
\end{align*}
We use the accelerated BART (XBART) implementation of this algorithm detailed 
in \citet{krantsevich2022}. BART \citep{chipman2010} has been shown to be an 
effective and easily applicable non-parametric regression technique that 
requires few assumptions in order to capture complex relationships that can 
otherwise confound effect estimation. We follow \cite{hahn2020} in our choice
of BART priors,
\begin{align*}
  \alpha_{\mu_0} = 0.95,&~
  \alpha_{\tau} = 0.25, \\
  \beta_{\mu_0} = 2,&~
  \beta_{\tau} = 3.
\end{align*}
This choice prefers a more simple treatment effect model, $\tau(x_i)$, that is
less likely to branch, and more likely to have shallower trees 
than the prognostic model. Similarly, we use 200 trees for the prognostic model,
and 50 for the treatment. We take 500 burn-in sweeps, and then 2000 sweeps
to estimate the posterior BART distributions.

ATE is estimated in the same way as for the linear model in Equation 
(\ref{eq:bayesATE}), but where the BART posterior is used for the 
treatment effect distribution.

\subsection{Model selection and model evaluation}

For the non-Bayesian models we separate the evaluation of the model class and
estimation of the ATE and CATE parameters in two procedures. We evaluate the
predictive capacity of each model class using nested cross-validation. The
procedure is represented in Figure \ref{fig:exp_model}. Here, our aim is to
compare the predictive performance of three model classes: LASSO, Ridge and
Gradient Boosted Regression (GBR). Our second procedure is to estimate the ATE
and CATE parameters. The procedure is represented in Figure
\ref{fig:exp_param}. We use bootstrap sampling (with replacement) to generate
uncertainty estimates for the parameters, which we obtain over several draws of
the same model class, but with model parameter re-fitting.

Focusing on the first procedure, we apply nested cross-validation to evaluate
which model class performs best. In a first step, as Figure 1 shows, we
pre-process the full dataset (containing 3,400 variables) to generate a dataset
with a smaller set of highly predictive features (containing 91 variables). We
apply a supervised machine learning approach with a LASSO model to select our
top 91 predictors of the outcome of interest using outcomes measured in 2006.
Note that in our later estimations of the treatment effect, the outcome is
measured in 2019. We implement this intermediary step in order to reduce the
correlation between variables and eliminate redundant information.

We assume that the top 91\footnote{We were aiming for approximately 100
features, and 91 was the closest we could get the LASSO estimator to select by
changing the value of its regularisation strength.} features that are most
predictive of the outcome in 2006 correlate with the features that would be
most predictive of the outcome in 2019. By choosing to apply this
pseudo-supervised ML approach on the same outcome variable, but measured at a
different time point, we obtain a good indication of the features that are
useful for a model to perform well. Improved model performance here will also
mean that the selected features are likely to represent the important
confounders. We have chosen 2006 to ensure there is no overlap with 2019
outcomes to avoid overfitting issues with subsequent models.\footnote{We do not
compromise predictive performance when we use the selected subset of features
as opposed to the full set of features. For example, the predictive performance
from a Gradient Boosted Tree model that predicts earnings in 2006, using 5-fold
nested cross-validation, is statistically similar between models that use the
91 feature set and the full, 3,400 feature set (with Root Mean-Squared Errors
(RMSEs) of 484.251 and 482.286, respectively). This is a negligible loss in
predictive performance. There is a slightly larger associated loss between the
restricted and full feature sets from models predicting earnings in 2019 (RMSEs
of 843.548 and 831.931, respectively), but this is still not statistically
significant.}

Using the top 91 predictors, we then apply nested cross validation to evaluate
the predictive capacity of each model class (LASSO, Ridge, GBR). First, we
split the data into train and test folds with an 80-20 split. Within the 80
percent train fold we perform 5-fold cross-validation in order to train and
evaluate the performance of each configuration of hyperparameters. We do this
separately for the outcome surface using the treated observations and the
outcome surface using the control observations. From this, we select the models
with the best mean predictive scores. We then evaluate the predictive
performance of the selected model on the holdout test.

We repeat this process ten times (10-outer scores) for each model class. This
allows us to evaluate the performance based on the mean and standard deviation
of these scores. Note that thus far, we have not evaluated any particular
configuration of the model, rather the performance of the model class on random
(without replacement) subsets of data. The nested cross validation procedure
protects us against overfitting when reporting predictive performance, as the
model selection and validation happens on different data.

Table \ref{tab:ncvhos} shows that the GBR is the best performing model class.
It yields the highest out-of-sample R-squared and the lowest MSE. This is true
for both the outcome surfaces separately.

As the DR-learner model relies on the same treatment and control outcome surfaces estimated in the T-learner, we do not repeat Table \ref{tab:ncvhos} for the DR results. A further component of the DR model, however, is the propensity score. Here, we implement a regularised logistic regression to predict the likelihood of being treated (to obtain a further degree). Specifically, we use cross validation to fit a Logistic regression and obtain the predictions from the original sample. The holdout performance of the fitted Logistic regression model yields an area under the ROC curve of 0.71.

\subsubsection*{Inference via bootstrapping}

Once we have selected the best performing model class, we turn to the
estimation of the parameters and their associated uncertainty. We use a
bootstrapped validation proceedure to capture the uncertainty arising from
model hyperparameter selection in addition to that from estimating parameters
of a fixed model from noisy, finite data. 

%Here, we estimate the same model class on different realisations of the data. 

A common approach to inference in the causal machine learning literature is to
use cross-fitting \citep{cherno2018} or sample splitting \citep{athey2019}.
These methods ensure that the standard errors on the estimators are not
underestimated because they avoid using the same data point to both select
hyperparameters of the model and to estimate the parameters of the outcome or
effect surfaces. 
% When the same data point is used to perform both tasks then the standard
% errors would not reflect both the uncertainty stemming from model selection
% and that which stems from estimating parameters on noisy, finite data. 
The result of using the same data for model selection and effect estimation is
that our standard errors would suffer from pre-test bias since the model may
suffer from overfitting.

Sample splitting and cross-fitting are appropriate when the sample size is
large. An issue with studies that rely on survey-based data is that sample
sizes are often not large enough to efficiently use these methods. For example,
there may not enough data to split the dataset into separate train and test
datasets for each model such that each of these splits would cover all the
common and uncommon values of the $X$-features that are observed in the full
sample. Consequently, the ML models may not find representative functional
forms for $\mu_0(x)$ and $\mu_1(x)$.
% If we were to use a training dataset that was insufficiently sized or
% non-representative, it would be difficult for the ML models to effectively map
% the $X$-features to the outcome surfaces, $\mu_1(x)$ and $\mu_0(x)$. There
% would also not be enough data in the test set to effectively estimate the
% parameters of the model configuration chosen in the train set.
As a result, our estimate treatment effects are likely to have a large degree
of uncertainty.

A suitable alternate procedure is to use bootstrapping. Bootstrap resampling
allows us to estimate variation in the point model parameter estimates. In this
way, we side-step the need to rely on the assumption of asymptotic normality,
and it is more efficient than sample splitting to generate standard errors. In
our bootstrapping procedure, we ensure that the standard errors reflect the
sources of uncertainty stemming from both the selection of the model and the
estimation of the model. As a result, we generate standard errors that avoid
any potential pre-test issues.

As a first step we obtain the 91 top predictors from the initial
pre-processing of the full dataset, shown in Figure \ref{fig:exp_param}. That
is, we train a supervised machine learning LASSO model to extract the features
that best predict earnings in 2006.

The second step involves training our models using the 91 top predictors on a
bootstrapped sample, $s$, to select the best models for $\mu^{(s)}_1(x)$ and
$\mu^{(s)}_0(x)$. Within this bootstrap sample, we divide the dataset into five
folds and perform cross-validation to select the best model configuration.
Similar to the cross-validation description above, our model configuration is
trained on subsets of the data, and then evaluated on holdout samples. We
modify the 5-fold cross validation to ensure bootstrap replicated training data
does not simultaneously appear in the training and validation set. We perform
this model selection step within the bootstrapping procedure to capture the
uncertainty coming from the selection of hyperparameters. If we simply
re-estimated the same model with a given set of hyperparameters in each
bootstrap model then the uncertainty is only over the model parameters, and not
the model choice (e.g.~the GBR tree depth).

%This process allows us to select the best model configurations for the outcome
%surfaces: $\mu _1(x)$ and $\mu _0(x)$.
Third, and once we have these predicted outcome surfaces, $\mu^{(s)}_1(x)$ and
$\mu^{(s)}_0(x)$, we are able to calculate the individual treatment effect,
$\tau(x_i)$, for each person, $i$, in the original sample (not the individuals
from the bootstrap sample) by substituting the values of their features into
the LASSO, Ridge or tree estimators for the outcome surfaces. We can obtain a
sample mean, $\bar{\tau}^{(s)}$, by averaging
$\frac{1}{n}\sum^n_{i{=}1}\tau^{(s)}(x_i)$ using the bootstrapped effect model.
We repeat this procedure over $S=100$ bootstrap samples. This provides an
empirical distribution of $\bar{\tau}$ and $\tau(x_i)$. The grand mean over the
bootstrap sample means, $\bar{\tau}_G = \frac{1}{S} \sum^S_{s{=1}}
\bar{\tau}^{(s)}$, will converge to the sample treatment effect mean. We use
$\bar{\tau}_G$ as an estimate of the ATE, and $\frac{1}{S} \sum^S_{s{=}1}
\tau^{(s)}(x_i)$ as an estimate of the individual CATE. The bootstrap resample
is the same size as the original sample because the variation of the ATE
depends on the size of the sample. Thus, to approximate this variation we need
to use resamples of the same size.

% The main goal of repeating over the bootstrap samples is to capture uncertainty
% coming from the model selection procedure (i.e.\\~the choice of
% hyperparameters) as well as the uncertainty over the model parameters.

% Each bootstrap sample gives us an estimate of the (C)ATE\todo{(C)ATE?}. Across
% 100 bootstrap samples, we have a distribution of (C)ATE estimates. We take the
% standard deviation of these 100 (C)ATE values to obtain the standard error. 

To obtain confidence intervals for the ATE and CATE estimates we use standard
empirical bootstrap confidence interval estimators \citep{efron1986}.

% To obtain the confidence intervals from these samples, we undertake the
% following procedure (e.g. for a 90\% confidence interval):
% \begin{enumerate}
% \item Multiply the (C)ATE by two 
% \item Take the 5th and 95th percentile of the (C)ATE
%   distribution empirically from the bootstrap samples
% \item Calculate the bounds of the confidence interval 
% 	\begin{enumerate}
% 	\item Upper bound = Step 1 - 5th percentile 
% 	\item Lower bound = Step 1 - 95th percentile
% 	\end{enumerate}
% \end{enumerate}

% To do this we approximate the critical values of the confidence interval as $\delta^* = \bar{x}^* - \bar{x}$ such that the confidence interval 
% \begin{align*}
% P(\delta_{0.95} \leq \bar{x} - \mu \leq \delta_{0.05} | \mu) \Leftrightarrow P(\bar{x} - \delta_{0.95} \geq \mu \geq \bar{x} - \delta_{0.05} | \mu) = 0.90
% \end{align*}
% can be reduced to $[2\bar{x} -\bar{x}^*_{0.05} , 2\bar{x}
% -\bar{x}^*_{0.95}]$.\todo{I think we need to change the notation here to use
% $\tau$ instead of x and $\mu$ since these are overloaded. Or if this is a
% standard empirical procedure, we can just reference it}

For the DR-learners, similar to the T-learner, we train $\mu_1(x)$ and
$\mu_0(x)$ models across 100 bootstrap samples and weight these outcome
surfaces by the propensity score model, $\rho(x)$, which is estimated using
logistic regression (as described previously). 

\subsubsection*{Inference for the Bayesian models}

The inference process for the Bayesian models a little different since the
hyper-paramters of the models are either fixed or selected automatically by the
learning algorithm (maximum likelihood type-II or MCMC). Bayesian inference
procedures tend to afford some protection against over-fitting since they are
parsimonious when choosing posterior distributions over model parameters that
vary from their prior distributions, which induces a natural model complexity
penalty\footnote{This point can be understood more thoroughly by examining the
evidence lower bound in variational Bayesian inference, see Chapter 10 of
\citet{bishop2006}.}. As such, we use all the available data to learn the model
posterior distributions, which we then sample from to form empirical estimates
of the (C)ATE as outlined in the previous section.

\section{Results}

There are clear economic benefits to gaining an additional qualification in later life (25 years or older). The effects remain strong up to a decade-and-a-half after course completion. Table \ref{tab:atebslvl} displays a gain of approximately \$88-110 per week in gross earnings across the T-learner approaches. In 2019, this was roughly 7-8 percent of the average gross weekly earnings of \$1256.20 for all Australian employees (ABS, 2019; 6345.0 Wage Price Index, Australia). 

The effect sizes from the GBR model are smaller than that of the two linear models. GBR better captures non-linearities. For example, age is likely to exhibit a highly non-linear relationship with earnings in 2019. Those who were aged 46 or above in 2001 will be aged 65 or above in 2019. This means they are more likely to have retired by 2019 compared to those who were aged below 46 in 2001. As a result, we may expect a shift down in earnings at age 46. 

Age fixed-effects alone are unlikely to capture the differential age effects across other variables such as across different occupations, or by gender, and earnings. The linear ML models include age fixed effects. However, they do not include interactions between age and other variables whereas GBR does include them.

To illustrate how GBR adequately captures non-linearities we re-estimated our results focusing on those who were aged 25-45 in 2001. This is the same as interacting a binary variable (for age 25-45) with every other feature in the model. In Appendix Figure \ref{fig:valadle46}, we see that the results across the models are now more similar than when we use the full sample.

The Doubly Robust (DR) models estimate smaller effects compared to the T-learner models. Table \ref{tab:atebslvl} displays a gain of approximately \$62-69 per week in gross earnings across the DR approaches. The estimated effect sizes are statistically different from zero. The confidence intervals for the DR estimates also exclude the point estimates from the T-Learner approach.

One reason the DR approach differs from the T-learner approach is that the
former uses additional information from the propensity score (i.e. we estimate
machine learning models to gain a better understanding of the treatment
assignment process, the students’ background, and the nature and complexity of
their situation that may have led them to pursue further education). Thus, the
doubly robust approach can improve upon the T-learner approach because it can
reduce misspecification error either through a correctly specified propensity
score model or through correctly specified outcome equations. Another feature
of the Doubly Robust approach is that it places a higher weight on observations
in the area where the relative count of treatment and control observations is
more balanced (i.e. the area of overlap). A benefit of this is that it can also
provide better extrapolations of the predicted outcomes. 

The Bayesian models estimate similar sized effects to the DR models for the
most part. However, they tend to have more uncertainty associated with their
estimates. They all remain significant with the 95\% confidence intervals
remaining above \$0. The hierarchical linear model and the Gaussian process
both estimate a gain of approximately \$61-\$63 per week in gross earnings,
with the Gaussian process being more certain in its estimate. Interestingly,
the Gaussian process prefers a much smoother and smaller treatment effect
component compared to its prognostic component -- the treatment kernel length
scale is long, and the kernel has a small amplitude and offset ($l_\tau = 243$,
$\sigma_\tau^2 = 0.0517^2$, and $\tau_0 = 0.0312^2$). Whereas the prognostic
kernel parameters stay relatively close to their initial settings ($l_{\mu_0} =
16$, and $\sigma^2_{\mu_0} = 1.42^2$). The Bayesian causal forest estimates a
slightly higher gain of \$84.50 per week in gross earnings, which is more inline
with the GBR T-learner. This suggests that the tree ensemble methods may be
able to more easily capture non-linear relationships than the other models.

%Turning to other measures of earnings, the value-add in earnings (taking the growth in earnings between 2001 and 2019 - all expressed in 2019 terms) is also higher for those who gained an additional qualification, compared to those who did not advance in their education, by approximately \$60-80 per week. As a proportion of the growth in earnings for all Australian employees between 2001 and 2019, this represents 18 percent of the overall growth in earnings. The results are contrasted graphically in Appendix Figure \ref{fig:earnings}.
%
%The similarity in the results between the value-add increase in earnings and the level increase in 2019 earnings suggests that, once we control for background characteristics, we have accounted for inherent or base level differences between those who do and do not obtain an additional qualification. 
%In other words, the difference in 2019 earnings reflects the causal effect of advancing education in later-life.

Proportionate changes in earnings can be measured by taking the log of the earnings measures. In Appendix Figure \ref{fig:valadlelog}, we see that the proportionate change in earnings was large at 50 percent. This is likely to be because of people entering the labour market as a result of the new qualification. We find that a new qualification increases the likelihood of employment by approximately 8 percent. See Figure \ref{fig:empl}.

As previously mentioned, the ML models estimate smaller returns than the returns estimated in DD-FE or cross-sectional models (OLS with and without controls) where features have been selected based on theory or previous empirical learnings. For example, the `OLS Baseline model' uses the features in models estimated in Chesters (2015). The DD-FE eliminates all selection effects that are fixed over time. Figure \ref{fig:method} displays the estimated returns from six different approaches.

A potential reason for the smaller results estimated in the ML models is that the additional features included, as well as the non-linear specifications of the features, more effectively account for selection into treatment. The smaller results suggest individuals positively select into further study i.e. the characteristics that lead one to complete further study are positively correlated to future earnings. Once we control for this upward selection bias, we thus estimate smaller returns to further education.

The smaller estimated results relative to the DD-FE model are likely to stem from the inclusion of key time-varying variables such as the `change in total gross income' in the ML models, as well as other non-linear specifications. For example, the ML models allow the treatment effects to vary in a highly flexible fashion across different parts of the feature distributions rather than making linear extrapolations.

This points to a benefit of using ML models, compared to conventional models, because they can more effectively identify confounders. We show evidence of the types of confounders missed in conventional models in Table \ref{tab:mlvars}, as well as the direction of the bias stemming from their omission. 

In addition, we show evidence that models which allow for more flexible functional-form specifications lead to differences in the ATE. Within our ML models, the GBR tree ensemble tended to perform better (in terms of the nested cv results) compared to the linear-based models. The former yielded a slightly smaller ATE compared to the LASSO and Ridge results, for example, and they were also consistent with results from the Bayesian Causal Forest.

%The largest earnings gains are associated with acquiring an undergraduate degree or above (including Graduate certificates and Graduate diplomas). As displayed in Figure \ref{fig:degree}, only modest gains are associated with secondary or post-secondary certificates or diplomas. More generally, significant earnings gains are achieved when returning students acquire qualifications above those previously held. In fact, the earnings gains are much higher if students advance their qualification status rather than gaining an additional degree at the same or lower level.
%
%The magnitude of the returns to education also differs depending on the type of subject studied. Obtaining a degree in a technical subject is likely to yield a return that is twice as high as obtaining a degree in a nontechnical degree. The returns are statistically significant for the former but estimated with a high variance for the latter. Technical subjects are STEM subjects, medicine, and health-related fields whereas nontechnical subjects encompass those in the creative arts, arts, humanities, and social science disciplines. These results are also displayed in Figure \ref{fig:degree}.

\section{Sub-group analysis}

Qualification advancements may not benefit individuals in the same way. In this section we analyse if there is heterogeneity in the treatment impacts. We use a data-driven approach to select the sub-groups.  
%
%Last, Figure \ref{fig:wellbeing} shows that there are limited gains in improved mental health or well-being (life satisfaction or job satisfaction) from obtaining an additional qualification.
%
%%% HTEs Graphs %%% 

Specifically, we identify the important variables for which we expect to see the largest changes in the treatment effects. This involves using a Permutation Importance procedure.

% Figure \ref{fig:htelasso} 
% Figure \ref{fig:hteridge} 

\subsection{Permutation importance feature selection method}

We use a permutation importance selection method \citep{breiman2001,molnar2020}
to evaluate the relative importance of individual features. Our aim here is to
understand where the heterogeneous treatment effects are most pronounced. In
other words, we aim to identify the sub-groups for which the treatment effects
differ most significantly. In selecting the important features our objective is
to understand how to partition the data by the treatment effects as opposed to
predicting the outcomes themselves. 

The permutation importance proceedure involves testing the performance of a
model after permuting the order of samples of each individual feature, thereby
keeping the underlying distribution of that feature intact but breaking the
predicitve relationship learned by the model with that feature. The model
performance we are interested in, as previously mentioned, is the one that maps
the features to the individual treatment effects. 

Following the approach described above, we compute the individual treatment effects. Note that we train the model on the bootstrapped sample but estimate the individual
treatment effects using the feature values for individuals from the original
sample. Thus, for every individual we have a distribution of values of their
individual treatment effects.

After obtaining the individual treatment effects, we train another model that
maps the features to the individual treatment effects. We use cross-validation
to select our hyperparameters and obtain the optimal model. 

Using the original data, we take a single column among the features and
permute the order of the data and calculate a new set of individual treatment
effects. We compare the new and original individual treatment effects (based on
the permuted data and those from the non-permuted data) and calculate the Mean
Squared Errors (MSE). 

We repeat this for all the features, permuting them individually and evaluating
how they change the prediction of the individual treatment effect target.
Features that yield the largest MSEs are likely to be more important than those
features with lower MSEs since permuting those features breaks the
most informative predictive relationships. 

We then repeat the above steps across all the bootstrap samples. Note that a
different bootstrap sample will change the value of the individual treatment
effects since we train different outcome surfaces for $\mu^{(s)}_0(x)$ and
$\mu^{(s)}_1(x)$ for each bootstrap sample.

We embed the permutation importance selection method in a bootstrapping
procedure in order to capture hyperparameter uncertainty. For example, a
different `tree depth' could be chosen between different bootstrap samples.
This would affect the type of non-linear/interaction relationships that would
be captured by the models, which in turn would affect which features turn out
to be important. 

Finally, we obtain an average MSE for each feature, averaged across all
bootstrap samples. This average value allows us to rank the features by their
importance. Again, those with the largest average MSE values are the most
important. We can also evaluate the uncertainty of this estimate since we
obtain a distribution of MSE values across the different bootstrap samples.

Figure \ref{fig:featgbrDR} displays the top ten features (based on the
permutation importance procedure described above) and a residual category for
all the other features. The features that are most important are: weekly gross
wages on the main job and income- or wealth-related variables. Together, this
class of income/wealth variables accounts for 40\% of the importance of all
variables. We focus on these selected features since our Nested CV approach
pointed to the better predictive performance of the GBR model over the linear
models.

Other important features include those related to employment, including
occupational status, employment expectations, and employment history. The
demographic background of the individual, namely their age, is also important. 

%Appendix Figure \ref{fig:featgbr} reiterates the importance of features related to employment, including occupational status, employment expectations, and employment history. The demographic background of the individual, namely their age, is also important. 

Figure \ref{fig:dengbrlevDR} displays the distribution of the MSE values across
the bootstrap samples for the GBR model. It displays the distributions for the
top 3 features. The feature with the highest importance score: weekly gross
wage in the main job. This suggests that in some of the bootstrap samples,
where the MSE is larger, the individual treatment effects from the permuted
data differ greatly from the original individual treatment effects.

The results from the T-learner model (using GBR) shows a similar story to the
results from the permutation importance procedure using the DR model. Overall,
as Appendix Figure \ref{fig:featgbr} shows, income and employment-related
variables are the most salient in explaining treatment effect heterogeneity.

%However, the LASSO and RIDGE models also demonstrate the importance of other demographic characteristics such as the number of and presence of young children in the household, as well as economic-related variables such as previous educational attainment, amount remaining on the home loan and work aspirations. 

% update dengbrlev fig                                  Done
% create for the DR measure                          Figure \ref{fig:dengbrlevDR} 
% add density_GBR_le_100_top3mid3            Figure \ref{fig:dengbrlevtopmid3} 
% add HTE from DR                                       Figure \ref{fig:htedr}   
% add HTE from DR stacked with T-Learner   Figure \ref{fig:htemldr} 

Continuing to focus on the results from the Doubly Robust model, Figure
\ref{fig:htedr} shows that there is heterogeneity in the treatment impacts. We
have identified the features that were considered most important according to
the permutation procedure. For each feature, we divide the sample into two
groups. For continuous variables, we take the median value and divide the
sample into those who are above and below this median value. 

Weekly personal income has a large impact on the effect size. Those with below
median income in 2001 derive more benefits than those with above median income,
possibly because high income earners hit an earnings ceiling. Younger people in
2001 also derive more returns, as they may have had more time to accumulate
returns. This result aligns with findings from previous studies
\citep{polidano2016,dorsett2016,perales2017}. Weekly personal income and age
are likely to be highly correlated -- with older individuals tending to earn a
higher personal income. We cannot say which variable is the main driver of the
heterogeneous treatment effects and there may also be interaction effects between them.

We also investigate if there are heterogeneous treatment effects according to
commonly used variables in Figure \ref{fig:htedr}. Females reap slightly higher
returns compared to males although this is not statistically significant.
Similar treatment effects apply to those with and without a resident children,
although the effect sizes widen in favour of parents with older children in the
household. 
%
%The T-learner (GBR) model is mostly consistent with the Doubly Robust model in
%the estimates of the CATEs. Figure \ref{fig:htegbr} shows some differences
%based on age, weekly income and the household income. One reason for the
%larger effects in the Doubly Robust model is that it places more emphasis in
%the areas of overlap compared to the T-learner: there may be few treated
%observations in the extremes of the continuous variables of age and income. 
\

Acquiring an additional qualification may increase earnings through a number of
potential mechanisms. We find evidence that, in Figure \ref{fig:empl} for
example, it increases the chance that individuals move from being unemployed or
out of the labour force to being employed. The increase in employment is
approximately 8 percentage points and is statistically significant. We also
find evidence pointing to workers switching occupations or industries. This
suggests that further education in later life can support the economic goals of
a larger workforce as well as a more mobile one.

\subsection{Sensitivity Analysis}

For sensitivity analysis, we repeated the T-learner estimations using feature
inputs values taken from individuals two years before they began study. Thus, we
examine if our main results are sensitive to changes in the mapping equations
for the treatment and control outcome equations when features are measured
closer to the event of study, compared to taking input values in 2001. We also
measured outcomes four years after study began. This means that the timing
between when the feature input values are measured, when a further degree
commenced and was completed, as well as when the outcomes are measured, are all
closer together. This necessarily leads us to estimate the short-term returns
of obtaining a further degree. 

Our results from the sensitivity analysis are similar to that of the main
results. Specifically, the gains in gross earnings from a further degree in the
sensitivity analysis are: \$74 per week (Ridge), \$117 per week (LASSO) and
\$93 per week (GBR). The key take-away from these results is that the average
treatment effects in the main analysis are not sensitive to whether our
features use 2001 as the input year or use the two years before study. 

Furthermore, the main results are not sensitive to when outcomes are measured
i.e. the returns measured four years after the start of a study spell are
comparable to the returns averaged over 2 to 17 years after study completion.
This may point to the fact that the returns to further study are accrued in the
immediate years following the completion of the degree. It also suggests the
returns may not atrophy over time, especially since the majority of people who
did complete a degree in the main analysis did so in the earlier years of the
survey (Figure \ref{fig:yearcompdeg}). Unfortunately, our sample sizes are not
sufficient to explore heterogeneity in treatment effects by the year of
completion. 

The importance of employment-related features such as earnings (individual and
household), wages, and hours worked are reiterated in the sensitivity analysis
using the panel structure of the data. Namely, when we define our outcomes 4
years after the start of a study spell and where we define features two years
before study started, we also see similar results to that of the main results.
However, in Figure \ref{fig:featgbr_panel}, it is clear that the `trend' or
`growth' in the values of features such as individual earnings, hours worked
and household income are also important. This finding of dynamic selection is
echoed in the literature \citep{jacobson2005, dynarski2016, dynarski2018}. 

In Figure \ref{fig:featgbr_panel}, the feature mental health is also picked.
This result may reflect the fact that the timing of the measurement of
features, treatment and outcomes are all closer together compared to the main
results. This means that mental health is an important factor in explaining the
heterogeneity in relatively `short-term' treatment effects.

\section{Conclusions}

Using a machine learning based methodology and data from the rich and
representative Household Income and Labour Dynamics Australia survey we have
shown that completing an additional degree later in life can add \$60-80 (AUD,
2019) per week to an individual's gross earnings. This represents roughly 7-8
percent of the weekly gross earning for the average worker in Australia. Our
machine learning methodology has also uncovered sources of heterogeneity in
this effect.
%, in particular those less than 45 years old tend to reap more benefits than
%others.

Our methodology has allowed us to exploit the full set of background
information on individuals from the HILDA survey, beginning with more than
3,400 variables, to control our analysis. 
% In particular, the key benefits of our automated feature selection process
% include the potential to minimise confounding bias while minimising
% overfitting issues. It can do this by identifying features (and functional
% forms) that may have been previously overlooked or unanticipated.
We find that our automated feature selection method selects a set of
controls/features that include those that have theoretical foundations and/or
align with those chosen in past empirical studies. However, we also choose
features that have been traditionally overlooked. These include variables such
as household debt, wealth, housing, and geographic mobility variables. Other
important predictors include the ages of both resident and non-resident
children: non-resident children aged 15 or above matter and resident children
aged 0-4 are important.

Qualification advancements do not benefit Australian workers in the same way:
those with lower weekly earnings appear to benefit more from later-life study
than those with higher earnings. One possible reason is that ceiling effects
limit the potential returns from additional education. We also find that
younger Australians (less than 45 years of age) benefit more than their older
counterparts. Again, a ceiling effect phenomenon may apply since age is highly
correlated to weekly earnings.

% ****Dan, do you have other theories about why we see this result? 

Acquiring an additional qualification may increase earnings through a number of potential mechanisms. We find evidence that it increases the chance that individuals move from being unemployed or out of the labour force to being employed. We also find evidence pointing to workers switching occupations or industries. This suggests that further education in later-life can support the economic goals of a larger workforce as well as a more mobile one.

%%%%%%%%%%%%%%%%%%%%%%%%%%%%%%%%%%%%%%%%%%%%%
%%%%%%%%%%%%%%%%      FIGURES      %%%%%%%%%%%%%%%%%%%%%
%%%%%%%%%%%%%%%%%%%%%%%%%%%%%%%%%%%%%%%%%%%%%

\begin{landscape}   

\section{Tables and Figures}

%\begin{longtable}{l p{4cm} l p{3.5cm}}      
% \begin{longtable}{p{0.5\textwidth} l c c c}     
% \begin{longtable}{l l c c c}
% \newcolumntype{C}[1]{>{\centering\arraybackslash}p{#1}} % top-aligned and centered horizontally 

\newcolumntype{C}[1]{>{\centering\arraybackslash}m{#1}} % center columns horizontally and vertically 

% Table 1    
\begin{longtable}{p{0.65\textwidth} C{0.2\textwidth} C{0.2\textwidth} C{0.2\textwidth}}
\caption{Summary Statistics} \\ 
\hline 
Variable label & Variable name & Mean & SD \\
\hline
\endfirsthead
\multicolumn{4}{c}{\textit{Continued from previous page}} \\
\hline
Variable label & Variable name & Mean & SD \\
\hline
\endhead
\hline \multicolumn{4}{r}{\textit{Continued on next page}} \\
\endfoot
\hline
\endlastfoot    
\textbf{Outcomes} & & &\\
Annual Earnings individual in 2019 	& 	y\_wscei	 & 	614.730	 & 	1044.717	 \\
Imputed wages & 		 & 		 & 		 \\
Change in annual earnings between 2001 and 2019  	& 	y\_dwscei	 & 	129.029	 & 	980.754	 \\
&  & &  \\
\textbf{Treatment Indicators} 	& 		 & 		 & 		 \\
Highest level of educ changed between 2001 and 2017 	& 	reduhl	 & 	0.097	 & 	0.296	 \\
Extra degree attained in 2002 to 2017 	& 	redufl	 & 	0.257	 & 	0.437	 \\
Extra degree Bachelor and/or above 	& 	bachab	 & 	0.072	 & 	0.259	 \\
Below bachelor 	& 	bbach	 & 	0.209	 & 	0.406	 \\
Technical degree 	& 	techdeg	 & 	0.151	 & 	0.358	 \\
Qualitative degree* 	& 	qualdeg	 & 	0.080	 & 	0.272	 \\
&  & & \\
\textbf{Covariates (features)} 	& 		 & 		 & 		 \\
\textbf{\textit{Demographics}} 	& 		 & 		 & 		 \\
Sex 	& 	hgsex	 & 	1.536	 & 	0.499	 \\
Section of State 	& 	hhsos	 & 	0.690	 & 	1.046	 \\
Age 	& 	hgage1	 & 	46.025	 & 	12.832	 \\
Age of youngest person in HH 	& 	hhyng	 & 	27.115	 & 	21.886	 \\
No.  persons aged 0-4 years in HH 	& 	hh0\_4	 & 	0.257	 & 	0.589	 \\
No.  persons aged 10-14 years in HH 	& 	hh10\_14	 & 	0.274	 & 	0.606	 \\
Age when first left home 	& 	fmagelh	 & 	21.502	 & 	11.230	 \\
Living circumstances 	& 	hgms	 & 	1.997	 & 	1.708	 \\
English fluency 	& 	hgeab	 & 	1.604	 & 	0.262	 \\
Unemployment rate in region	& 	hhura	 & 	6.884	 & 	1.075	 \\
\textbf{\textit{Education}} 	& 		 & 		 & 		 \\
Highest year of school completed/attending 	& 	edhists	 & 	2.383	 & 	1.439	 \\
Bachelor degree (without honours) obtained 	& 	edqobd	 & 	0.211	 & 	0.330	 \\
Masters degree obtained 	& 	edqoms	 & 	0.041	 & 	0.160	 \\
Doctorate obtained 	& 	edqodc	 & 	0.011	 & 	0.085	 \\
No.  qualifications unknown 	& 	edqunk	 & 	0.078	 & 	0.403	 \\
\textbf{\textit{Employment}} 	& 		 & 		 & 		 \\
Occupation 	& 	jbmo61	 & 	3.772	 & 	1.825	 \\
Years in paid work 	& 	ehtjbyr	 & 	21.963	 & 	11.907	 \\
Tenure with current employer 	& 	jbempt	 & 	8.505	 & 	7.369	 \\
Type of work schedule 	& 	jbmday	 & 	3.785	 & 	2.612	 \\
Current work schedule 	& 	jbmsch	 & 	2.255	 & 	1.819	 \\
Casual worker 	& 	jbcasab	 & 	1.797	 & 	0.291	 \\
Hours/week worked at home 	& 	jbmhrh	 & 	12.372	 & 	7.174	 \\
Hours/week travelling to and from work 	& 	lshrcom	 & 	3.052	 & 	3.716	 \\
Satisfaction with employment opportunities 	& 	losateo	 & 	6.693	 & 	2.557	 \\
Occupational status - current main job 	& 	jbmo6s	 & 	50.177	 & 	19.199	 \\
No.  persons employed at place of work 	& 	jbmwpsz	 & 	3.746	 & 	1.961	 \\
Age intends to retire 	& 	rtiage1	 & 	345.709	 & 	230.208	 \\
Age retired/intends to retire 	& 	rtage	 & 	113.904	 & 	130.211	 \\
Prob.  of losing job in next 12 months 	& 	jbmploj	 & 	15.196	 & 	35.018	 \\
Prob.  of accepting similar/better job 	& 	jbmpgj	 & 	59.585	 & 	26.196	 \\
Looked for work in last 4 weeks 	& 	jsl4wk	 & 	1.272	 & 	0.411	 \\
Years unemployed and looking for work	& 	ehtujyr	 & 	0.464	 & 	1.647	 \\
Hours per week worked in last job 	& 	ujljhru	 & 	34.990	 & 	6.922	 \\
Industry of last job 	& 	ujljin1	 & 	9.373	 & 	1.822	 \\
\textbf{\textit{Work preferences}} 	& 		 & 		 & 		 \\
Total hours per week would choose to work 	& 	jbprhr	 & 	34.378	 & 	6.407	 \\
Importance of work situation to your life 	& 	loimpew	 & 	6.854	 & 	2.908	 \\
\textbf{\textit{Childcare}} 	& 		 & 		 & 		 \\
Child looks after self 	& 	chu\_sf	 & 	0.128	 & 	0.144	 \\
Uses child care while at work	& 	cpno	 & 	1.257	 & 	0.139	 \\
Parent provides child care 	& 	cpu\_me 	 & 	0.434	 & 	0.151	 \\
\textbf{\textit{Work-family balance}} 	& 		 & 		 & 		 \\
Do fair share of looking after children 	& 	pashare	 & 	2.411	 & 	0.671	 \\
Miss out on home/family activities 	& 	pawkmfh	 & 	3.904	 & 	1.069	 \\
Working makes me a better parent 	& 	pawkbp	 & 	4.038	 & 	0.979	 \\
\textbf{\textit{Family}} 	& 		 & 		 & 		 \\
No.  dependent children aged 5-9 	& 	hhd5\_9	 & 	0.261	 & 	0.584	 \\
No.  dependent children aged 10-14 	& 	hhd1014	 & 	0.269	 & 	0.604	 \\
No.  non-resident children 	& 	tcnr	 & 	0.993	 & 	1.373	 \\
Sex of non-resident child 	& 	ncsex1	 & 	1.509	 & 	0.320	 \\
Likely to have a child in the future 	& 	icprob	 & 	1.188	 & 	0.374	 \\
\textbf{\textit{Finances}} 	& 		 & 		 & 		 \\
Owned a home previously 	& 	hspown	 & 	1.368	 & 	0.424	 \\
Amount outstanding on home loans 	& 	hsmgowe	 & 	96803.720	 & 	43547.610	 \\
Time until home loan paid off	& 	hsmgfin	 & 	2011.858	 & 	4.157	 \\
Food expenses outside the home 	& 	xposml	 & 	36.982	 & 	42.522	 \\
SEIFA (level of economic resources) 	& 	hhec10	 & 	5.463	 & 	2.897	 \\
Taxes on total income 	& 	txtottp	 & 	7476.727	 & 	14035.510	 \\
Change in total gross income since 1 year ago 	& 	wslya	 & 	2231.465	 & 	1950.065	 \\
Had an incorporated business 	& 	bifinc	 & 	1.715	 & 	0.199	 \\
Had a non-LLC or unincorporated business 	& 	bifuinc	 & 	1.259	 & 	0.193	 \\
\textbf{\textit{Income}} 	& 		 & 		 & 		 \\
HH current weekly gross wages - all jobs 	& 	hiwscei	 & 	992.666	 & 	918.261	 \\
Current weekly gross wages - main job 	& 	wscme	 & 	468.062	 & 	556.185	 \\
HH financial year gross wages 	& 	hiwsfei	 & 	52472.490	 & 	49458.180	 \\
Financial year gross wages 	& 	wsfe	 & 	25463.770	 & 	30265.630	 \\
Financial year regular market income	& 	tifmktp	 & 	30734.790	 & 	33618.860	 \\
Financial year disposable total income 	& 	tifditp	 & 	27477.160	 & 	22701.270	 \\
Imputation flag: current weekly gross wages - all jobs 	& 	wscef	 & 	0.070	 & 	0.256	 \\
Imputation flag: current weekly gross wages - other jobs 	& 	wscoef	 & 	0.044	 & 	0.205	 \\
Imputation flag: financial year gross wages 	& 	wsfef	 & 	0.071	 & 	0.256	 \\
\textbf{\textit{Other sources of income}} 	& 		 & 		 & 		 \\
Receive superannuation/annuity payments 	& 	oifsup	 & 	0.059	 & 	0.232	 \\
Receive redundancy and severance payments 	& 	oifrsv	 & 	0.002	 & 	0.038	 \\
Receive other irregular payment 	& 	oifirr	 & 	0.001	 & 	0.027	 \\
Receive government pensions or allowances 	& 	bncyth	 & 	0.004	 & 	0.027	 \\
Receive Disability Support Pension 	& 	bnfdsp	 & 	0.151	 & 	0.181	 \\
Receive other regular public payments 	& 	oifpub	 & 	0.000	 & 	0.019	 \\
Financial year regular private income 	& 	tifprin	 & 	77.299	 & 	1409.625	 \\
Financial year investments 	& 	oifinvp	 & 	1951.052	 & 	10569.050	 \\
Financial year dividends 	& 	oidvry	 & 	744.263	 & 	4651.593	 \\
Financial year interest 	& 	oiint	 & 	666.116	 & 	3448.494	 \\
Financial year regular private pensions 	& 	oifpp	 & 	967.101	 & 	5055.004	 \\
Financial year business income (loss) 	& 	bifn	 & 	185.652	 & 	3274.511	 \\
Financial year business income (profit) 	& 	bifip	 & 	2597.792	 & 	13649.410	 \\
Financial year irregular transfers from non-resident parents 	& 	oifnpt	 & 	35.067	 & 	1305.812	 \\
Financial year public transfers 	& 	bnfapt	 & 	2865.540	 & 	4717.042	 \\
Financial year government non-income support payments 	& 	bnfnis	 & 	1025.031	 & 	2237.987	 \\
HH financial year public transfers 	& 	hifapti	 & 	5542.675	 & 	7937.136	 \\
HH financial year business income 	& 	hibifip	 & 	4880.589	 & 	18393.360	 \\
\textbf{\textit{Health}} 	& 		 & 		 & 		 \\
Imputation flag: current weekly public transfers 	& 	bncapuf	 & 	0.044	 & 	0.204	 \\
Imputation flag: financial year investments 	& 	oifinf	 & 	0.124	 & 	0.330	 \\
Imputation flag: financial year dividends 	& 	oidvryf	 & 	0.079	 & 	0.270	 \\
Imputation flag: financial year rental income 	& 	oirntf	 & 	0.071	 & 	0.257	 \\
Imputation flag: financial year business income 	& 	biff	 & 	0.071	 & 	0.258	 \\
Health limits vigorous activities 	& 	gh3a	 & 	2.108	 & 	0.718	 \\
How much pain interfered with normal work 	& 	gh8	 & 	1.704	 & 	0.971	 \\
Health condition/disability developed last 12 months 	& 	helthyr	 & 	1.870	 & 	0.151	 \\
Tobacco expense in average week 	& 	lstbca	 & 	37.771	 & 	10.690	 \\
\textbf{\textit{Housing}} 	& 		 & 		 & 		 \\
Years at current address 	& 	hsyrcad	 & 	9.541	 & 	10.226	 \\
External condition of dwelling 	& 	docond	 & 	1.970	 & 	0.870	 \\
No dwelling security 	& 	dosecno	 & 	0.552	 & 	0.497	 \\
No.  homes lived in last 10 years 	& 	mhn10yr	 & 	3.456	 & 	1.107	 \\
Moved to be near place of work 	& 	mhreawp	 & 	0.084	 & 	0.111	 \\
Moved because I was travelling	& 	mhrearo	 & 	0.009	 & 	0.038	 \\
\textbf{\textit{Attitudes}} 	& 		 & 		 & 		 \\
Importance of religion 	& 	loimprl	 & 	4.612	 & 	3.483	 \\
Working mothers care more about work success 	& 	atwkwms	 & 	3.729	 & 	1.807	 \\
Mothers who don't need money shouldn't work 	& 	atwkmsw	 & 	3.951	 & 	1.982	 \\
\textbf{\textit{Identifiers}} 	& 		 & 		 & 		 \\
Family number person 02 	& 	hhfam02	 & 	NA	 & 	NA	 \\
Relationship to person 03 	& 	rg03	 & 	NA	 & 	NA	 \\
ID of other responder for HH Questionnaire 	& 	hhp2	 & 	NA	 & 	NA
\label{tab:sumstats}
\end{longtable}  
\parbox{1.5\textwidth}{\footnotesize *Definition of technical and qualitative degree: Technical: STEM, Architecture, Agriculture and Environment, Medicine, Other Health-related Studies and Nursing, Management and Commerce and Law. Non-technical: Education, Society and Culture (includes economics!), Creative Arts, and Food, Hospitality and Personal Services. }\\

% Table 2 
\clearpage    
\begin{longtable}{p{0.5\textwidth} C{0.2\textwidth} C{0.2\textwidth} C{0.2\textwidth} C{0.2\textwidth}}
\caption{ML variables omitted by OLS Baseline model} \\ 
\hline 
Variable label & Variable name & Relationship with re-education (redufl) & Relationship with outcome (y\_wscei) & Bias direction in OLS models \\
\hline
\endfirsthead
\multicolumn{5}{c}{\textit{Continued from previous page}} \\
\hline
Variable label & Variable name & Relationship with re-education (redufl) & Relationship with outcome (y\_wscei) & Bias direction in OLS models \\
\hline
\endhead
\hline \multicolumn{5}{r}{\textit{Continued on next page}} \\
\endfoot
\hline
\endlastfoot    
\textbf{\textit{Education}} 	& 		 & 		 & 		 & 		\\
Doctorate obtained 	& 	edqodc	 & 	-	 & 	+	 & 	-	\\
\textbf{\textit{Employment}} 	& 		 & 		 & 		 & 		\\
Tenure with current employer 	& 	jbempt	 & 	-	 & 	-	 & 	+	\\
Current work schedule 	& 	jbmsch	 & 	-	 & 	-	 & 	+	\\
Casual worker 	& 	jbcasab	 & 	-	 & 	+	 & 	-	\\
Occupational status - current main job 	& 	jbmo6s	 & 	+	 & 	+	 & 	+	\\
No.  persons employed at place of work 	& 	jbmwpsz	 & 	+	 & 	+	 & 	+	\\
Prob.  of accepting similar/better job 	& 	jbmpgj	 & 	+	 & 	+	 & 	+	\\
Years unemployed and looking for work	& 	ehtujyr	 & 	+	 & 	-	 & 	-	\\
\textbf{\textit{Work-life balance}} 	& 		 & 		 & 		 & 		\\
Total hours per week would choose to work 	& 	jbprhr	 & 	+	 & 	+	 & 	+	\\
Parent provides child care 	& 	cpu\_me 	 & 		 & 		 & 	-	\\
Do fair share of looking after children 	& 	pashare	 & 	-	 & 	+	 & 	-	\\
Miss out on home/family activities 	& 	pawkmfh	 & 	+	 & 	+	 & 	+	\\
\textbf{\textit{Income}} 	& 		 & 		 & 		 & 		\\
Current weekly gross wages - main job 	& 	wscme	 & 	+	 & 	+	 & 	+	\\
Imputation flag: current weekly gross wages - all jobs 	& 	wscef	 & 	+	 & 	+	 & 	+	\\
Change in total gross income since 1 year ago 	& 	wslya	 & 	+	 & 	+	 & 	+	\\
Financial year investments 	& 	oifinvp	 & 	-	 & 	-	 & 	+	\\
Financial year business income (profit) 	& 	bifip	 & 	-	 & 	-	 & 	+	\\
Amount outstanding on home loans 	& 	hsmgowe	 & 	+	 & 	+	 & 	+	\\
Imputation flag: financial year dividends 	& 	oidvryf	 & 	+	 & 	-	 & 	-	\\
Imputation flag: financial year rental income 	& 	oirntf	 & 	+	 & 	+	 & 	+	\\
Imputation flag: financial year business income 	& 	biff	 & 	+	 & 	-	 & 	-	\\
\textbf{\textit{Health}} 	& 		 & 		 & 		 & 		\\
Health limits vigorous activities 	& 	gh3a	 & 	+	 & 	+	 & 	+	\\
Tobacco expense in average week 	& 	lstbca	 & 	-	 & 	-	 & 	+	\\
\textbf{\textit{Identifiers}} 	& 		 & 		 & 		 & 		\\
ID of other responder for HH Questionnaire 	& 	hhp2	 & 	-	 & 	-	 & 	+	
\label{tab:mlvars}
\end{longtable}  
%\parbox{1.5\textwidth}{\footnotesize *Notes can be entered here}\\            

% Table 3
% Nested CV – Holdout sample
\clearpage
\begin{table}[H]
\centering
\small
\caption{Nested CV Holdout Sample: Level Earnings}
\begin{tabular}{cccccccc}
\toprule
Model   &  Outcome surface  &  Negative MSE  &  NMSE Std  &  R-squared  &  R-squared Std  & ATE   & ATE\_std   \\
\midrule                         
\multirow{2}{*}{\hfil GBR}  &  Treated  &  -886515  &  452077 &  0.22  &  0.06  &  \multirow{2}{*}{\hfil 68.2}  &  \multirow{2}{*}{\hfil  28.4 }	 \\
&  Control  &  -659056  &  107251	 &  0.36 & 0.07  & & \\
\midrule																						
\multirow{2}{*}{\hfil LASSO} & Treated & -955958	 & 361911	 & 0.15	 & 0.09	 & \multirow{2}{*}{\hfil 94.1}	 & \multirow{2}{*}{\hfil 14.5}	 \\
& Control	 & -710521 & 	178030	 & 0.32	 & 0.05	 & & \\
\midrule  																								
\multirow{2}{*}{\hfil Ridge}	 & Treated	 & -966849	& 434518	 & 0.16	 & 0.08	 &  \multirow{2}{*}{\hfil  97.8}  &  \multirow{2}{*}{\hfil 14.5}	 \\
&  Control  &  -712374  &  174033  &  0.32  &  0.04  & & \\
\bottomrule
\end{tabular}
\par\medskip
\parbox{1.1\textwidth}{\footnotesize Notes: 5 fold CV performed on 80\% train sample. All statistics presented in this table are based on the 20\% holdout sample. Ten outer folds are used. See Figure \ref{fig:exp_model} for more details.} 
\label{tab:ncvhos}
\end{table}

% Table 4
% Bootstrapped ATEs
\begin{table}[H]
\centering
\small
\caption{Average Treatment Effects: Level Earnings. Comparison across models.}
\begin{tabular}{p{0.5\textwidth} C{0.1\textwidth} C{0.1\textwidth} C{0.2\textwidth}}
\toprule
Model  &  N  &  ATE  &  CI (ATE)  \\
\midrule 	   						
OLS (S-learner)	 & 	5441	 & 	64.41	 & 	[8.16, 120.66]	 \\
T-learner (GBR)	 & 	5441	 & 	88.38	 & 	[30.72, 137.15]	 \\
T-learner (LASSO)	 & 	5441	 & 	110.08	 & 	[4.01, 182.49]	 \\
T-learner (Ridge)	 & 	5441	 & 	108.95	 & 	[46.84, 183.05]	 \\
Doubly Robust (GBR)	 & 	5441	 & 	68.85	 & [50.91, 82.07] \\
Doubly Robust (LASSO)	 & 	5441	 & 	54.64	 & 	[27.97, 72.74]	 \\
Doubly Robust (Ridge)	 & 	5441	 & 	61.74	 & 	[45.7, 78.86]	 \\
% Bayesian Ridge	 & 	5441	 & 	45.23	 & 	[0.60, 89.86]	 \\
%Bayesian Ridge (counterfactual sampling)	 & 	5441	 & 	44.64	 & 	[2.24, 87.04]	 \\
Hierarchical Linear Model	 & 	5441	 & 	63.22 & [0.63, 121.70]  \\
Gaussian Process  & 	5441	 & 	61.01	 & 	[12.63, 109.51]	 \\
Bayesian Causal Forests  & 5441 & 84.51 & [26.28, 141.17] \\
% Hierarchical Bayesian Linear Model & 	5441	 & 	45.74	 & [-9.50, 113.53]  \\ Gamma(1,1) priors
\bottomrule
\end{tabular}
\par\medskip
\parbox{1.0\textwidth}{\footnotesize Notes: Sample of 25 or older respondents who had completed a degree at any point between 2002 and 2017. Total completions: 1,383.}
\label{tab:atebslvl}
\end{table}

% \includegraphics[scale=0.7]{_figures/figname.pdf}
% \includegraphics[width=1\textwidth]{_figures/figname.pdf}

% Method
%Explainers
\clearpage
\begin{figure}[htbp]
\centering
\caption{Selecting and Evaluating Model Class}
  \label{fig:exp_model}
    \includegraphics[scale=0.65]{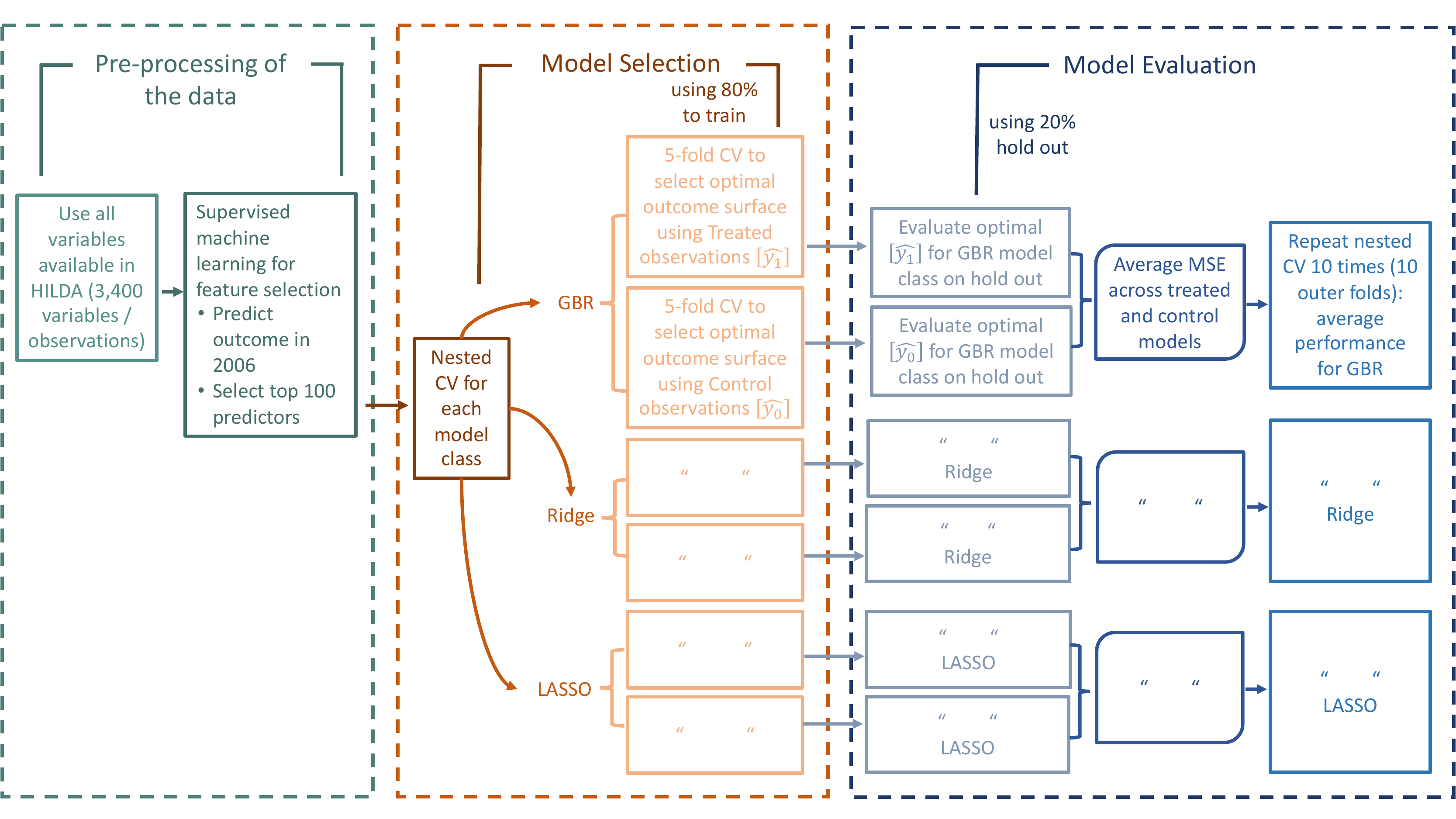}
\end{figure}

\begin{figure}[htbp]
\centering
\caption{Generating Uncertainty Parameters}
  \label{fig:exp_param}
    \includegraphics[scale=0.65]{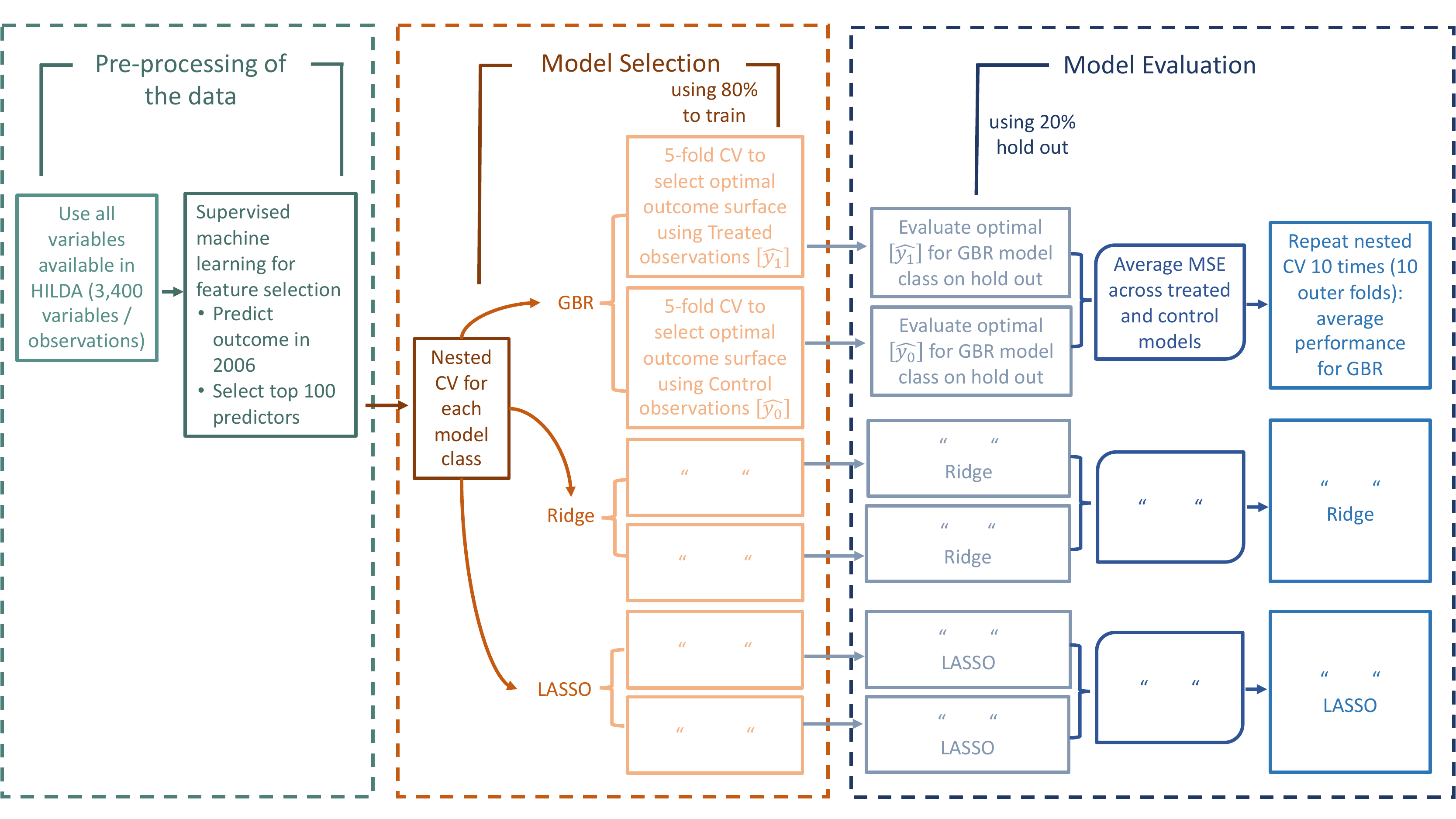}
\end{figure}

\end{landscape}

% Descriptive figures
\begin{figure}[H]
\centering
\caption{Timing of Completion}
\vspace{0.5cm}
  \label{fig:yearcomp}
    \includegraphics[width=0.75\textwidth]{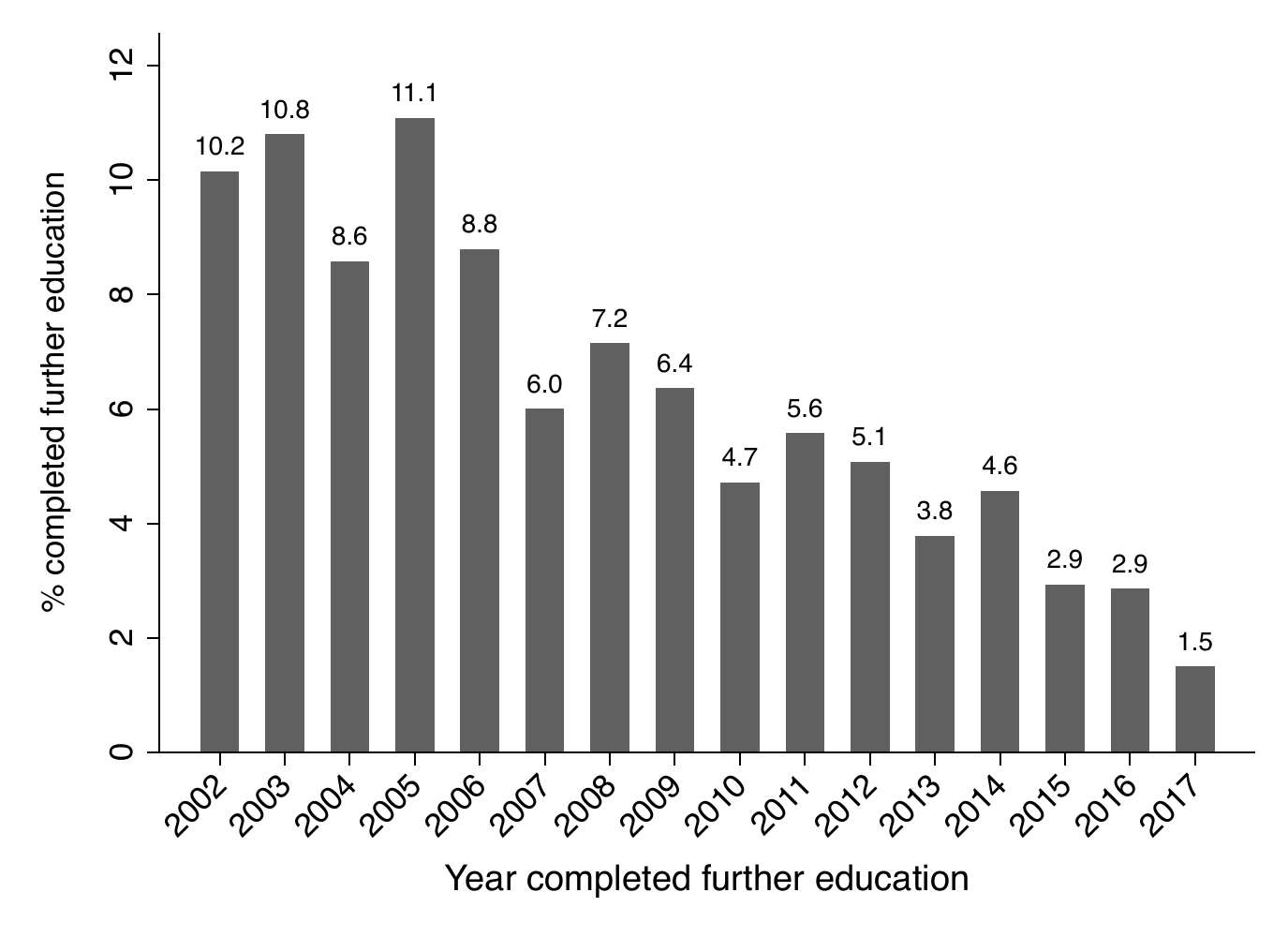}
    \parbox{1\textwidth}{\footnotesize{\textit{Notes}: Sample of 25 or older respondents who had completed a degree at any point between 2002 and 2017. Total completions: 1,383.}}
\end{figure}

\begin{figure}[H]
\centering
\caption{Degree completions by age}
\vspace{0.5cm}
  \label{fig:degbyage}
    \includegraphics[width=0.75\textwidth]{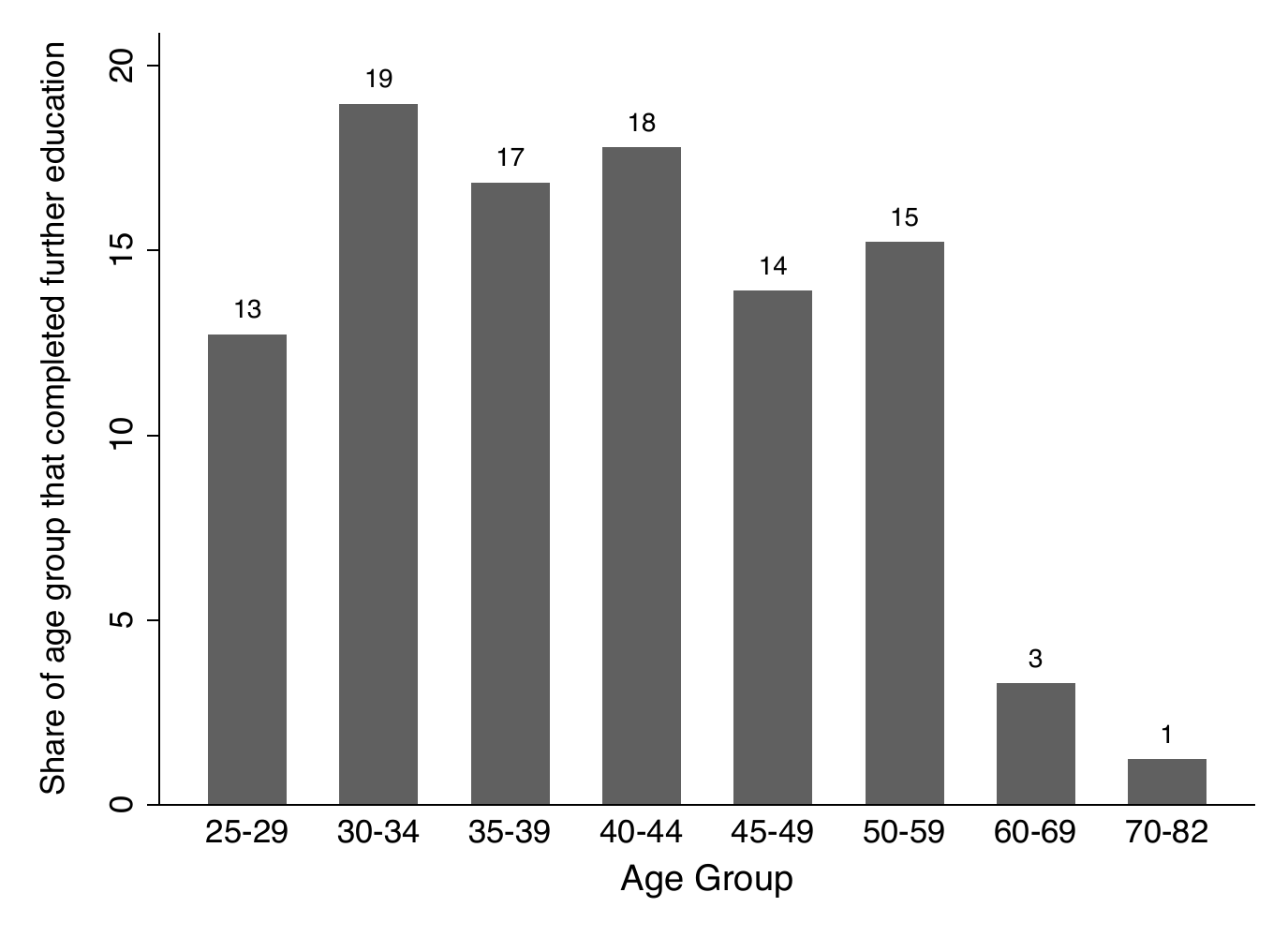}
\end{figure}

\begin{figure}[H]
\centering
\caption{Timing of Completion by Type of Degree}
\vspace{0.5cm}
  \label{fig:yearcompdeg}
    \includegraphics[width=0.75\textwidth]{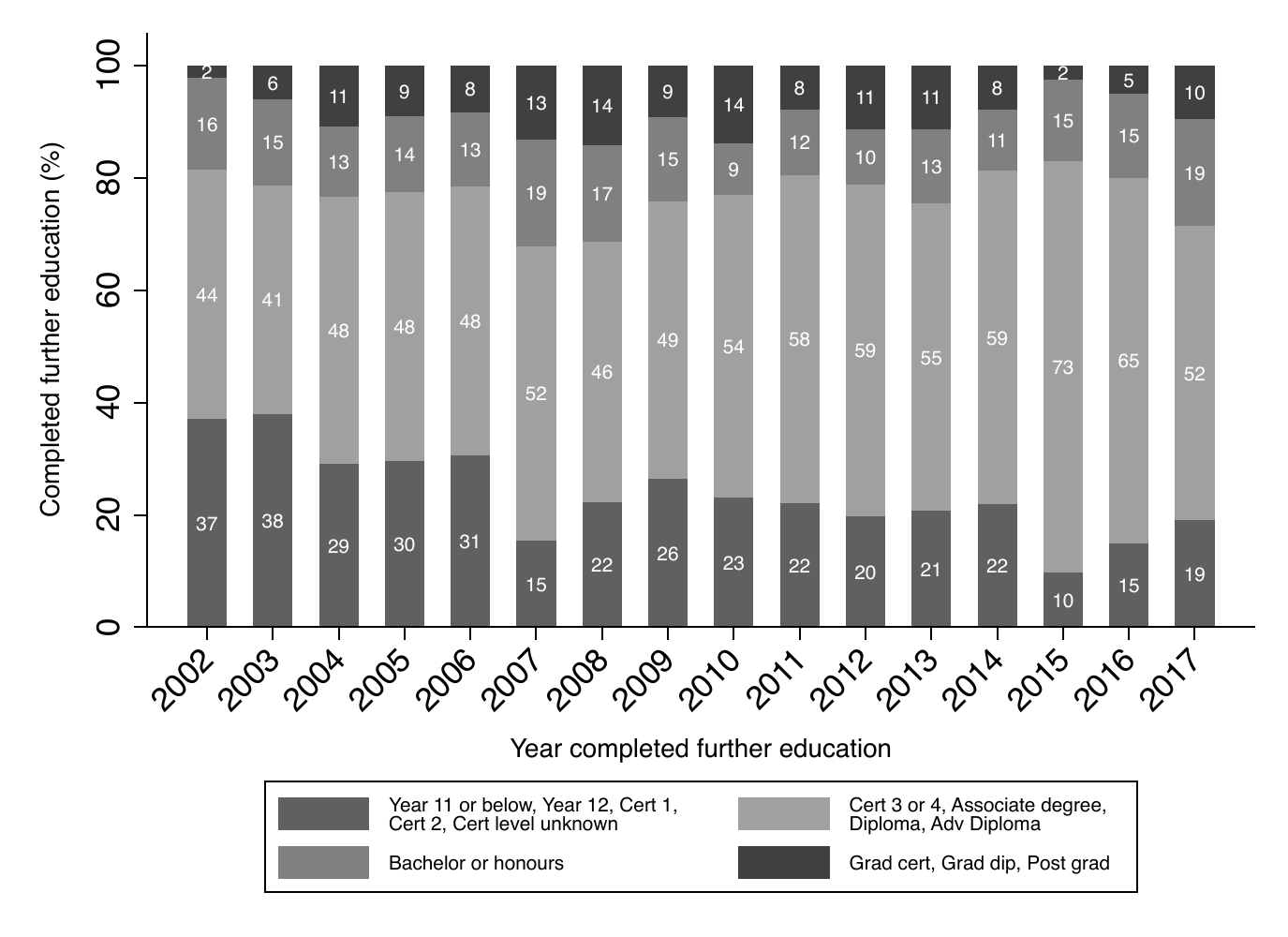}
\parbox{1\textwidth}{\footnotesize{\textit{Notes}: Sample of 25 or older respondents who had completed a degree at any point between 2002 and 2017. Total completions: 1,383.}}
\end{figure}

%\begin{figure}[H]
%\centering
%\caption{Timing of Completion by Type of Degree and Gender}
%\vspace{0.5cm}
%  \label{fig:yearcompdeggen}
%    \includegraphics[width=0.75\textwidth]{_figures/year_completed_by_degree_gender.pdf}
%\parbox{1\textwidth}{\footnotesize{\textit{Notes}: Sample of 25 or older respondents who had completed a degree at any point between 2002 and 2017. Total completions: 1,383.}}
%\end{figure}

\begin{figure}[H]
\centering
\caption{Degree completions by sex}
\vspace{0.5cm}
  \label{fig:degbysex}
    \includegraphics[width=0.75\textwidth]{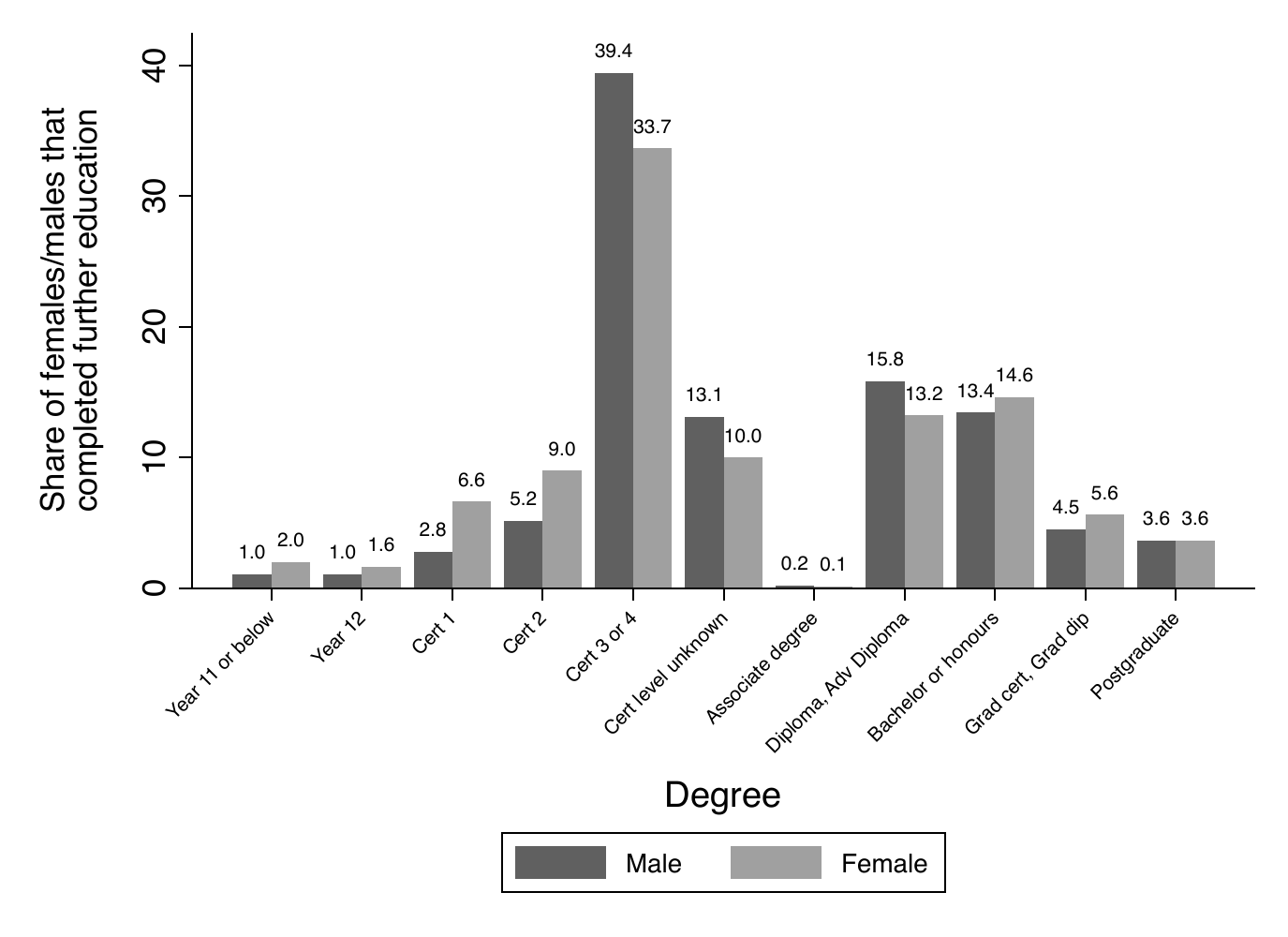}
\end{figure}

\begin{figure}[H]
\centering
\caption{Earnings and Employment by year}
\vspace{0.5cm}
  \label{fig:yearearnempl}
    \includegraphics[width=0.75\textwidth]{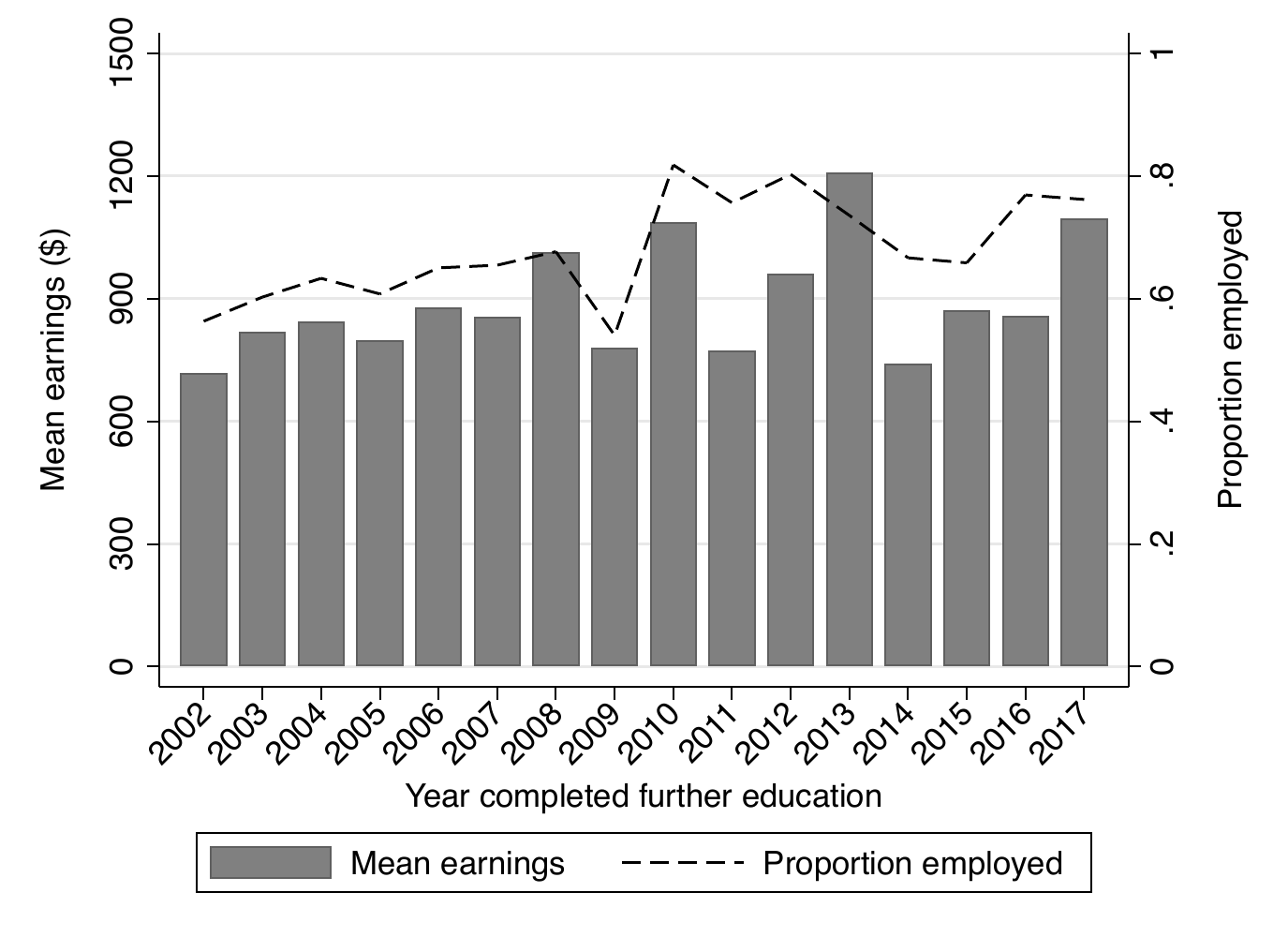}
\end{figure}

%\begin{figure}[H]
%\centering
%\caption{Earnings and Employment by year and sex}
%\vspace{0.5cm}
%  \label{fig:sexearnempl}
%    \includegraphics[width=0.75\textwidth]{_figures/mean_earnings_empl_bysex.pdf}
%\end{figure}

% \clearpage
%\begin{figure}[H]
%\centering
%\caption{Important Features in Heterogeneous Treatment Effects Estimation using T-Learner (GBR): Level Earnings}
%\vspace{0.5cm}
%  \label{fig:featgbr}
%    \includegraphics[width=0.75\textwidth]{_figures/influenceP_GBR_le_100.pdf}
%\parbox{1\textwidth}{\footnotesize{\textit{Notes}: Sample of 25 or older who had completed a degree at any point between 2002 and 2017. Total number of observations 5,441.}}
%\end{figure}
%
%\begin{figure}[H]
%\centering
%\caption{Top 3 Features Distribution of Importance using T-Learner: Level Earnings}
%\vspace{0.5cm}
%  \label{fig:dengbrlev}
%    \includegraphics[width=0.75\textwidth]{_figures/density_GBR_le_100_top3.pdf}
%\parbox{1\textwidth}{\footnotesize{\textit{Notes}: Sample of 25 or older who had completed a degree at any point between 2002 and 2017. Total number of observations 5,441.}}
%\end{figure}
%
%\begin{figure}[H]
%\centering
%\caption{Top and Middle 3 Features Distribution of Importance using T-Learner: Level Earnings}
%\vspace{0.5cm}
%  \label{fig:dengbrlevtopmid3}
%    \includegraphics[width=0.75\textwidth]{_figures/density_GBR_le_100_top3mid3.pdf}
%\parbox{1\textwidth}{\footnotesize{\textit{Notes}: Sample of 25 or older who had completed a degree at any point between 2002 and 2017. Total number of observations 5,441.}}
%\end{figure}

\begin{figure}[H]
\centering
\caption{Important Features in Heterogeneous Treatment Effects Estimation using DR: Level Earnings}
\vspace{0.5cm}
  \label{fig:featgbrDR}
    \includegraphics[width=0.75\textwidth]{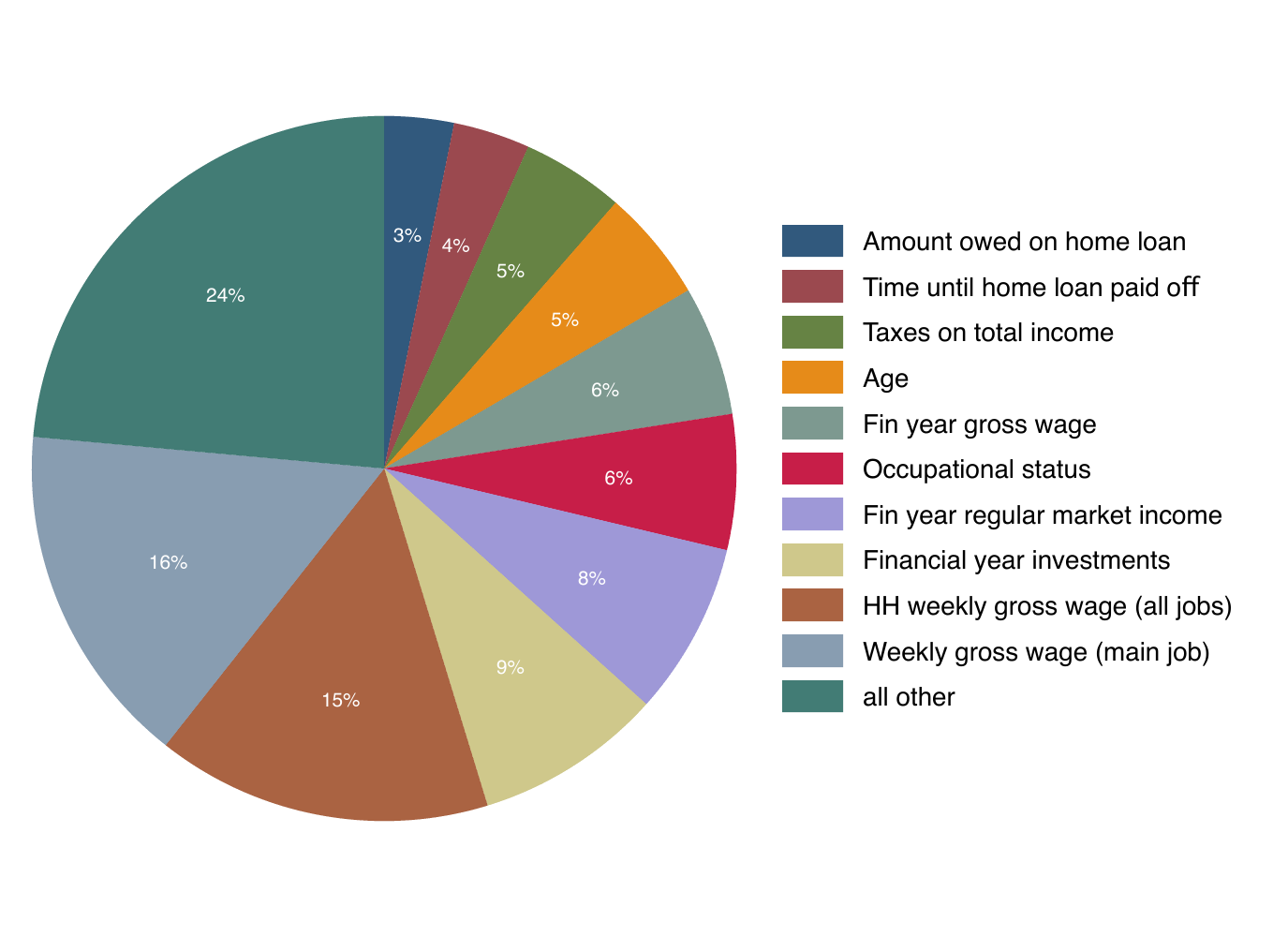}
\parbox{1\textwidth}{\footnotesize{\textit{Notes}: Sample of 25 or older who had completed a degree at any point between 2002 and 2017. Total number of observations 5,441.}}
\end{figure}

\begin{figure}[H]
\centering
\caption{Top 3 Features Distribution of Importance using DR: Level Earnings}
\vspace{0.5cm}
  \label{fig:dengbrlevDR} 
    \includegraphics[width=0.7\textwidth]{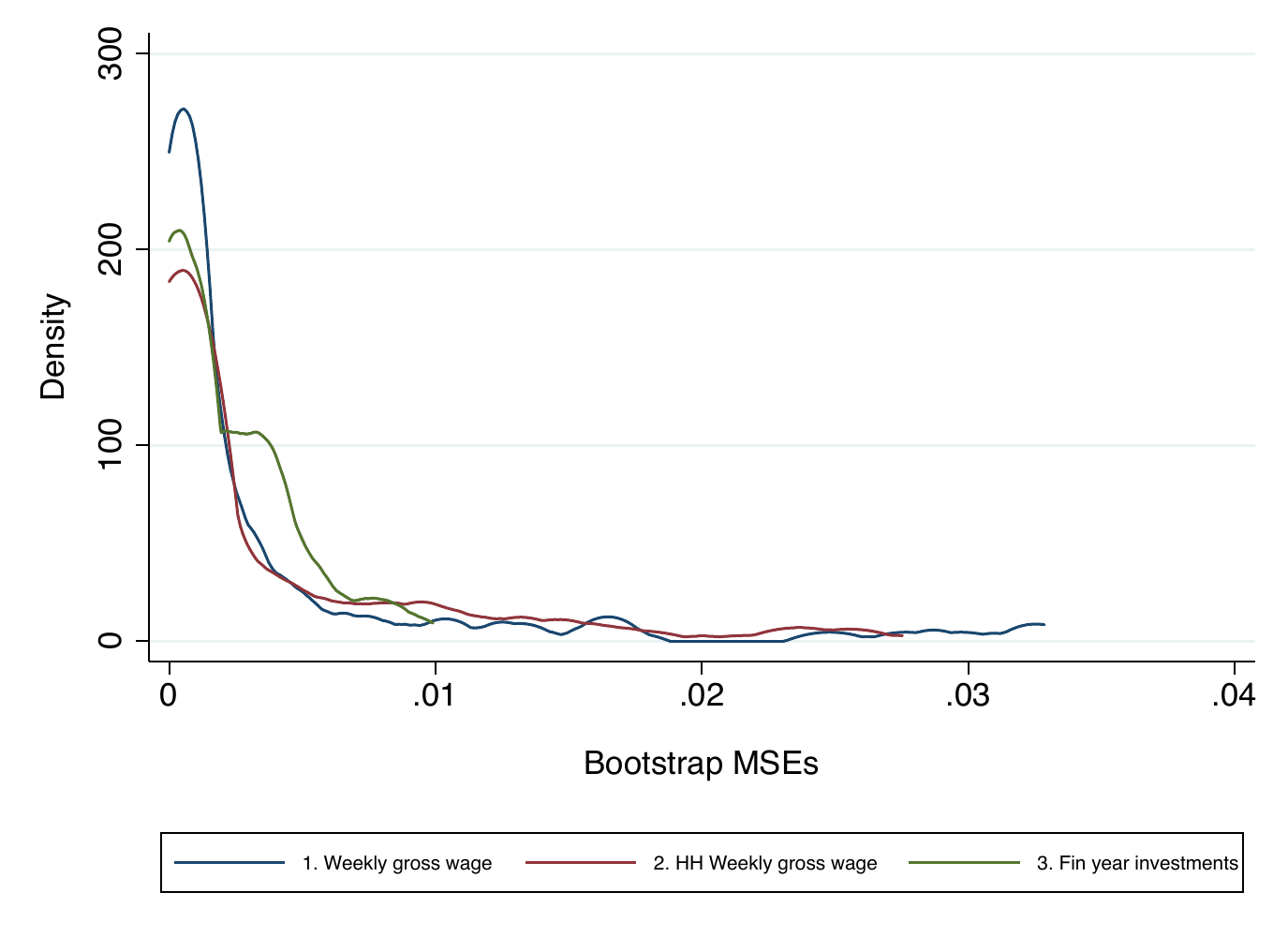}
\parbox{1\textwidth}{\footnotesize{\textit{Notes}: Sample of 25 or older who had completed a degree at any point between 2002 and 2017. Total number of observations 5,441.}}
\end{figure}

%\begin{figure}[H]
%\centering
%\caption{Top and Middle 3 Features Distribution of Importance using DR: Level Earnings}
%\vspace{0.5cm}
%  \label{fig:dengbrlevtopmid3}
%    \includegraphics[width=0.75\textwidth]{_figures/density_GBR_le_100_top3mid3_DR.pdf}
%\parbox{1\textwidth}{\footnotesize{\textit{Notes}: Sample of 25 or older who had completed a degree at any point between 2002 and 2017. Total number of observations 5,441.}}
%\end{figure}

% Results

\begin{figure}[H]
\centering
\caption{Comparison of Treatment Effects across Different Methods}
\vspace{0.5cm}
  \label{fig:method}
    \includegraphics[width=0.7\textwidth]{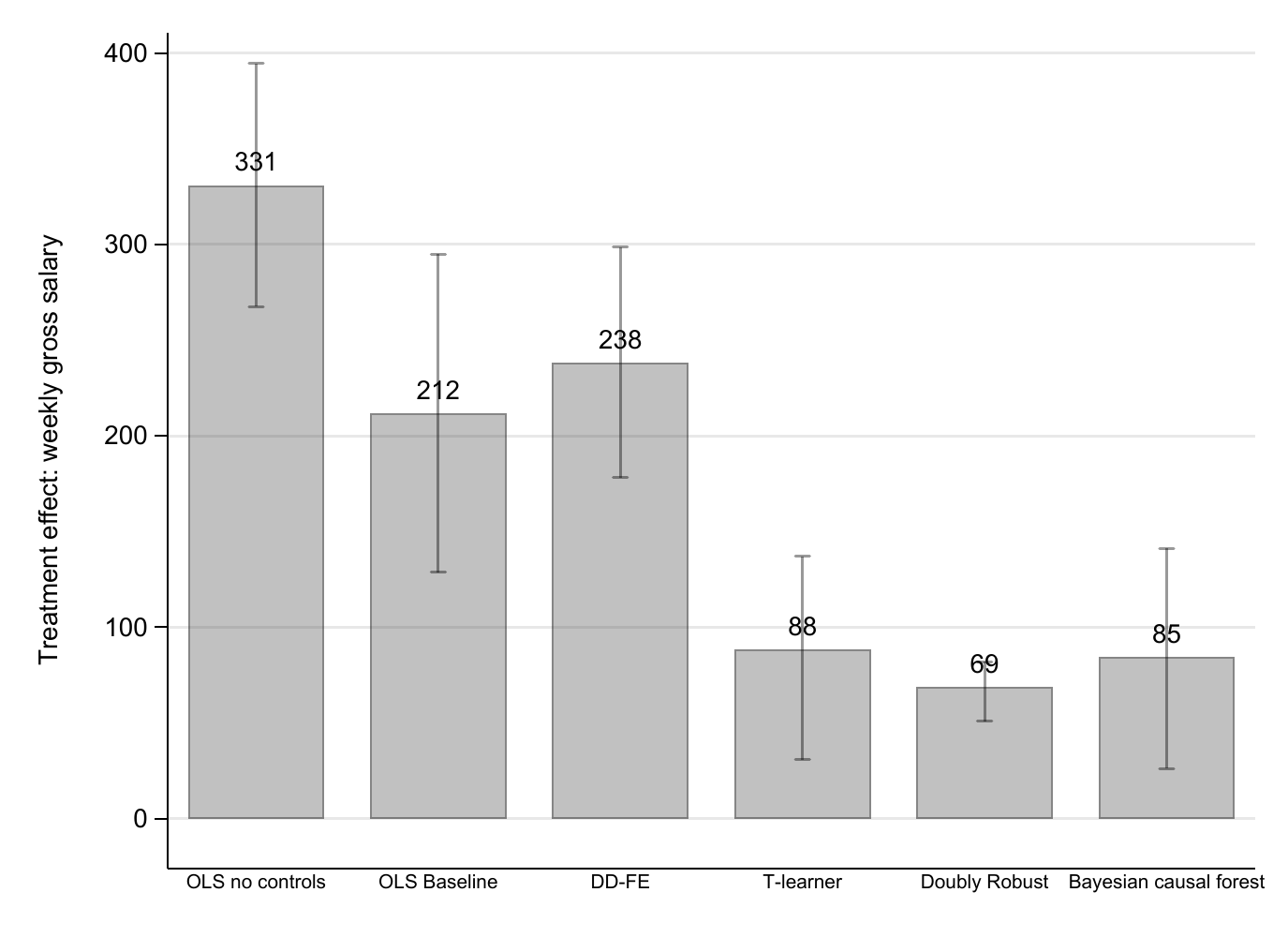}
\parbox{1\textwidth}{\footnotesize{\textit{Notes}: Unless stated otherwise, the method uses a sample of 25 or older respondents who had completed a degree at any point between 2002 and 2017. Total number of observations 5,441. The OLS Baseline model uses the features manually selected in models by Chesters (2015). The Difference-in-Difference Fixed Effects (DD-FE) model uses the same individuals as the other methods but follows them over two waves: 2001 and 2019 (i.e. there are 10,882 person-wave observations); person and wave fixed effects included. The T-learner and Doubly Robust results are based on the Gradient Boosted Regression. The last bar is based on the Bayesian Causal Forest.}}
\end{figure}

\begin{figure}[H]
\centering
\caption{Other Employment Outcomes}
\vspace{0.5cm}
  \label{fig:empl}
    \includegraphics[width=0.75\textwidth]{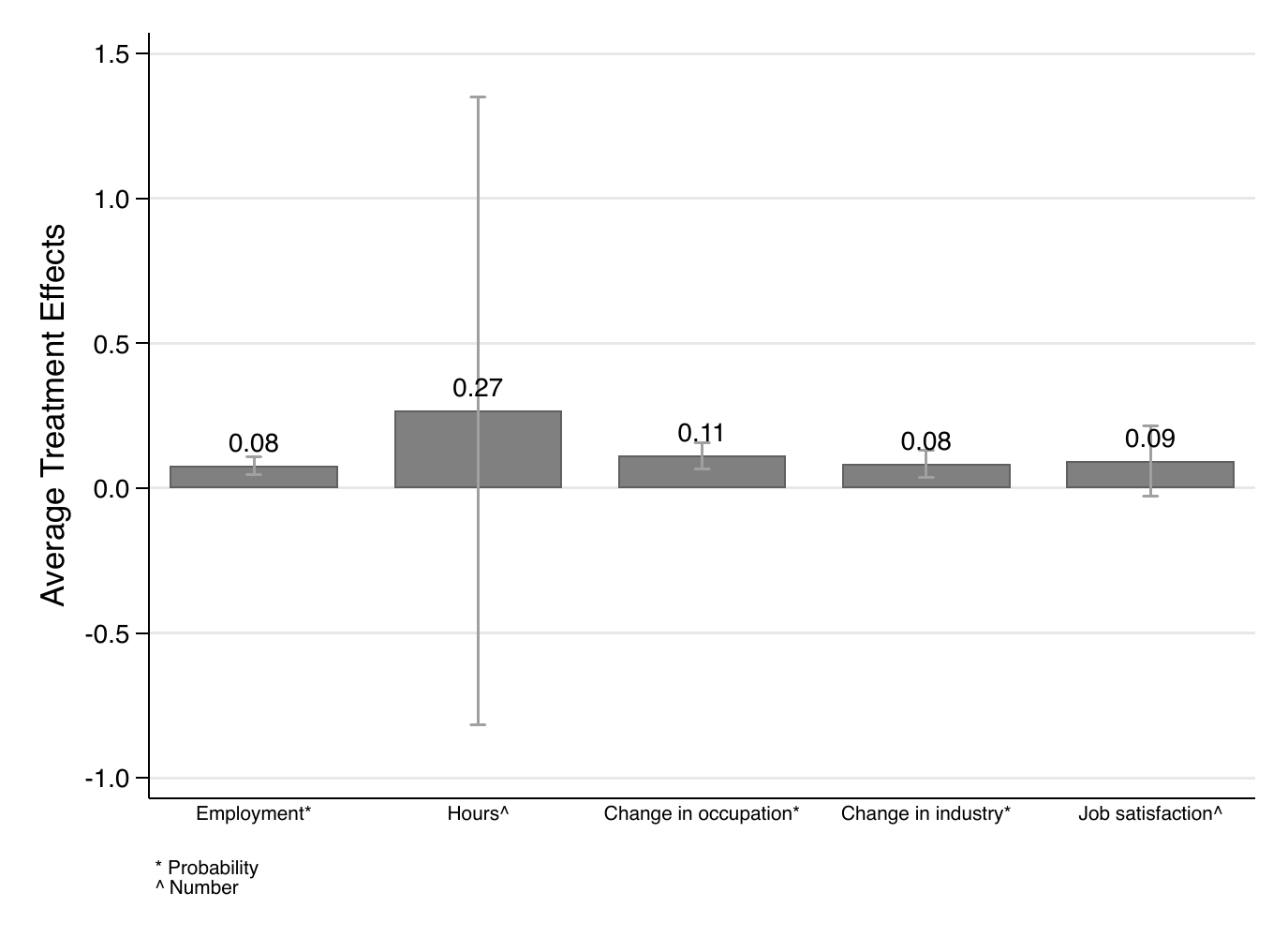}
\parbox{1\textwidth}{\footnotesize{\textit{Notes}: The impact of a new qualification. Sample of people who are 25 or older in 2001. Observation sizes vary depending on the outcome variable. All results are estimated using the LASSO algorithm.}}
\end{figure}

\begin{figure}[H]
\centering
\caption{Earnings HTEs: DR}
\vspace{0.5cm}
  \label{fig:htedr}
    \includegraphics[width=0.75\textwidth]{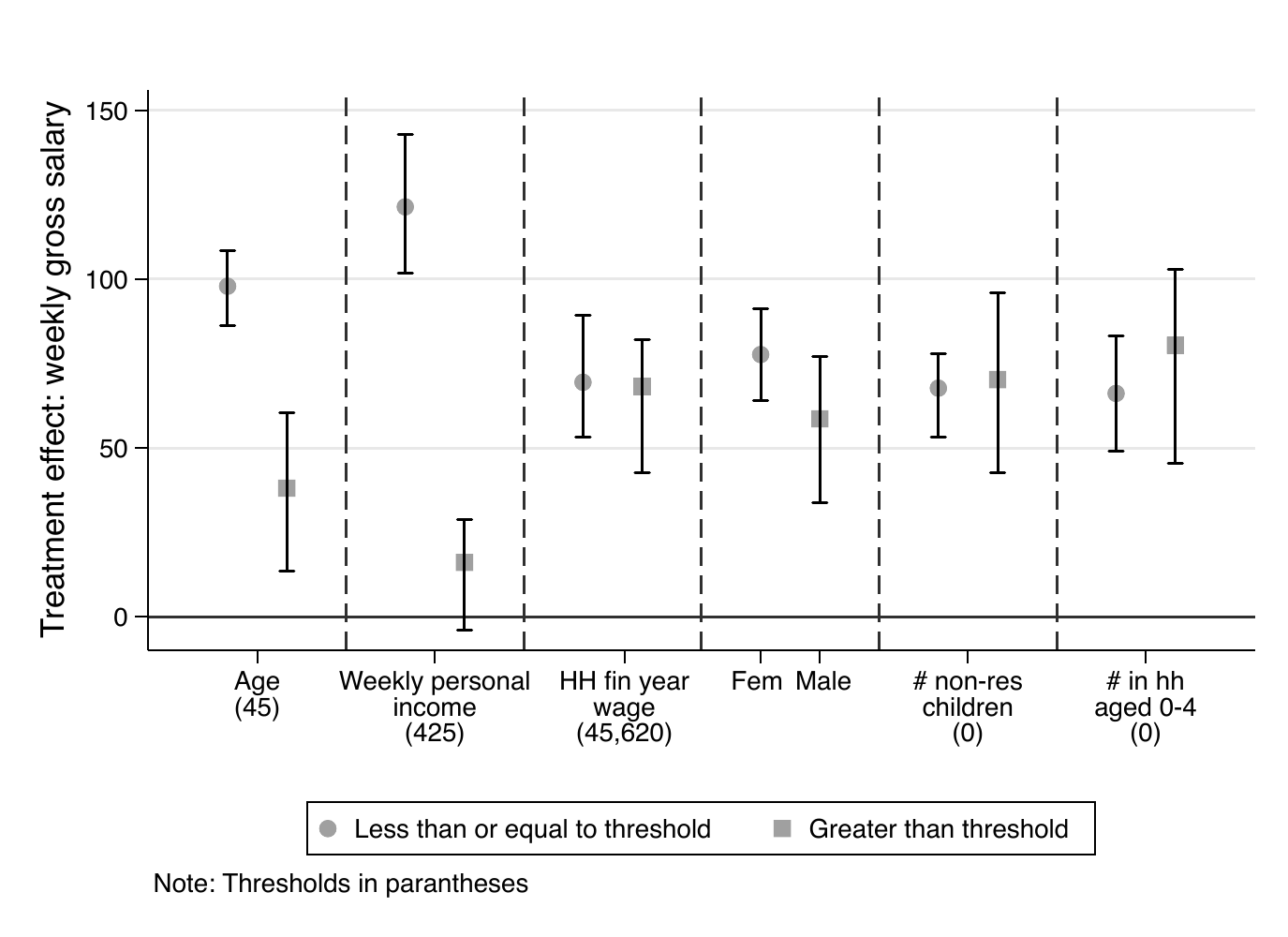}
\parbox{1\textwidth}{\footnotesize{\textit{Notes}: Sample of 25 or older who had completed a degree at any point between 2002 and 2017. Total number of observations 5,441.}}
\end{figure}

%\begin{figure}[H]
%\centering
%\caption{Earnings HTEs: ML and DR}
%\vspace{0.5cm}
%  \label{fig:htemldr}
%    \includegraphics[width=0.75\textwidth]{_figures/hte_earnings_GL_100_MLvDR.pdf}
%\parbox{1\textwidth}{\footnotesize{\textit{Notes}: }}
%\end{figure}

\clearpage
\bibliographystyle{aea}
\bibliography{reed.bib}
%\renewcommand*{\bibfont}{\small}
%\printbibliography 
%\appendix
%\normalsize
%\setcounter{table}{0}
%\renewcommand{\thetable}{A\arabic{table}}

\clearpage

\appendix

%\titleformat{\section}{\Large\bfseries}{\appendixname\ \thesection}{1em}{}
% Adds "Appendix" before the section number

% Define the "Appendix" chapter style
\titleformat{\section}[display]
  {\normalfont\Large\bfseries}{\appendixname\ \thesection}{0.2em}{}

% Define the "Online Appendix" chapter style
\titleformat{name=\section,numberless}[display]
  {\normalfont\Large\bfseries}{Online Appendix}{0.2em}{}

\section{Figures}
%
%\begin{figure}[H]
%\centering
%\caption{Timing of Completion by Age}
%\vspace{0.5cm}
%  \label{fig:yearcompage}
%    \includegraphics[width=0.75\textwidth]{_figures/year_completed_byage.pdf}
%\parbox{1\textwidth}{\footnotesize{\textit{Notes}: Sample of 25 or older respondents who had completed a degree at any point between 2002 and 2017. Total completions: 1,383.}}
%\end{figure}
%
%\begin{figure}[H]
%\centering
%\caption{Value-add in earnings}
%\vspace{0.5cm}
%  \label{fig:earnings}
%    \includegraphics[width=0.75\textwidth]{_figures/valadearn.pdf}
%\parbox{1\textwidth}{\footnotesize{\textit{Notes}: }}
%\end{figure}

\begin{figure}[H]
\centering
\caption{Value-add in earnings: 25-45 year-old sample}
\vspace{0.5cm}
  \label{fig:valadle46}
    \includegraphics[width=0.65\textwidth]{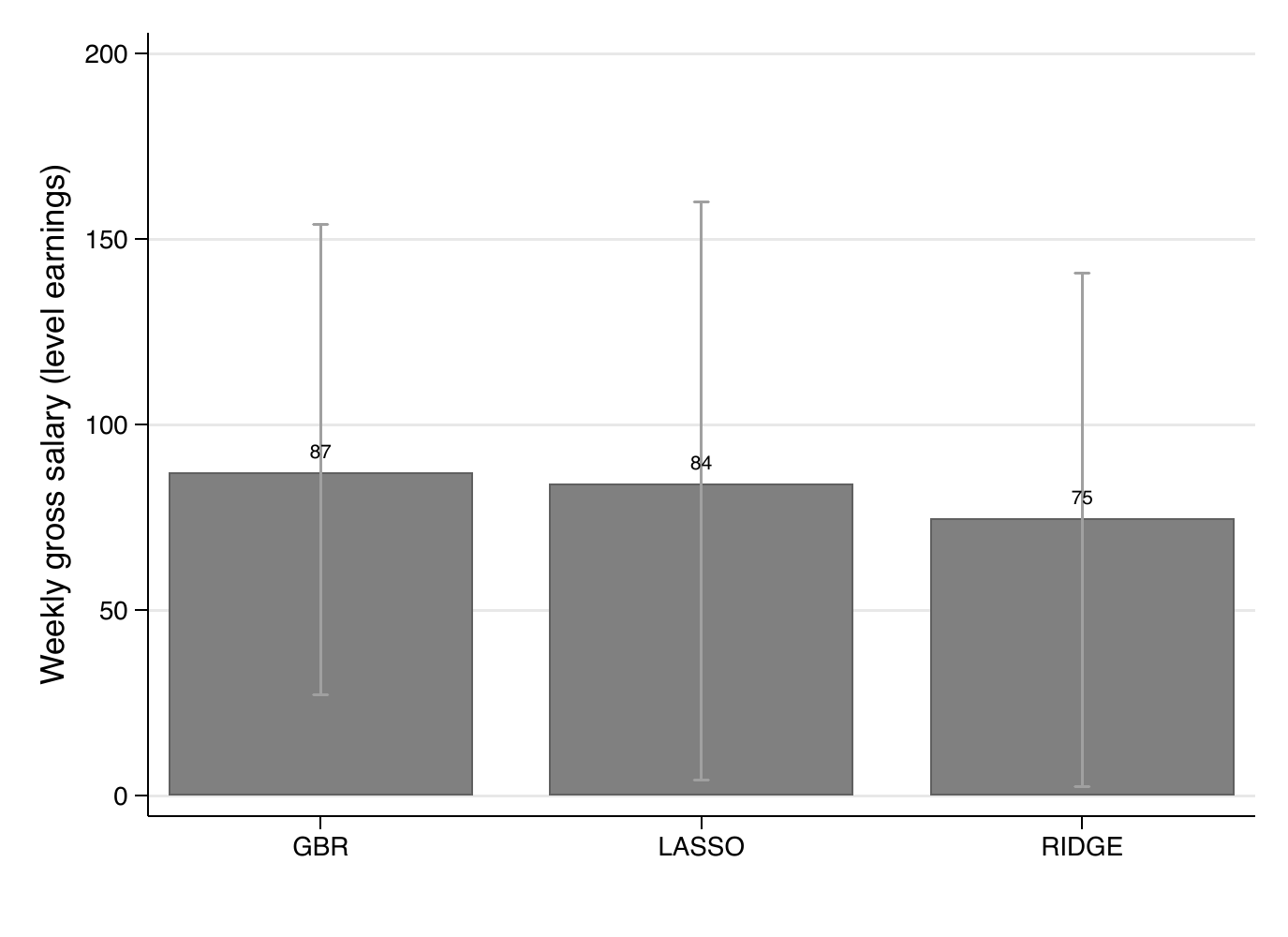}
\parbox{1\textwidth}{\footnotesize{\textit{Notes}: Sample of 25-45 who had completed a degree at any point between 2002 and 2017. Total number of observations 3,684.}}
\end{figure}

\begin{figure}[H]
\centering
\caption{Value-add in log earnings}
\vspace{0.5cm}
  \label{fig:valadlelog}
    \includegraphics[width=0.65\textwidth]{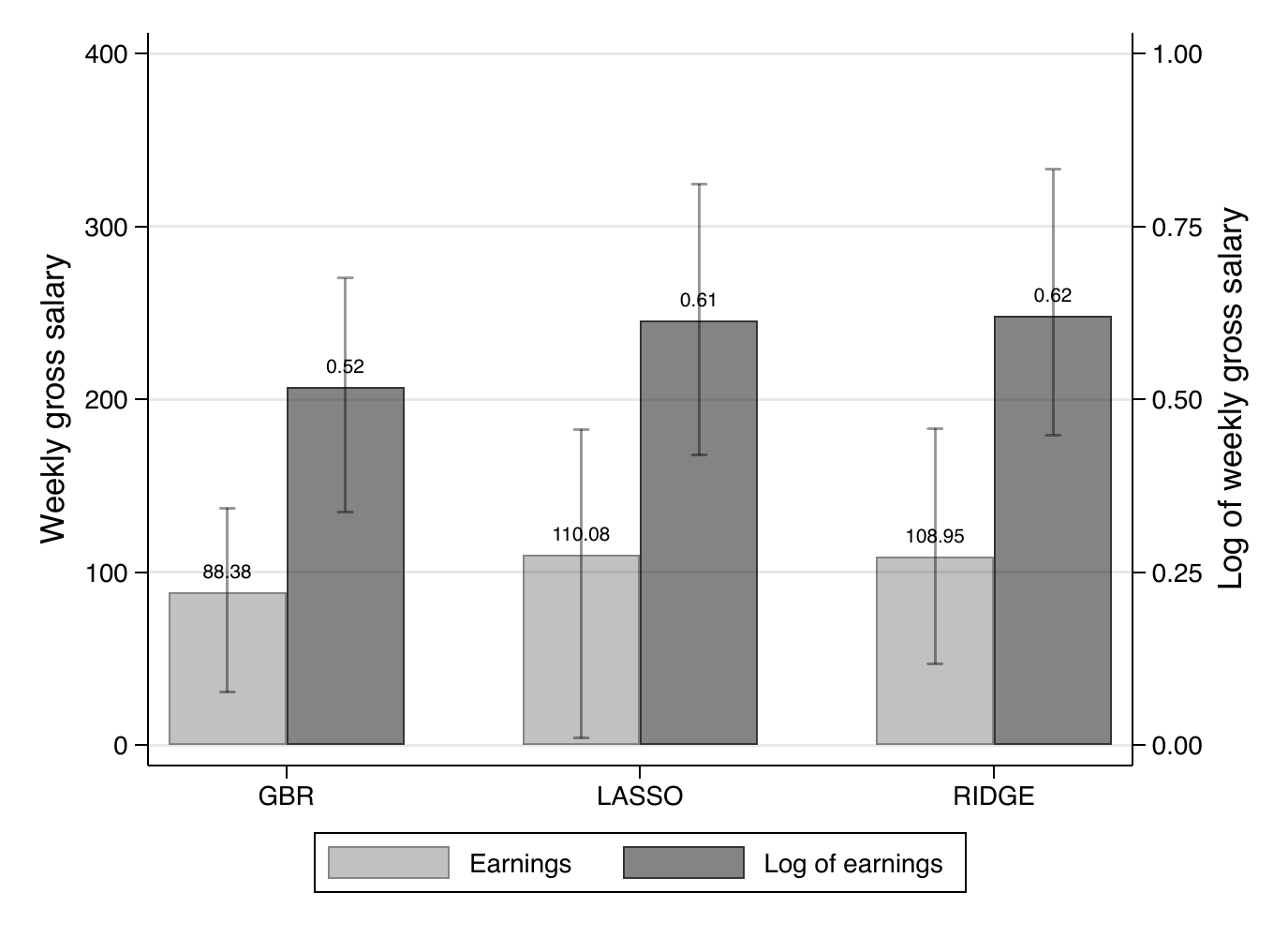}
\parbox{1\textwidth}{\footnotesize{\textit{Notes}: Sample of 25 or older who had completed a degree at any point between 2002 and 2017. Total number of observations 5,441.}}
\end{figure}
%
%% Pie charts RIDGE and LASSO
%\begin{figure}[H]
%\centering
%\caption{Important Features in Heterogeneous Treatment Effects Estimation using RIDGE: Level Earnings}
%\vspace{0.5cm}
%  \label{fig:featridge}
%    \includegraphics[width=0.75\textwidth]{_figures/influenceP_RIDGE_le_100.pdf}
%\parbox{1\textwidth}{\footnotesize{\textit{Notes}: Sample of 25 or older who had completed a degree at any point between 2002 and 2017. Total number of observations 5,441.}}
%\end{figure}
%
%\begin{figure}[H]
%\centering
%\caption{Important Features in Heterogeneous Treatment Effects Estimation using LASSO: Level Earnings}
%\vspace{0.5cm}
%  \label{fig:featlasso}
%    \includegraphics[width=0.75\textwidth]{_figures/influenceP_LASSO_le_100.pdf}
%\parbox{1\textwidth}{\footnotesize{\textit{Notes}: Sample of 25 or older who had completed a degree at any point between 2002 and 2017. Total number of observations 5,441.}}
%\end{figure}
%

\begin{figure}[H]
\centering
\caption{Important Features in Heterogeneous Treatment Effects Estimation using T-Learner (GBR): Level Earnings}
\vspace{0.5cm}
  \label{fig:featgbr}
    \includegraphics[width=0.7\textwidth]{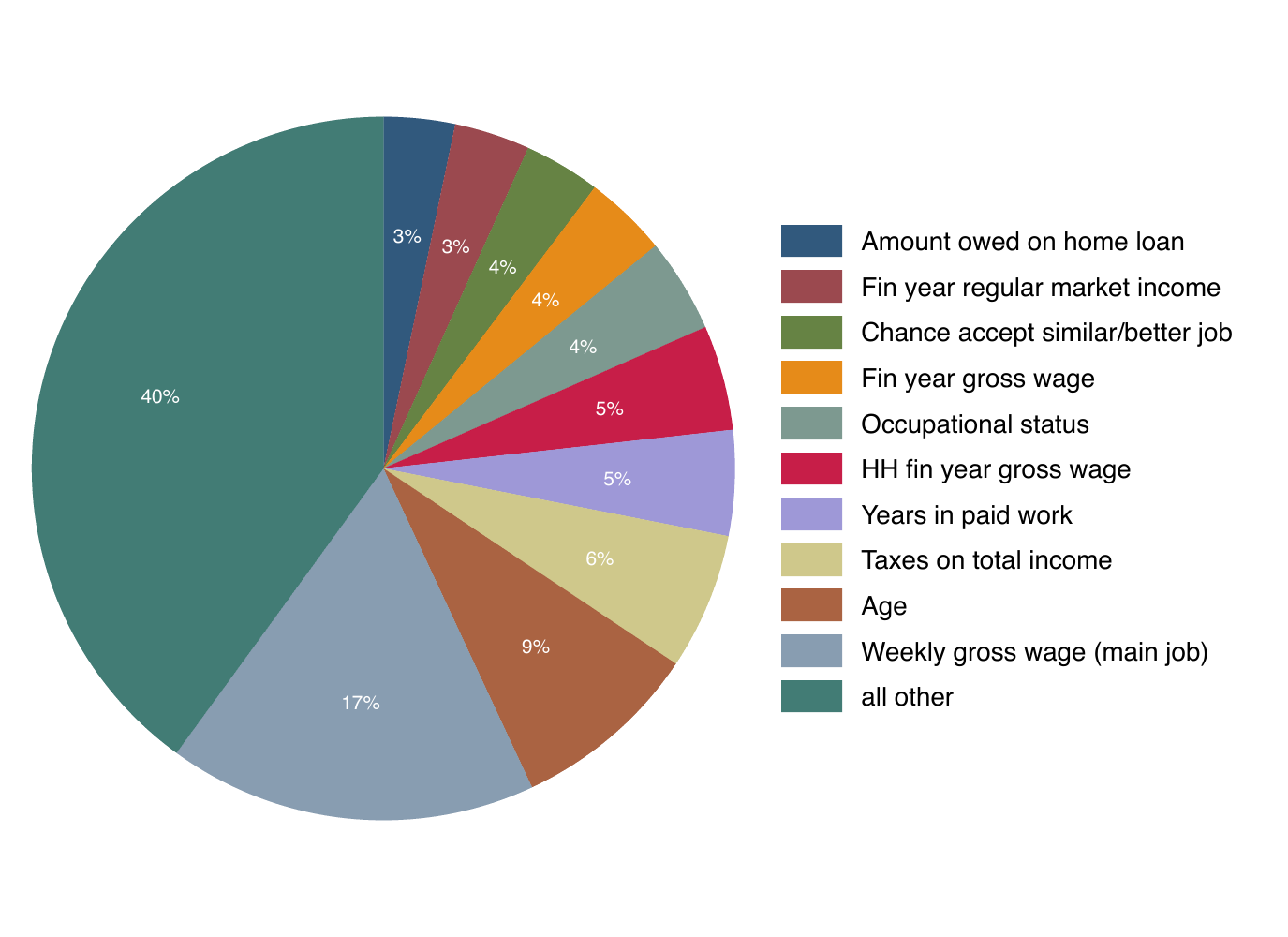}
\parbox{1\textwidth}{\footnotesize{\textit{Notes}: Sample of 25 or older who had completed a degree at any point between 2002 and 2017. Total number of observations 5,441.}}
\end{figure}

\begin{figure}[H]
\centering
\caption{Important Features in Heterogeneous Treatment Effects Estimation using panel sample (GBR): Level Earnings}
\vspace{0.5cm}
  \label{fig:featgbr_panel}
    \includegraphics[width=0.7\textwidth]{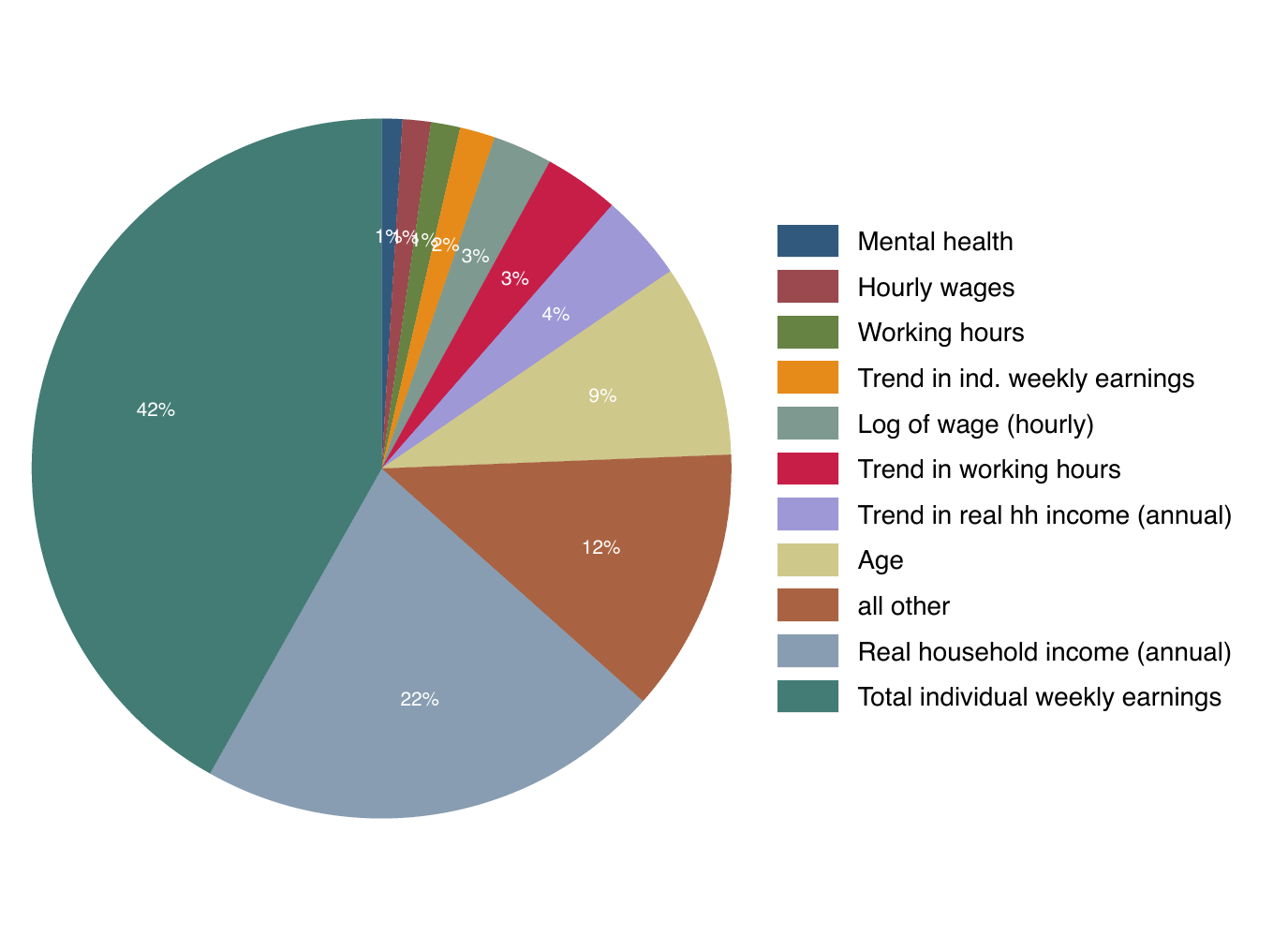}
\parbox{1\textwidth}{\footnotesize{\textit{Notes}: Sample of 21 or older individuals who had completed a degree at any point between 2003 and 2015, inclusive. Outcomes are defined 4 years after a study spell began and features are defined in both the two years preceding the start of a study spell. There were 1,814 individuals who started and completed a further educational degree, and 60,945 non-unique control observations who never completed a further degree.}}
\end{figure}

\clearpage

% Change the name of the appendix from "Appendix" to "Online Appendix"
\renewcommand{\appendixname}{Online Appendix}

\section{Main Sample}

\subsection{Sample Selection}

Our analysis sample includes everyone who was 25 or above and not currently studying in 2001, who are observed in both 2001 and 2019 in terms of the outcome and treatment variables. 

We delete any individuals who were currently studying in 2001 if:
\begin{itemize}
  \item They reported currently studying full part or part time for the main survey 
  \item According to the calendar, they have undertaken any full time or part time studies
  \item They are currently receiving Abstudy/Austudy payment or had received these last financial year 
  \item They have cited study as the reason for not looking for work 
\end{itemize}

1078 individuals were deleted after applying this sample exclusion.

\subsection{Variable description}

\subsubsection{Outcome Variables}
Weekly earnings from main job in 2019 (\textit{w19\textunderscore{}wscmei}) records the weekly earnings from the main job for the individual in 2019. 

Employed in 2019 (\textit{w19\textunderscore{}employed}) records whether the individual is employed in 2019 or not. 

Weekly earnings from all jobs in 2019 (\textit{w19\textunderscore{}earning}) records the weekly earnings from all jobs for the individual in 2019. 

Working hours in 2019 (\textit{w19\textunderscore{}wkhr}) records the total number of hours the individual works in all jobs in a week on average. Working hours are set to 0 for those not working. 

\subsubsection{Treatment Variables: Re-education}
Re-education based on highest attainment\footnote{The HILDA variable on highest attainment was constructed using three components: the age the individual left school, the highest education attainment in the previous wave and the current level of secondary school attained or currently studying for.} (\textit{reduhl}) records whether the individual has had re-education between 2002 and 2017, based on whether there was a change in the highest education level attained stated in the two years.  

Re-education completion based on detailed qualifications (\textit{redudl}) records whether the individual has completed any one of the following qualifications since last interviewed between 2002 and 2017\footnote{Refer to https://en.wikipedia.org/wiki/Australian\textunderscore{}Qualifications\textunderscore{}Framework for how most of these degrees are situated relative to each other in a hierarchy and the duration of these qualifications.}:

\begin{itemize}
  \item Trade certificate or apprenticeship
  \item Technicians cert/Advanced certificate
  \item Teaching qualification 
  \item Nursing qualification 
  \item Associate Degree
  \item Advance Diploma (3 years full time or equivalent)
  \item Bachelor degree but not honours
  \item Certificate I
  \item Certificate II
  \item Certificate III
  \item Certificate IV
  \item Certificate of unknown level
  \item Doctorate
  \item Diploma NFI
  \item Diploma (2 years full time or equivalent)
  \item Graduate Certificate
  \item Graduate Diploma 
  \item Honours 
  \item Masters 
  \item Other 
\end{itemize}  

Re-education completion based on both highest attainment and detailed qualifications (\textit{redufl}) records whether the individual has completed re-education based on both the variables \textit{reduhl} and \textit{redudl}. When either of these variables has a value of 1, this variable will take on the value of 1. 

\subsubsection{Input Variables}
For each variable, missing values (if any) have been set to zero and a new binary variable has been generated to indicate the observations that are missing. 

\emph{Demographics}

Female (\textit{p\textunderscore{}fem}) records whether the individual is female. 

Age group in 2001 records whether in 2001 the individual was:
\begin{itemize}
  \item Aged 25-34 (\textit{p\textunderscore{}age1})
  \item Aged 35-44 (\textit{p\textunderscore{}age2})
  \item Aged 45-54 (\textit{p\textunderscore{}age3})
  \item Aged 55-64 (\textit{p\textunderscore{}age4})
  \item Aged 65 and above (\textit{p\textunderscore{}age5})
 \end{itemize} 

Country of birth records whether or not an individual was born in:
\begin{itemize}
  \item Australia and not indigenous (\textit{p\textunderscore{}cob1}) 
  \item English speaking countries (\textit{p\textunderscore{}cob2})
  \item Non-English speaking countries (\textit{p\textunderscore{}cob3})
  \item Indigenous (\textit{p\textunderscore{}cob4})
\end{itemize}  
  
Poor English speaking abilities (\textit{p\textunderscore{}poeng}) records whether the individual has poor English speaking abilities. 

Remoteness records whether the individual lives in:
\begin{itemize}
  \item A major city (\textit{p\textunderscore{}urdg1})
  \item An inner region (\textit{p\textunderscore{}urdg2}) 
  \item Outer and remote areas or migratory in nature (\textit{p\textunderscore{}urdg3})
\end{itemize}  
  
Marital status in 2001 records whether in 2001 the individual was:
\begin{itemize}
  \item Married (\textit{p\textunderscore{}mar1})
  \item De facto (\textit{p\textunderscore{}mar2})
  \item Separated (\textit{p\textunderscore{}mar3})
  \item Divorced (\textit{p\textunderscore{}mar4})
  \item Widowed (\textit{p\textunderscore{}mar5})
  \item Single and never been married (\textit{p\textunderscore{}mar6})
\end{itemize}  
  
\emph{Parental Status}

Number of dependents in 2001 (\textit{p\textunderscore{}noch}) records the number of dependent children the individual had in 2001. 

\emph{Physical Health}

Severity of health conditions in 2001 records whether the individual had:
\begin{itemize}
  \item No health conditions (\textit{p\textunderscore{}ddeg1})
  \item A mild condition (\textit{p\textunderscore{}ddeg2})
  \item A moderate condition (\textit{p\textunderscore{}ddeg3})
  \item A severe condition (\textit{p\textunderscore{}ddeg4})
\end{itemize}  
  
\emph{Labour Force Variables}

Labour market status in 2001 records whether the individual was:
\begin{itemize}
  \item Employed (\textit{p\textunderscore{}lfs1})
  \item Unemployed (\textit{p\textunderscore{}lfs2})
  \item Not in the labour market (\textit{p\textunderscore{}lfs3})
\end{itemize} 

Extent of working hour match with preferences in 2001 records whether the match between the individual’s total weekly working hours across all jobs and their preferred number of working hours made them:
\begin{itemize}
  \item Not working (\textit{p\textunderscore{}whp1})
  \item Underemployed by at least 4 hours a week  (\textit{p\textunderscore{}whp2})
  \item Roughly Matched: Preferred and Actual Hours Worked differ by less than 4 hours a week (\textit{p\textunderscore{}whp3})
  \item Overemployed by at least 4 hours a week  (\textit{p\textunderscore{}whp4})
\end{itemize}  
  
Employee type in 2001 records whether the individual was:
\begin{itemize}
  \item Not working (\textit{p\textunderscore{}emp1})
  \item An employee (\textit{p\textunderscore{}emp2})
  \item An employee of own business (\textit{p\textunderscore{}emp3})
  \item Self Employed (\textit{p\textunderscore{}emp4})
  \item Unpaid family worker (\textit{p\textunderscore{}emp5})
\end{itemize}  
  
Contract type in 2001 records whether the individual was:
\begin{itemize}
  \item Not working (\textit{p\textunderscore{}con1})
  \item On a fixed term contract (\textit{p\textunderscore{}con2})
  \item On a casual contract (\textit{p\textunderscore{}con3})
  \item On a permanent contract (\textit{p\textunderscore{}con4})
  \item On other types of contracts (\textit{p\textunderscore{}con5})
\end{itemize}  
  
Occupation in 2001 records whether the individual was working as:
\begin{itemize}
  \item Not working (\textit{p\textunderscore{}occ1})
  \item Armed forces (\textit{p\textunderscore{}occ2})
  \item Legislators, Senior Officials and Managers (\textit{p\textunderscore{}occ3})
  \item Professionals (\textit{p\textunderscore{}occ4})
  \item Technicians and Associate Professionals (\textit{p\textunderscore{}occ5})
  \item Clerks (\textit{p\textunderscore{}occ6})
  \item Service Workers and Shop and Market Sales Workers (\textit{p\textunderscore{}occ7})
  \item Skilled Agriculture and Fishery Workers (\textit{p\textunderscore{}occ8})
  \item Craft and Related Trades Workers (\textit{p\textunderscore{}occ9})
  \item Plant and Machine Operators and Assemblers (\textit{p\textunderscore{}occ10})
  \item Elementary Occupations (\textit{p\textunderscore{}occ11})
\end{itemize}  
  
Household income in 2001 (\textit{p\textunderscore{}rehdi}) records the real value of the individual’s total household income indexed at 2012 price levels and adjusted for household size.
 
Partner labour force status in 2001 records whether the individual had:
\begin{itemize}
  \item No partner or no resident partner (\textit{p\textunderscore{}plfs1})
  \item A partner who was employed (\textit{p\textunderscore{}plfs2})
  \item A partner who was unemployed (\textit{p\textunderscore{}plfs3})
  \item A partner who was not in the labour force (\textit{p\textunderscore{}plfs4})
\end{itemize}  
  
\emph{Parental information}

Father’s country of birth records whether or not the individual’s father was born in:
\begin{itemize}
  \item Australia (\textit{p\textunderscore{}fcob1}) 
  \item English speaking countries (\textit{p\textunderscore{}fcob2})
  \item Non-English speaking countries or indigenous (\textit{p\textunderscore{}fcob3})
\end{itemize}  
 
Mother’s country of birth records whether or not the individual’s mother was born in:
\begin{itemize}
  \item Australia (\textit{p\textunderscore{}mcob1}) 
  \item English speaking countries (\textit{p\textunderscore{}mcob2})
  \item Non-English speaking countries or indigenous (\textit{p\textunderscore{}mcob3})
\end{itemize}  
 
Father’s education records whether the individual’s father’s highest education, as reported in 2005, was:
\begin{itemize}
  \item None (\textit{p\textunderscore{}fedu1})
  \item Primary (\textit{p\textunderscore{}fedu2})
  \item Below secondary (\textit{p\textunderscore{}fedu3})
  \item Secondary (\textit{p\textunderscore{}fedu4})
  \item Post-secondary, non-university (\textit{p\textunderscore{}fedu5})
  \item Post-secondary, university (\textit{p\textunderscore{}fedu6}) 
\end{itemize}
 
Mother’s education records whether the individual’s mother’s highest education, as reported in 2005, was:
\begin{itemize}
  \item None (\textit{p\textunderscore{}medu1})
  \item Primary (\textit{p\textunderscore{}medu2})
  \item Below secondary (\textit{p\textunderscore{}medu3})
  \item Secondary (\textit{p\textunderscore{}medu4})
  \item Post-secondary, non-university (\textit{p\textunderscore{}medu5})
  \item Post-secondary, university (\textit{p\textunderscore{}medu6}) 
\end{itemize}  
 
Father undertaken post-school qualification through employer or non-tertiary means (\textit{p\textunderscore{}fpsm}) records whether the individual’s father had undertaken his highest qualification through employers or other channels other than tertiary education, as reported in 2005. 

Mother undertaken post-school qualification through employer or non-tertiary means (\textit{p\textunderscore{}mpsm}) records whether the individual’s mother had undertaken his highest qualification through employers or other channels other than tertiary education, as reported in 2005. 

Father’s Employment at age 14 records whether the individual’s father was working when they were aged 14, in the following categories:
\begin{itemize}
  \item Father deceased or not living with respondent (\textit{p\textunderscore{}femp1})
  \item Father not employed (\textit{p\textunderscore{}femp2})
  \item Father employed (\textit{p\textunderscore{}femp3})
\end{itemize}  
 
Mother’s Employment at age 14 (\textit{p\textunderscore{}memp}) records whether the individual’s mother was working when they were aged 14, in the following categories:
\begin{itemize}
  \item Mother deceased or not living with respondent (\textit{p\textunderscore{}memp1})
  \item Mother not employed (\textit{p\textunderscore{}memp2})
  \item Mother employed (\textit{p\textunderscore{}memp3})
\end{itemize}  
 
Father substantially unemployed growing up records whether the individual’s father had been unemployed for 6 months or more when they were growing up, in the following categories:
\begin{itemize}
  \item Father not living with respondent (\textit{p\textunderscore{}fsue1})
  \item Father not substantially unemployed (\textit{p\textunderscore{}fsue2})
  \item Father substantially unemployed (\textit{p\textunderscore{}fsue3})
\end{itemize} 

Father’s Occupation records whether at age 14 the individual’s father was last known working as:
\begin{itemize}
  \item Father not in household (\textit{p\textunderscore{}focc1})
  \item Armed forces (\textit{p\textunderscore{}focc2})
  \item Legislators, Senior Officials and Managers (\textit{p\textunderscore{}focc3})
  \item Professionals (\textit{p\textunderscore{}focc4})
  \item Technicians and Associate Professionals (\textit{p\textunderscore{}focc5})
  \item Clerks (\textit{p\textunderscore{}focc6})
  \item Service Workers and Shop and Market Sales Workers (\textit{p\textunderscore{}focc7})
  \item Skilled Agriculture and Fishery Workers (\textit{p\textunderscore{}focc8})
  \item Craft and Related Trades Workers (\textit{p\textunderscore{}focc9})
  \item Plant and Machine Operators and Assemblers (\textit{p\textunderscore{}focc10})
  \item Elementary Occupations (\textit{p\textunderscore{}focc11})
\end{itemize}  
 
Mother’s Occupation records whether at age 14 the individual’s mother last known working as:
\begin{itemize}
  \item Moher not in household (\textit{p\textunderscore{}focc1})
  \item Armed forces (\textit{p\textunderscore{}mocc2})
  \item Legislators, Senior Officials and Managers (\textit{p\textunderscore{}mocc3})
  \item Professionals (\textit{p\textunderscore{}mocc4})
  \item Technicians and Associate Professionals (\textit{p\textunderscore{}mocc5})
  \item Clerks (\textit{p\textunderscore{}mocc6})
  \item Service Workers and Shop and Market Sales Workers (\textit{p\textunderscore{}mocc7})
  \item Skilled Agriculture and Fishery Workers (\textit{p\textunderscore{}mocc8})
  \item Craft and Related Trades Workers (\textit{p\textunderscore{}mocc9})
  \item Plant and Machine Operators and Assemblers (\textit{p\textunderscore{}mocc10})
  \item Elementary Occupations (\textit{p\textunderscore{}mocc11})
\end{itemize}  
  
\emph{Non-cognitive variables}

Well-being in 2001 (\textit{p\textunderscore{}losat}) records the life satisfaction score, which ranges from 0 to 10, of the individual reported in 2001. A higher score means the individual is more satisfied with his/her life.

Attitude towards having job in 2001 (\textit{p\textunderscore{}jbwk}) records the average score of attitude towards having a job reported by the individual in 2001 across two items (\textit{p\textunderscore{}jadnm} and \textit{p\textunderscore{}jahpj}), in a scale ranging from 1 to 7, with a higher score indicating a more favourable attitude towards having a job. 

Enjoy job without needing money in 2001 (\textit{p\textunderscore{}jadnm}) records the extent the individual agreed with the statement that the person would enjoy having a job even if they did not need the money in 2001, in a scale ranging from 1 to 7, with a higher score indicating more agreement.

Important to have paying job in 2001 (\textit{p\textunderscore{}jahpj}) records the extent the individual agreed with the statement that in order to be happy in life it is important to have a paying job in 2001, in a scale ranging from 1 to 7, with a higher score indicating more agreement.

\emph{Prior Year Outcome variables}

Mental health in 2001 (\textit{p\textunderscore{}mh01}). This is the transformed mental health scores from the aggregation of mental health items of the SF-36 Health Survey, as reported by the individual in 2001. It ranges from 0 to 100, with higher scores indicating better mental health.   

Mental health in 2001 below norm (\textit{p\textunderscore{}mb01}) records whether the individual’s mental health scores for 2001 was below the average of mental health scores across our analytical sample for that year. 

Working hours in 2001 (\textit{p\textunderscore{}wh01}) records the number of hours the individual works across all jobs in a week on average. Working hours are set to 0 for those not working.

Hourly Wages in 2001 (\textit{p\textunderscore{}hrw01}) records the average hourly wage of the individual’s main job in 2001. Hourly wages are set to 0 for those not working and set to missing for those reporting working more than 100 hours a week. 

\subsubsection{Variables that are not included in the model}

The unique person identifier (\textit{xwaveid}).

Completed re-education after 2017 based on highest education (\textit{rehllt}) records whether the individual had only completed their re-education after 2017, comparing their education level in 2017 and 2019.

Completed re-education after 2017 based on detailed qualifications (\textit{redllt}) records whether the individual has completed any one of the following qualifications since last interviewed between 2018 and 2019:

\begin{itemize}
  \item Trade certificate or apprenticeship
  \item Technicians cert/Advanced certificate
  \item Teaching qualification 
  \item Nursing qualification 
  \item Associate Degree
  \item Advance Diploma (3 years full time or equivalent)
  \item Bachelor degree but not honours
  \item Certificate I
  \item Certificate II
  \item Certificate III
  \item Certificate IV
  \item Certificate of unknown level
  \item Doctorate
  \item Diploma NFI
  \item Diploma (2 years full time or equivalent)
  \item Graduate Certificate
  \item Graduate Diploma 
  \item Honours 
  \item Masters 
  \item Other 
\end{itemize}  
  
Completed re-education after 2017 based on both highest attainment and detailed qualifications (\textit{refllt}) records whether the individual has completed re-education after 2017 based on both the variables \textit{rehllt} and \textit{redllt}. When either of these variables has a value of 1, this variable will take on the value of 1. 

\emph{Timing of Education Completion}

Year of first re-education completion records the year of the first reported instance of re-education completion as provided by the detailed qualification variables and include the following categories:
\begin{itemize}
  \item 2002 (\textit{p\textunderscore{}rcom1})
  \item 2003 (\textit{p\textunderscore{}rcom2})
  \item 2004 (\textit{p\textunderscore{}rcom3})
  \item 2005 (\textit{p\textunderscore{}rcom4})
  \item 2006 (\textit{p\textunderscore{}rcom5})
  \item 2007 (\textit{p\textunderscore{}rcom6})
  \item 2008 (\textit{p\textunderscore{}rcom7})
  \item 2009 (\textit{p\textunderscore{}rcom8})
  \item 2010 (\textit{p\textunderscore{}rcom9})
  \item 2011 (\textit{p\textunderscore{}rcom10})
  \item 2012 (\textit{p\textunderscore{}rcom11})
  \item 2013 (\textit{p\textunderscore{}rcom12})
  \item 2014 (\textit{p\textunderscore{}rcom13})
  \item 2015 (\textit{p\textunderscore{}rcom14})
  \item 2016 (\textit{p\textunderscore{}rcom15})
  \item 2017 (\textit{p\textunderscore{}rcom16})
  \item 2018 (\textit{p\textunderscore{}rcom17})
  \item 2019 (\textit{p\textunderscore{}rcom18})
\end{itemize}  
  
Locus of control in 2003 (\textit{p\textunderscore{}cotrl}) records the transformed composite score\footnote{See Buddlemeyer and Powdthavee (2015) for details of the transformation.} for locus of control items reported by the individual in 2003, the first year in HILDA for which this information becomes available. The transformation results in a variable that is ranged between 7 and 49. Locus of control measures the degree to which individuals attribute outcomes to internal versus external factors or the extent their welfare are in their own control compared to external circumstances. A higher score indicates having a more external locus of control, which is considered as a favourable personality trait. 

Frequency of reading books in 2012 (\textit{p\textunderscore{}rdf}) records the frequency the individual reads books in 2012, the first year in HILDA for which this information becomes available. This is a proxy for love of learning\footnote{HILDA contains a question on reading newspapers and magazines but we feel that reflects a care for or understanding of current issues more than a love of learning.}. This is a categorical variable encompassing the following frequencies:
\begin{itemize}
  \item Every day or most days (\textit{p\textunderscore{}rdf1}) 
  \item Several times a week (\textit{p\textunderscore{}rdf2}) 
  \item About once a week (\textit{p\textunderscore{}rdf3}) 
  \item 2 or 3 times a month (\textit{p\textunderscore{}rdf4}) 
  \item About once a month (\textit{p\textunderscore{}rdf5}) 
  \item Less than once a month (\textit{p\textunderscore{}rdf6}) 
  \item Never (\textit{p\textunderscore{}rdf7}) 
\end{itemize}

\clearpage

\section{Panel Sample: Sensitivity Analysis}

\subsection{Sample Selection}

\textbf{Treated sample:} For any person in HILDA who ever reported \textit{starting} a degree (determined by taking a person who switches from reporting “not currently studying” in one wave to “currently studying” in the next wave) and/or \textit{completing} a degree, we select their first study event as a treatment observation if it satisfies three other conditions.

They are: (1) at least 21 years old in the starting year of study\footnote{Note that we expanded the age range in this sensitivity analysis to ensure sufficient treatment observations for the estimation of the treatment outcome surfaces.}, (2) they were present in the two years before the start of study (in order to have information on their feature values), (3) there were not currently studying in any of the two years before the starting year of further study (to avoid reverse-causation issues), (4) they completed their further degree and (5) they were present in the survey and had a non-missing outcome 4 years after the start of study. 

If a study event does not satisfy these conditions, we look to the next study event that satisfies these conditions or (if unavailable) delete the person from our sample completely. Conditions (3) and (5) together mean that we analyse a sample of individuals who started their degrees anytime between 2003 and 2015.

In our treated group, 1,814 individuals \textbf{started and completed a further educational degree}. 

\textbf{Control sample:} These are those who had never started re-education throughout HILDA. From these control observations, we assign a time stamp to them for the year the control person theoretically started to study. We do this for every year from 2003 to 2019. This implies that never re-educated individuals can be duplicated and used multiple times. For example, if a control individual is observed throughout the years 2001 to 2016, then they will be a control for the separate treated individuals that started re-education in 2003, in 2004, 2005 and up to 2017 i.e. the control individual will be duplicated 15 times.

There are 60,945 control observations i.e. individuals who never completed a further degree. However, as described above, these are non-unique observations in the sense that a control individual can be duplicated up to 15 times. 

\subsection{Variable description}

\subsubsection{Outcome Variables}
Weekly earnings from main job in fourth year after the individual started their re-education (\textit{f4\textunderscore{}wscmei}) records the weekly earnings from the main job for the individual in the fourth year after the individual started their re-education. 

\subsubsection{Treatment Variables: Re-education}
Re-education completion based on both highest attainment and detailed qualifications (\textit{redufl}) records whether the individual has completed re-education based on a comparison of the highest education attainment and the number of qualifications gained across waves 1 and 17. If either of these have gone up, \textit{redufl} takes a value of 1 and 0 otherwise. 

\subsubsection{Input Variables}

\underline{Characteristics in the Year Prior to Re-education Start}

\emph{Demographics}

Gender (\textit{p1\textunderscore{}hgsex}) records the gender of the individual. The value of 1 denotes males whereas the value 2 denotes females. 

Age (\textit{p1\textunderscore{}hgage}) records the age of the individual in the year prior to re-education start.
 
Country of birth (\textit{p1\textunderscore{}anbcob}) records whether or not an individual was born in:
\begin{itemize}
  \item Australia (value=1) 
  \item English speaking countries (value=2)
  \item Non-English speaking countries (value=3)
\end{itemize}  

Indigenous Status (\textit{p1\textunderscore{}anatsi}) records whether or not an individual is:
\begin{itemize}
  \item Not indigenous (value=1)
  \item Aboriginal (value=2)
  \item Torres Islander (value=3)
  \item Both Aboriginal and Torres Islander (value=4) 
\end{itemize}  

Poor English speaking abilities (\textit{p1\textunderscore{}poeng}) records whether the individual has poor English speaking abilities in the year prior to re-education start. 

State of residence (\textit{p1\textunderscore{}hhstate}) records the state of residence of the individual in the year prior to re-education start:
\begin{itemize}
  \item NSW (value=1)
  \item VIC (value=2)
  \item QLD (value=3)
  \item SA (value=4)
  \item WA (value=5)
  \item TAS (value=6)
  \item NT (value=7)
  \item ACT (value=8) 
\end{itemize}  

Remoteness (\textit{p1\textunderscore{}hhsos}) records whether, in the year prior to re-education start, the individual lives in:
\begin{itemize}
  \item A major city (value=0)
  \item An inner region (value=1)
  \item Outer and remote areas (value=2) 
  \item migratory in nature (value=3)
\end{itemize}  

Marital status (\textit{p1\textunderscore{}mrcurr}) records whether, in the year prior to re-education start, the individual was:
\begin{itemize}
  \item Married (value=1)
  \item De facto (value=2)
  \item Separated (value=3)
  \item Divorced (value=4)
  \item Widowed (value=5)
  \item Single and never been married (value=6)
\end{itemize}  

Household size (\textit{p1\textunderscore{}hhsize}) records the total number of individuals living in the same household as the individual (including the individual) in the year prior to re-education start.

Sexual orientation (\textit{p1\textunderscore{}lgtb}) records that the individual’s sexual orientation is not heterosexual.  The variable is constructed from the Sexual Identity question that is only asked in waves 12 and 16. We combine answers from both waves to create a binary indicator for the individual ever reporting a sexual identity that is not heterosexual, treating sexual orientation as a fixed trait for a given individual. 
  
\emph{Parental Status}

Number of dependents (\textit{p1\textunderscore{}totalkids}) records the number of children under 15 the individual had in the household in the year prior to re-education start. 

Having children (\textit{p1\textunderscore{}anykid}) records the individual had any dependents in the household in the year prior to re-education start. 

Children under 5 (\textit{p1\textunderscore{}kidu5}) records the individual had children under 5 in the household in the year prior to re-education start. 

Age of youngest (\textit{p1\textunderscore{}rcyng}) records the age of the youngest children living with the respondent in the year prior to re-education start (including adult children). 

\emph{Physical Health}

Severity of health conditions (\textit{p1\textunderscore{}disdeg}) records whether, in the year prior to re-education start, the individual had:
\begin{itemize}
  \item No health conditions (value=0)
  \item A mild condition (value=1)
  \item A moderate condition (value=2)
  \item A severe condition (value=3)
\end{itemize}  
  
\emph{Labour Force Variables}

Labour market status (\textit{p1\textunderscore{}lfs}) records whether the individual was:
\begin{itemize}
  \item Employed (value=1)
  \item Unemployed (value=2)
  \item Not in the labour market (value=3)
\end{itemize} 

Extent of working hour match with preferences (\textit{p1\textunderscore{}whpref}) records whether, in the year prior to re-education start, the match between the individual’s total weekly working hours across all jobs and their preferred number of working hours made them:
\begin{itemize}
  \item Underemployed by at least 4 hours a week (value=1)
  \item Roughly Matched: Preferred and Actual Hours Worked differ by less than 4 hours a week (value=2)
  \item Overemployed by at least 4 hours a week (value=3)
\end{itemize}  

Employee type (\textit{p1\textunderscore{}emptype}) records whether, in the year prior to re-education start, the individual was:
\begin{itemize}
  \item An employee (value=1)
  \item An employee of own business (value=2)
  \item Self Employed (value=3)
  \item Unpaid family worker (value=4)
\end{itemize}  

Contract type (\textit{p1\textunderscore{}contype}) records whether, in the year prior to re-education start, the individual was:
\begin{itemize}
  \item On a fixed term contract (value=1)
  \item On a casual contract (value=2)
  \item On a permanent contract (value=3)
  \item On other types of contracts  (value=4)
\end{itemize}  

Occupation (\textit{p1\textunderscore{}occ}) records whether, in the year prior to re-education start, the individual was working as:
\begin{itemize}
  \item Armed forces (value=0)
  \item Legislators, Senior Officials and Managers (value=1)
  \item Professionals (value=2)
  \item Technicians and Associate Professionals (value=3)
  \item Clerks (value=4)
  \item Service Workers and Shop and Market Sales Workers (value=5)
  \item Skilled Agriculture and Fishery Workers (value=6)
  \item Craft and Related Trades Workers (value=7)
  \item Plant and Machine Operators and Assemblers (value=8)
  \item Elementary Occupations (value=9)
\end{itemize}  

Union membership (\textit{p1\textunderscore{}union}) records whether the individual was a union member in the year prior to re-education start.

Real household income (\textit{p1\textunderscore{}rhdi}) records the real value of the individual’s total household income indexed at 2012 price levels and adjusted for household size in the year prior to re-education start.

Partner labour force status (\textit{p1\textunderscore{}plfs}) records whether, in the year prior to re-education start, the individual:
\begin{itemize}
  \item Had no partner or no resident partner (value=0)
  \item Had a partner who was employed (value=1)
  \item Had a partner who was unemployed (value=2)
  \item Had a partner who was not in the labour force (value=3)
\end{itemize}  

Years in paid work (\textit{p1\textunderscore{}ehtjb})  records the total number of years in paid work the individual has spent in the year prior to re-education start.

Percent finding as least as good a job (\textit{p1\textunderscore{}jbmpgj})  records, for employees, the percentage that they will find as least as good a job as they currently have in their own estimation in the year prior to re-education start

Occupational scale (\textit{p1\textunderscore{}jbmo6s}) records the Australian Socioeconomic Index 2006 ranking of the individual’s occupation in the year prior to re-education start. It ranges from 0 to 100, with higher scores indicating higher occupational status. 

Tenure with employer (\textit{p1\textunderscore{}jbempt}) records the total years spent with the current employer for the individual in the year prior to starting re-education.

\emph{Parental information}

Father’s country of birth (\textit{p1\textunderscore{}fcob}) records whether or not the individual’s father was born in:
\begin{itemize}
  \item Australia (value=1) 
  \item English speaking countries (value=2)
  \item Non-English speaking countries or indigenous (value=3)
\end{itemize}  

Mother’s country of birth (\textit{p1\textunderscore{}mcob}) records whether or not the individual’s mother was born in:
\begin{itemize}
  \item Australia (value=1)
  \item English speaking countries (value=2)
  \item Non-English speaking countries or indigenous (value=3)
\end{itemize}  

Father’s education records whether the individual’s father’s highest education, as reported in 2005, was:
\begin{itemize}
  \item None (value=1)
  \item Primary (value=2)
  \item Below secondary (value=3)
  \item Secondary (value=4)
  \item Post-secondary, non-university (value=5)
  \item Post-secondary, university (value=6)
\end{itemize}  

Mother’s education records whether the individual’s mother’s highest education, as reported in 2005, was:
\begin{itemize}
  \item None (value=1)
  \item Primary (value=2)
  \item Below secondary (value=3)
  \item Secondary (value=4)
  \item Post-secondary, non-university (value=5)
  \item Post-secondary, university (value=6)
\end{itemize}  

Father undertaken post-school qualification through employer or non-tertiary means (\textit{p\textunderscore{}fpsm}) records whether the individual’s father had undertaken his highest qualification through employers or other channels other than tertiary education, as reported in 2005. 

Mother undertaken post-school qualification through employer or non-tertiary means (\textit{p\textunderscore{}mpsm}) records whether the individual’s mother had undertaken his highest qualification through employers or other channels other than tertiary education, as reported in 2005. 

Father’s Employment at age 14 (\textit{p1\textunderscore{}femp}) records whether the individual’s father was working or not when they were aged 14. 

Mother’s Employment at age 14 (\textit{p1\textunderscore{}memp}) records whether the individual’s mother was working or not when they were aged 14.

Father substantially unemployed growing up (\textit{p1\textunderscore{}fsue}) records whether the individual’s father had been unemployed or 6 months or more when they were aged 14. 

Father’s Occupation (\textit{p1\textunderscore{}focc}) records whether at age 14 the individual’s father was last known working as:
\begin{itemize}
  \item Armed forces (value=0)
  \item Legislators, Senior Officials and Managers (value=1)
  \item Professionals (value=2)
  \item Technicians and Associate Professionals (value=3)
  \item Clerks (value=4)
  \item Service Workers and Shop and Market Sales Workers (value=5)
  \item Skilled Agriculture and Fishery Workers (value=6)
  \item Craft and Related Trades Workers (value=7)
  \item Plant and Machine Operators and Assemblers (value=8)
  \item Elementary Occupations (value=9)
\end{itemize}  

Mother’s Occupation (\textit{p1\textunderscore{}mocc}) records whether at age 14 the individual’s mother last known working as:
\begin{itemize}
  \item Armed forces (value=0)
  \item Legislators, Senior Officials and Managers (value=1)
  \item Professionals (value=2)
  \item Technicians and Associate Professionals (value=3)
  \item Clerks (value=4)
  \item Service Workers and Shop and Market Sales Workers (value=5)
  \item Skilled Agriculture and Fishery Workers (value=6)
  \item Craft and Related Trades Workers (value=7)
  \item Plant and Machine Operators and Assemblers (value=8)
  \item Elementary Occupations (value=9)
\end{itemize}  

\emph{Income Support}

On income support (\textit{p1\textunderscore{}onis}) records the individual was on income support in the year prior to starting re-education

On Newstart (\textit{p1\textunderscore{}onnsa}) records the individual was on Newstart Allowance in the year prior to starting re-education

On Age Pension (\textit{p1\textunderscore{}onap}) records the individual was on Age Pension in the year prior to starting re-education

On DSP (\textit{p1\textunderscore{}ondsp}) records the individual was on Disability Support Pension in the year prior to starting re-education

On Carer Payment (\textit{p1\textunderscore{}oncp}) records the individual was on Carer Payment in the year prior to starting re-education

On Widow Allowance/Wife Pension (\textit{p1\textunderscore{}onww}) records the individual was on Widow Allowance/Wife Pension in the year prior to starting re-education

On Youth Allowance (\textit{p1\textunderscore{}onya}) records the individual was on Youth Allowance in the year prior to starting re-education

On Mature Age Allowance (\textit{p1\textunderscore{}onma}) records the individual was on Mature Age Allowance in the year prior to starting re-education

On Mature Age Partner Allowance (\textit{p1\textunderscore{}onmap}) records the individual was on Mature Age Partner Allowance in the year prior to starting re-education

On Ab/Austudy (\textit{p1\textunderscore{}onsdy}) records the individual was on Ab/Austudy in the year prior to starting re-education

On Bereavement Allowance (\textit{p1\textunderscore{}onba}) records the individual was on Bereavement Allowance in the year prior to starting re-education

On Sickness Allowance/Speical Benefits (\textit{p1\textunderscore{}onsab}) records the individual was on Sickness Allowance/Speical Benefits in the year prior to starting re-education

On Partner Allowance (\textit{p1\textunderscore{}onpa}) records the individual was on Partner Allowance in the year prior to starting re-education

On Parenting Payments (\textit{p1\textunderscore{}onpp}) records the individual was on Parenting Payments in the year prior to starting re-education

\emph{Housing situation}

Mortgage balance (\textit{p1\textunderscore{}hsmgowe}) records the amount still owing on the mortgage that the individual had in the year prior to re-education start. For those without a mortgage or not home owner, the mortgage balance is set to 0. 

Non home owners (\textit{p1\textunderscore{}renter}) records whether the individual was renting or not living in their own homes in the year prior to re-education start.

\emph{Prior Year Outcomes}

Weekly income from all jobs (\textit{p1\textunderscore{}earning}) records the weekly earnings from all jobs for the individual in the year prior to the individual starting their re-education. 

Weekly income from main job (\textit{p1\textunderscore{}wscmei}) records the weekly earnings from the main job for the individual in the year prior to the individual starting their re-education.
 
Weekly working hours (\textit{p1\textunderscore{}wkhr}) records the total number of hours the individual works in all jobs in a week on average in the year prior to the individual started their re-education. Working hours are set to 0 for those not working. 

Real hourly wage (\textit{p1\textunderscore{}rlwage}) records the real hourly wage of the individual in the year prior to the individual starting their re-education, indexed at 2012 price levels. Hourly wages are set to 0 for those not working and set to missing for those reporting working more than 100 hours a week. All wages have then been adjusted up by \$1 to preserve sample size for the logarithm transformation. 

Log hourly wage (\textit{p1\textunderscore{}lnwage}) records the log of  \textit{p1\textunderscore{}rlwage}. 

Mental health (\textit{p1\textunderscore{}ghmh}) records the transformed mental health scores from the aggregation of mental health items of the SF-36 Health Survey, as reported by the individual in the year prior to the individual started their re-education. It ranges from 0 to 100, with higher scores indicating better mental health.   

Life satisfaction (\textit{p1\textunderscore{}losat}) records the life satisfaction score reported by the individual in the year prior to the individual started their re-education. It ranges from 0 to 10, with higher scores indicating higher life satisfaction. 

\underline{Delta variables}

For all the variables described in the preceding section titled Characteristics in the Year Prior to Re-education Start, we create a further set of change or delta variables. Specifically, each delta variable is the subtracting of the value of a given characteristic in the two years prior to starting re-education from the value of this characteristic in the year prior to re-education start. 

All delta variables are denoted by the d\textunderscore{} prefix. 

\underline{Education-related variables}

Level of re-education completed: Bachelor and above (\textit{bachab}) records whether the individual had completed re-education at bachelor and above. The variable is set to 0 for the control group and missing for those who had completed certificates. 

Level of re-education completed: Below Bachelor (\textit{bbach}) records whether the individual had completed re-education that is below bachelor level. The variable is set to 0 for the control group and missing for those who had completed a bachelor or higher qualification. 

Main field of study: technical degree (\textit{techdeg}) records whether the individual’s main field of study was a technical degree. The variable is set to 0 for the control group and missing for those whose main field of study was a qualitative degree. Technical degrees include:
\begin{itemize}
  \item Natural and physical sciences
  \item Information technology
  \item Engineering and related technologies
  \item Architecture and building
  \item Agriculture, environment and related studies
  \item Medicine
  \item Nursing
  \item Other health-related (e.g. Pharmacy, Dental studies, Rehabilitation therapies, Optical science, Veterinary studies) 
  \item Management and commerce (e.g. Accounting, Business, Sales and marketing, Banking and finance, Office studies) 
  \item Law
\end{itemize}  

Main field of study: qualitative degree (\textit{qualdeg}) records whether the individual’s main field of study was a qualitative degree. The variable is set to 0 for the control group and missing for those whose main field of study was a technical degree. Qualitative degrees include:
\begin{itemize} 
  \item Education
  \item Society and culture (e.g. Economics, Political science, Social work, History, Psychology, Languages, Religion, Sport)
  \item Creative arts
  \item Food, hospitality and personal services
  \item Other
\end{itemize}  

Study duration (\textit{fsddur}) records the total number of waves an individual had spent studying from the start of their first study event counted in our sample. 

Starting Study intensity (\textit{csftsd}) records whether the individual was studying full time or not when they started their re-education. 

Finishing Study intensity (\textit{fsftsd}) records whether the individual was studying full time or not when they completed their re-education. 

\underline{Other variables}

Number of waves in HILDA (\textit{numwave}) records the number of waves in which the respondent has submitted a valid response for the HILDA survey. 

\subsubsection{Variables that are not included in the model}

The unique person identifier (\textit{xwaveid})

Wave started re-education (\textit{icswave}) 

Wave completed re-education (\textit{ifswave}) 

Control group indicator (\textit{control}) 

Started but did not complete re-education between 2003-2017 (\textit{ncomp}) 

Starting year of re-education imputed (\textit{impute}) is a binary indicator for individuals for which we observe their re-education completion but they never reported ever starting re-education and so we had to impute a starting wave for these individuals.  

Started re-educaton in wave 2018/19 (\textit{latestart}) is an indicator for those individuals who had started their re-education in 2018 or 2019.

\end{document}